\documentclass[fleqn,usenatbib]{mnras}
\usepackage{newtxtext,newtxmath}

\usepackage[T1]{fontenc}

\DeclareRobustCommand{\VAN}[3]{#2}
\let\VANthebibliography\thebibliography
\def\thebibliography{\DeclareRobustCommand{\VAN}[3]{##3}\VANthebibliography}

\usepackage{graphicx}	
\usepackage{amsmath}	
\usepackage[table]{xcolor}
\usepackage{orcidlink}

\newcommand{\halpha}{H$\alpha$}

\def\fAGN{f_{\rm AGN}}

\def\fEdd{\lambda_{\rm Edd}}
\def\fduty{f_{\rm duty}}
\def\Mhalo{M_{\rm H}}
\def\Mstar{M_{\star}}
\def\Mdyn{M_{\rm dyn}}

\def\Mbh{M_{\rm BH}}
\def\Lbol{L_{\rm bol}}
\def\Luv{L_{\rm UV}}
\def\Muv{M_{\rm UV}}
\def\Auv{A_{\rm UV}}
\def\Av{A_{\rm V}}

\def\SMF{\Phi(\Mstar)}
\def\BHMF{\Phi(\Mbh)}
\newcommand{\msun}{{\rm M}_{\odot}}

\defcitealias{Reines_2015}{RV15}
\defcitealias{Pacucci2023_LRDs}{P23}
\defcitealias{2025LiTipOf}{L25}
\defcitealias{Kokorev2024_PhotoConcensusOfLRDs}{K+24}
\defcitealias{2025GreenearXiv250905434G}{G+25}

\title[High-z demography of AGN]{Too many or too massive? Investigating the high-$z$ demography of active SMBHs from JWST}

\author[]{
Daniel Roberts$^{\orcidlink{0009-0009-7662-0445},1}$\thanks{E-mail: d.m.roberts@soton.ac.uk},
Francesco Shankar$^{\orcidlink{0000-0001-8973-5051},1}$,
Vieri Cammelli$^{2,3,4,5,6}$,
Fabio Fontanot$^{4,6}$,
\newauthor
~Alessandro Trinca$^{7,8,9}$,
Laura Bisigello$^{10}$,
Elena Dalla Bonta$^{11,10,12}$,
Hao Fu$^{\orcidlink{0009-0002-8051-1056},13,1}$,
Roberto Gilli$^{14}$,
\newauthor
~Andrea Grazian$^{10}$,
Luca Graziani$^{9,8,15}$,
Andrea Lapi$^{16,17,18,19}$,
Nicola Menci$^{8}$,
Jan Scholtz$^{20,21}$,
\newauthor
~Karthik Mahesh Varadarajan$^{1}$
\\
$^{1}$School of Physics and Astronomy, University of Southampton, Highfield, Southampton, SO17 1BJ, UK\\
$^{2}$Department of Physics, Informatics \& Mathematics, University of Modena \& Reggio Emilia, via G. Campi 213/A, 41125, Modena, Italy\\
$^{3}$Dipartimento di Fisica, Sezione di Astronomia, Universit{\'a} degli Studi di Trieste, via G.B. Tiepolo 11, I-34131, Trieste, Italy\\
$^{4}$INAF - Astronomical Observatory of Trieste, via G.B. Tiepolo 11, I-34143, Trieste, Italy\\
$^{5}$Department of Space, Earth \& Environment, Chalmers University of Technology, SE-412 96 Gothenburg, Sweden\\
$^{6}$IFPU - Institute for Fundamental Physics of the Universe, Via Beirut 2, I-34151 Trieste, Italy\\
$^{7}$Como Lake Center for Astrophysics, DiSAT, Universit{\`a} degli Studi dell’Insubria, via Valleggio 11, 22100, Como, Italy\\
$^{8}$INAF - Osservatorio Astronomico di Roma, Via Frascati 33, 00040 Monte Porzio Catone, Italy\\
$^{9}$Dipartimento di Fisica, ``Sapienza'' Universit{\`a} di Roma, Piazzale Aldo Moro 2, 00185 Roma, Italy\\
$^{10}$INAF - Osservatorio Astronomico di Padova, Vicolo dell'Osservatorio 5, I-35122 Padova, Italy\\
$^{11}$Dipartimento di Fisica e Astronomia ``G. Galilei'', Universit\'{a} di Padova, Vicolo dell’Osservatorio 3, I-35122 Padova, Italy\\
$^{12}$Jeremiah Horrocks Institute, University of Central Lancashire, Preston, PR1 2HE, UK\\
$^{13}$Center for Astronomy and Astrophysics and Department of Physics, Fudan University, Shanghai 200438, People’s Republic of China\\
$^{14}$Osservatorio di Astrofisica e Scienza dello Spazio di Bologna Via Gobetti 93/3, I-40129 Bologna, Italy\\
$^{15}$INFN - Sezione Roma1, Dipartimento di Fisica, ``Sapienza'' Universit{\`a} di Roma, Piazzale Aldo Moro 2, 00185, Roma, Italy\\
$^{16}$SISSA, Via Bonomea 265, 34136 Trieste, Italy\\
$^{17}$IFPU, Via Beirut 2, 34014 Trieste, Italy\\
$^{18}$INFN-Sezione di Trieste, via Valerio 2, 34127 Trieste, Italy\\
$^{19}$IRA-INAF, Via Gobetti 101, 40129 Bologna, Italy\\
$^{20}$Kavli Institute for Cosmology, University of Cambridge, Madingley Road, Cambridge, CB3 0HA, UK\\
$^{21}$Cavendish Laboratory, University of Cambridge, 19 JJ Thomson Avenue, Cambridge CB3 0HE, UK\\
}

\date{Accepted XXX. Received YYY; in original form ZZZ}
\pubyear{\the\year{}}

\begin{document}

\label{firstpage}
\pagerange{\pageref{firstpage}--\pageref{lastpage}}
\maketitle


\begin{abstract}
Recent JWST observations have unveiled a numerous population of low-luminosity active galactic nuclei (AGN) at $4\lesssim z\lesssim10$, with space densities roughly an order of magnitude above pre-JWST estimates, and many of these AGN have masses orders of magnitude above the local black hole mass–stellar mass ($\Mbh-\Mstar$) scaling relations. We investigate the consistency of these observations within a data-driven framework that links the galaxy stellar mass function to the supermassive black hole (SMBH) mass function and AGN luminosity functions using different $\Mbh-\Mstar$ relations and the observed Eddington-ratio distribution. By comparing our predictions against observed AGN luminosity functions at $z\sim5.5$ we find that observations can be reproduced either by highly-elevated $\Mbh-\Mstar$ relations paired with low duty cycles ($\fAGN \sim 0.08$), or moderate relations with higher duty cycles ($\fAGN \sim 0.5$). Through the So\l{}tan argument, we find that $\Mbh-\Mstar$ relations that are modestly above the local relation for AGN produce consistency between multiple tracers of the SMBH demography at $z\sim5.5$, while more extreme normalisations would require a weakly-evolving luminosity function at $z\geq5.5$. Continuity-equation modelling shows that initially high $\Mbh-\Mstar$ relations predict a strong two-phase evolutionary scenario and very steep low-mass SMBH mass functions in tension with several current estimates, while more moderate relations generate local SMBH mass functions in better agreement with present determinations and near-constant scaling relations. Our results favour a scenario where SMBHs at $z \sim 5$ on average lie modestly above local AGN scaling relations, with elevated but physically plausible duty cycles. Future wide-field clustering and demographic studies will help break the remaining degeneracies between SMBH scaling relations and AGN duty cycles at early cosmic times. 

\end{abstract}




\section{Introduction}

Observational evidence indicates that supermassive black holes (SMBHs) lie at the centre of most -- if not all -- massive galaxies, with masses up to several billion times the mass of the Sun. These SMBHs are typically identified during phases when they are actively accreting gas, shining as active galactic nuclei \citep[AGN;][]{1969LyndenBellNatur.223..690L,Soltan1982_QuasarMasses}. AGN have been observed throughout cosmic time and into the epoch of reionisation, with the most distant quasars observed at $z\gtrsim7.5$ \citep[e.g.][]{2018Natur.553..473B,2020ApJ...897L..14Y,2021ApJ...907L...1W}. Up until recently, observations of high-$z$ AGN were biased to the most luminous, and most massive quasars \citep[][]{2020InayoshiARA&A..58...27I,2023FanARA&A..61..373F}. However, the {\it James Webb Space Telescope} \citep[JWST;][]{JWST} has not only pushed back this frontier in the hunt for quasars to $z\sim10$ \citep[e.g. UHZ1;][]{2024NatAs...8..126B}, but extended the observable parameter space for AGN in both mass and luminosity to $\sim2$ orders of magnitude lower than previous quasar surveys at the same redshift, facilitating the discovery of a numerous population of lower-luminosity AGN at $z\gtrsim4$ \citep[e.g.][]{2023ApJ...959...39H,maiolino2024jadesdiversepopulationinfant,2023ScholtzarXiv231118731S,2025LabbeApJ...978...92L}. This population of lower-luminosity AGN is composed of both broad line (BL, or type-1) AGN, primarily identified from the broad Balmer emission \citep[e.g.][]{2023ApJ...959...39H,maiolino2024jadesdiversepopulationinfant,2025IgnasarXiv250403551J,Greene2024_LRDs,Mathee2024_LRDs,2024arXiv240403576K}, as well as narrow line (NL, or type-2) AGN, identified from high-ionisation emission lines \citep[e.g.][]{2023BrinchmannMNRAS.525.2087B,2023ScholtzarXiv231118731S,2024ChisholmMNRAS.534.2633C,2024UblerMNRAS.531..355U,2024MazzolariarXiv240815615M}, thereby, allowing for a more complete census of the SMBH population across the past $\sim13~{\rm Gyr}$.

At low redshift, the masses of the central SMBHs ($\Mbh$) display tight correlations with a number of host-galaxy properties, such as stellar mass of the spheroidal component \citep[$M_{\star,{\rm bulge}}$; ][]{1995KormendyARA&A..33..581K,Magorrian1998,2004HaringApJ...604L..89H}, the stellar velocity dispersion \citep[$\sigma_{\star}$; ][]{Ferrarese2000_MbhSigma,2000GebhardtApJ...539L..13G,Gultekin2009_MbhSigma}, and total stellar mass \citep[$\Mstar$; ][hereafter \citetalias{Reines_2015}]{Reines_2015}. The correlation of $\Mbh$ with properties that extend beyond the SMBH's sphere of influence has often been interpreted as evidence for a degree of interplay, or coevolution, between the SMBH and its host \citep[see][for a comprehensive review]{KormendyAndHo2013}. Many observational studies have suggested there is little -- if any -- evolution in the SMBH--galaxy scaling relations with redshift \citep[e.g.][]{2011CisternasApJ...741L..11C,2013SalvianderApJ...764...80S,2013SchrammApJ...767...13S,2021IzumiApJ...914...36I,sun2024evidencesignificantevolutionmbulletm,2025TanakaMstMbhApJ...979..215T}. For example, \citet{2013SalvianderApJ...764...80S} find no evidence of evolution in the $\Mbh-\sigma_{\star}$ relation of quasars at $z\lesssim1.2$ \citep[in agreement with theoretical works such as][]{SilkAndReese1998_QuasarsAndGalFormation,2009ShankarMbhSigApJ...694..867S,2018MNRAS.478.5063H}. Whereas \citet{sun2024evidencesignificantevolutionmbulletm} find the $\Mbh-\Mstar$ relation to be approximately constant up to $z\sim4$ \citep[consistent with theoretical works such as][]{2017MNRAS.464.2840A,2020ShankarMNRAS.493.1500S,2023MNRAS.518.2123Z,2024ApJ...976....6Z,2025HaoBHQarXiv251026305F}. Beyond this, measurements of the host dynamical mass ($\Mdyn$) determined from the gas dynamics have been used to probe the SMBH--galaxy connection into the epoch of reionisation \citep[e.g.][]{2013WangApJ...773...44W,2015WillottApJ...801..123W,2016VenemansApJ...816...37V,2018DecarliApJ...854...97D,2020PensabeneA&A...637A..84P}. \citet{2015WillottApJ...801..123W} and \citet{2017WillottApJ...850..108W} find $z\gtrsim6$ quasars to be consistent with the local relation, indicating that the $\Mbh-\Mdyn$ relation may have been in place since $z\sim7$ \citep[][]{2021IzumiApJ...914...36I}. More recently, \citet{2024TripodiHYPERIONA&A...689A.220T} using the HYPERION sample \citep[][]{2023ZappacostaHYPERIONA&A...678A.201Z} have suggested that high-$z$ quasars on average lie above the local $\Mbh-\Mdyn$ relation, possibly hinting that, at least in quasars, SMBH growth may precede that of its host.

As the new discoveries by JWST extend the mass range of SMBHs we can observe in this epoch, and with the ability to detect stellar light in the epoch of reionisation \citep[e.g.][]{2023DingStellarLightNatur.621...51D}, a pool of evidence is forming indicating that the $\Mbh-\Mstar$ relation may deviate significantly from the local estimates at $z\gtrsim4$. BH masses obtained from single-epoch virial estimators using the broad Balmer lines have found the high-$z$ AGN population to be over-massive with respect to local estimates of the $\Mbh-\Mstar$ relation, with BH-to-stellar mass ratios of order $\Mbh/\Mstar\sim0.1$, which is approximately two orders of magnitude above the local relation of \citetalias{Reines_2015}, while also displaying larger scatter in the $\Mstar-\Mbh$ plane \citep[][hereafter \citetalias{Pacucci2023_LRDs}]{Pacucci2023_LRDs}. This tendency for SMBHs to be systematically above the local relation has been observed in both type-1 \citep[e.g.][]{2023Ubler,maiolino2024jadesdiversepopulationinfant,2024Natur.636..594J,2023ApJ...959...39H} and type-2 AGN \citep[e.g.][]{2024ChisholmMNRAS.534.2633C}. Yet, these objects appear to lie along the local $\Mbh-\sigma_{\star}$ relation \citep[][]{maiolino2024jadesdiversepopulationinfant,2024ChisholmMNRAS.534.2633C}, possibly corroborating the findings of residual analysis that the $\Mbh-\sigma_{\star}$ relation is more fundamental than the $\Mbh-\Mstar$ relation \citep[][]{2007BernardiResidualsApJ...660..267B,Shankar_2016,2017MNRAS.468.4782B,2020FrP.....8...61M,2025arXiv250317478N,2025ShankarMNRAS.tmp..713S}. The linchpin of these mass estimates is the assumption that the observed widths of the broad Balmer lines are tracing the motion of the broad line region (BLR) clouds. If instead the broad lines are the result of scattering \citep[e.g.][]{2025RusakovarXiv250316595R,2025NaiduBHstararXiv250316596N,2026SneppenarXiv260118864S}, then the tension with the local $\Mbh-\Mstar$ relation may be alleviated, but a new tension with the local $\Mbh-\sigma_{\star}$ relation, which is believed to be more fundamental, would be introduced.

This significant deviation from the local $\Mbh-\Mstar$ relation may be suggesting that the growth of SMBHs outpaces that of their host galaxies. However, it is still unclear whether these potentially ultra-massive objects, and thus their evolutionary histories, are truly representative of the total population of AGN or just the massive end of the underlying $\Mbh$ distribution at these redshifts \citep[e.g.][]{2025GerisarXiv250622147G,2025BrooksarXiv251119609B}. Some works (e.g. \citealt{2025LiTipOf}, hereafter \citetalias{2025LiTipOf}; \citealt{2025SilvermanarXiv250723066S}; \citealt{2025RenarXiv250902027R}) have suggested the high-$z$ mean $\Mbh-\Mstar$ relation should lie noticeably below the one put forward by \citetalias{Pacucci2023_LRDs} when taking into account the effects of a larger intrinsic scatter and selection biases \citep[][]{2007LauerBias}, while others argue that selection biases alone cannot account for such a large offset with the local $\Mbh-\Mstar$ relation \citep[e.g.][]{2025arXiv250303675S}. 

Yet, it is not only the over-massive nature of these faint AGN that challenge our preconceived notions of the high-$z$ SMBH demography from pre-JWST observations. The number densities inferred are far greater than previously thought. For example, the UV luminosity function (UV LF) of the faint AGN \citep[e.g.][]{maiolino2024jadesdiversepopulationinfant,2023ApJ...959...39H,2025IgnasarXiv250403551J,2023ScholtzarXiv231118731S,2024GrazianApJ...974...84G} in the range $\Muv\gtrsim-20~{\rm mag}$ lies approximately an order of magnitude above extrapolation of the quasar UV LFs of \citet{2020NiidaApJ...904...89N} and \citet{2022GrazianQUBRICsApJ...924...62G,2023GrazianRubiconApJ...955...60G}. However, the increase in AGN number density with respect to previous observations is not necessarily in tension with the X-ray background, as another peculiarity of the faint AGN is that they are both radio \citep[][]{2024MazzolariarXiv241204224M} and X-ray weak \citep[][]{maiolino2024jwstmeetschandralarge} -- possibly intrinsically so -- which in turn may be the result of an underdeveloped corona \citep[][]{2024YueApJ...974L..26Y} or super-Eddington accretion \citep[][]{2024ApJ...976L..24M,2025MadauarXiv250109854M,2024InayoshiXrayarXiv241203653I,2024PacucciApJ...976...96P,2024LambridesarXiv240913047L}.

A particularly curious sub-population of the high-$z$ BL AGN population uncovered by JWST are the ``little red dots'' (LRDs) found at $4\lesssim z\lesssim10$ \citep[e.g.][see \citealt{2025InayoshiLRDReviewarXiv251203130I} for a review]{Mathee2024_LRDs,2023FurtakApJ...952..142F,2025LabbeApJ...978...92L,Kokorev2024_PhotoConcensusOfLRDs,Greene2024_LRDs,akins2024LRDs,2024arXiv240403576K}. These compact red objects are characterised by a ``v-shaped'' spectral energy distribution (SED), with blue rest-frame UV slopes, heavily reddened rest-frame optical slopes, and an inflection point at $\sim4000$ \AA{} (close to ${\rm H}_{\infty}$). The LRDs display broad Balmer lines and narrow [O III]5007~\AA{}, arguing for the presence of a BLR, and have hence been commonly classified as type-1 AGN. The LRDs are also observed to be extremely numerous, composing $\lesssim30\%$ of the total BL AGN population \citep[][]{2025HainlineApJ...979..138H} and lying approximately an order of magnitude above the quasar luminosity function \citep[e.g.][]{Kokorev2024_PhotoConcensusOfLRDs,Greene2024_LRDs,Mathee2024_LRDs,akins2024LRDs}. As in the wider BL AGN population observed by JWST, the LRDs appear significantly over-massive with respect to the local $\Mbh-\Mstar$ relation \citep[e.g.][]{inayoshi2024birthrapidlyspinningovermassive,2025DurodolaApJ...985..169D,2025IgnasDynamicalMassLRDarXiv250821748J} and are weak in both radio \citep[][]{2025LatifA&A...694L..14L,2025PergerA&A...693L...2P} and X-ray bands \citep[][]{2024YueApJ...974L..26Y,2025SacchiarXiv250509669S}. A growing body of evidence points to the SMBHs in these systems being enshrouded in dense gas \citep[][]{2025InayoshiApJ...980L..27I,2025Inayoshi_BBHarXiv250505322I,2025RusakovarXiv250316595R,2025NaiduBHstararXiv250316596N,2025DEugenioarXiv250311752D,2025deGraaffarXiv250316600D,2025DeGraaffarXiv251121820D,2025UmedaarXiv251204208U,2026AsadaarXiv260110573A,2026SneppenarXiv260118864S}, implying lower SMBH masses \citep[e.g.][]{2025NaiduBHstararXiv250316596N} and bolometric luminosities \citep[e.g.][]{2025GreenearXiv250905434G} than previously estimated. However, their exact nature is still debated, with many non-AGN scenarios proposed to explain their atypical properties \citep[e.g.][]{2024PGApJ...968....4P,2024SettonarXiv241103424S,2024BaggenApJ...977L..13B,2025aZwick_LRD_SMSarXiv250722014Z}. Therefore, while not all BL AGN are LRDs, not all LRDs are necessarily BL AGN, particularly in photometrically selected samples.

\citet{2025OJAp....8E...9J} recently suggested that a large fraction of AGN could be heavily obscured at $z\sim6$ by applying the So\l{}tan argument \citep[][]{Soltan1982_QuasarMasses} and making specific assumptions on the SMBH--galaxy scaling relations and galaxy stellar mass function (SMF). However, due to the dearth of these systems identified by JWST at $z<4$, the implications of a potentially large population of obscured AGN on the coevolution of SMBHs and their host galaxies remains unclear. However, there have been some low-$z$ systems identified with similar $\Mbh-\Mstar$ ratios \citep[e.g.][]{2023MezcuaApJ...943L...5M,2024MezcuaApJ...966L..30M}, as well as numerous LRD candidates \citep[e.g.][]{euclidcollaboration2025euclidquickdatarelease} and potential low-$z$ analogues \citep[e.g.][]{juodžbalis2024jadesrosettastone,2024MezcuaApJ...966L..30M,2024MNRAS.533.2948S,2025ApJ...980L..34L,2025BoormanarXiv250508885B,2025LoiaconoarXiv250612141L,2025JiarXiv250723774J,2025RinaldiarXiv250717738R}. These findings tend to suggest that this heavily obscured growth mode becomes less pervasive at $z\lesssim4$ \citep[][]{2025MaarXiv250408032M}, as expected if the LRDs are associated with the first AGN events after BH seed formation \citep[][]{2025arXiv250305537I,2025LoiaconoarXiv250612141L}.

In the meantime, theoretical models have been employed to investigate both the demographics and nature of the high-$z$ BL AGN observed by JWST, putting forward a wide range of hypotheses and predictions \citep[e.g.][]{2024TrincaLRDs,2025PorrasValverderaXiv250411566P,2025VolonteriA&A...695A..33V,2025LaChancearXiv250520439L,2025LaChancearXiv251213957L,2025Vieri,2025DayalPBHSAMarXiv250608116D,2025McClymontTHESANarXiv250613852M,2025QuadriarXiv250505556Q,2025HerreroCarrionarXiv251110725H}. Some models are able to simultaneously reproduce the over-massive nature and high space densities of the observed AGN by invoking mechanisms such as episodic super-Eddington accretion \citep[e.g. CAT][]{2024TrincaLRDs}, but most are not \citep[e.g.][]{2025PorrasValverderaXiv250411566P,2025Vieri}. 

The large masses and high number densities of the BL AGN population observed by JWST pose challenges to our understanding of the formation and evolution of SMBHs, both in the accumulation of such SMBH mass density in the first $\sim1~{\rm Gyr}$ of cosmic time and in defining a holistic model of the SMBH population from high to low redshifts. 

The aim of this work is twofold: 1) to probe the self-consistency and validity of such extreme high-$z$ mass and number density measurements from JWST by combining distinct data sets (namely the SMF, SMBH mass function, BHMF, SMBH--galaxy scaling relations, and the AGN LF), 2) to examine the implications of these high-$z$ initial conditions for the subsequent evolution of the SMBH population and connection to the low-z demography of SMBHs. In particular, in this work we will pin down the conditions on the SMBH scaling relations and fraction of galaxies hosting an AGN necessary at high-$z$ to reconcile the galaxy SMF with the AGN LFs and active BHMFs, as well as predicting the evolution of the SMBH mass density, BHMF, and the implied mean $\Mbh-\Mstar$ relation. It is pivotal to make sense of the reliability and self-consistency of these high-z data sets to in turn set robust and credible constraints on, for example, SMBH seed masses \citep[e.g.][]{2025JeonarXiv250814155J,2025CenciMNRAS.tmp.1302C,2025PacucciLRDsDirectCollapsearXiv250902664P}, obscured AGN fractions \citep[e.g.][]{2025OJAp....8E...9J}, or scaling relation with their hosts \citep[e.g.][]{2010ShankarApJ...718..231S}. We will show that whilst there is no unique explanation offered by current data, we can still identify classes of viable solutions that can be tested via independent tailored measurements (e.g. AGN clustering).

The paper is organised as follows, in section \ref{sec:Data} we describe the dataset we use and in section \ref{sec:Method} we outline our methodology to investigate the high-$z$ AGN demography. In section \ref{sec:Results} we first construct BHMFs from the SMF by adopting different $\Mbh-\Mstar$ relations and examine their consistency with the observational estimates of the high-$z$ active BHMF and set initial constraints on the viable duty cycles (\ref{ssec:Ceq}), we then convert these BHMFs to AGN LFs to pin down the conditions necessary to match the observational determinations and examine whether these are consistent with those inferred from the BHMF (\ref{ssec:ResbolLF} \& \ref{ssec:ResUVLF}), after that we test the consistency of the SMBH density obtained from these with independent estimates at both high and low-$z$ (\ref{ssec:rhoBH}). Finally, we forward model the BHMF to present day, self-consistently deriving the evolution of the $\Mbh-\Mstar$ relation (\ref{ssec:BHMF}). We conclude by discussing the implications of these results for our picture of SMBH--galaxy coevolution in section \ref{sec:Discussion}. Throughout this work we assume a standard $\Lambda{\rm CDM}$ cosmology with ${\rm H}_0 = 70~{\rm km\,s^{-1}\,Mpc^{-1}}$, $\Omega_{\rm m,0} = 0.3$, and $\Omega_{\Lambda,0} = 0.7$. Any distribution functions used in this work that are not computed in this cosmology have been rescaled to this cosmology. All magnitudes quoted in this work are expressed in the AB system \citep[][]{OkeAndGunn1983}.

\section{Data}\label{sec:Data}

\subsection{BL AGN Sample}\label{ssec:AGNsample}

\begin{table*}
    \centering
    \begin{tabular}{ccccccc}
        \hline
        Source & Object Type & Selection & $N_{\rm obj}$ used in & $N_{\rm obj}$ used to \\
        &&& $\Muv-\Lbol$ & inform $P(\lambda)$\\
        \hline\hline
        \citet{2023ApJ...959...39H} & BL AGN & Spectroscopic & 8 & 8 \\
        \citet{2025IgnasarXiv250403551J} & BL AGN & Spectroscopic & 6 & 6 \\
        \citet{2024arXiv240403576K} & LRDs & Spectroscopic & 11 & 0 \\
        \citet{Kokorev2024_PhotoConcensusOfLRDs} & LRDs & Photometric & 186 & 0 \\
        \citet{maiolino2024jadesdiversepopulationinfant} & BL AGN & Spectroscopic & 10 & 12 \\
        \citet{Mathee2024_LRDs} & LRDs & Spectroscopic & 19 & 0 \\
        \citet{2023Ubler} & BL AGN & Spectroscopic & 0 & 1 \\
        \hline
    \end{tabular}
    \caption{A brief summary of the data used from literature in this work. A full description of the data extracted from these works can be found in Appendix \ref{app:LRDsample}. The columns, from left to right, are: the paper the data originates from, the object type selected in the paper, whether the objects are photometrically or spectroscopically selected, the number of objects we use from their sample to derive a mapping between $\Lbol$ and $\Muv$, the number of objects we use from their sample to get a sense of the distribution of Eddington ratios. Here we only use the spectroscopically confirmed objects from \citet{2024arXiv240403576K}.}
    \label{tab:LRDnoSummary}
\end{table*}

Throughout this work we make use of two samples of AGN observed by JWST in the range $4.5 \leq z \leq6.5$ collected from several recent works \citep[][]{2023ApJ...959...39H,2023Ubler,Kokorev2024_PhotoConcensusOfLRDs,2024arXiv240403576K,Mathee2024_LRDs,maiolino2024jadesdiversepopulationinfant,2025IgnasarXiv250403551J}. First, we select a sample in the redshift range $4.5\lesssim z\lesssim6.5$ with measured Eddington ratios which we use as a baseline for our input Eddington ratio distribution function (ERDF; ${\rm P}(\fEdd)$; Figure \ref{fig:LRDfEddDist}). Second, we select a sample of objects in the redshift range $4.5\lesssim z\lesssim6.5$ with measured UV magnitudes and bolometric luminosities to construct an empirical mapping between the $\Muv$ and $\Lbol$ (Figure \ref{fig:kbolFit}). The samples adopted as a reference in this work are not exhaustive of all BL AGN detected by JWST, but they still represent the typical sources contributing to the $z\sim5.5$ luminosity functions that we compare to. We summarise the number of objects used from each of these works in Table \ref{tab:LRDnoSummary}. We refer the interested reader to the respective papers for the full methodology and simply summarise the selection criteria and derivation of the properties used in this work in Appendix \ref{app:LRDsample}.

\subsection{Reference Luminosity Function}\label{ssec:refLF}

\begin{figure}
    \centering
    \includegraphics[width=0.45\textwidth]{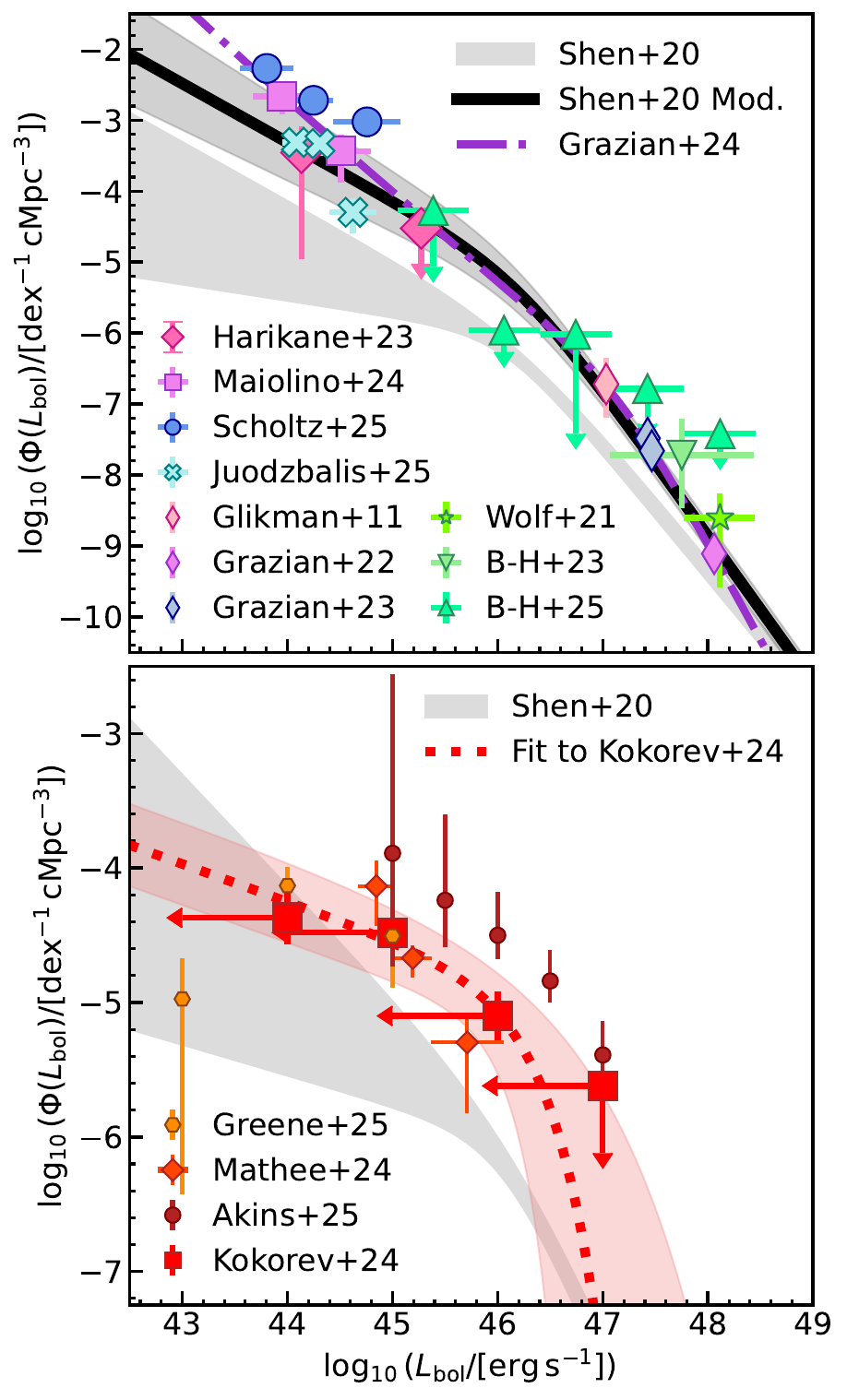}
    \caption{\emph{Top:} A comparison of our reference LF which is a modified version of the \citet{Shen_2020} global fit A to the high-$z$ data it was tuned to \citep[e.g.][]{2023ApJ...959...39H,maiolino2024jadesdiversepopulationinfant,2025IgnasarXiv250403551J,2011GlikmanApJ...728L..26G,2022GrazianQUBRICsApJ...924...62G,2023GrazianRubiconApJ...955...60G,2021WolfA&A...647A...5W,2023BarlowHallMNRAS.519.6055B,2025BarlowHallarXiv250616145B} 
    and the fiducial \citet{Shen_2020} global fit A. We also include the intermediate UV LF of \citet[][]{2024GrazianApJ...974...84G}. Here, the X-ray and UV-based estimated have been converted to the bolometric plane following \citet{Shen_2020}. A comparison between our reference LF and several theoretical models can be found in Fig. \ref{fig:SAMcomp_MFLF}. \\
    \emph{Bottom:} A comparison of the \citet{Kokorev2024_PhotoConcensusOfLRDs} LF (red square points) to our best fit Schechter function (red dotted line) and its $1\sigma$ uncertainty region (red shaded area), and the left-pointing arrows denote the $\sim1~{\rm dex}$ correction suggested by \citet{2025GreenearXiv250905434G}. We have also included the \citet{akins2024LRDs} LRD LF (dark red circular points), the LFs of \citet{2025GreenearXiv250905434G} (orange hexagonal points) and \citet{Mathee2024_LRDs} (dark orange diamond points), and the \citet{Shen_2020} quasar LF (grey shaded region).}
    \label{fig:K24bolLFfit}
\end{figure}

Through out this work we adopt a modified version of the \citet{Shen_2020} LF as our reference bolometric LF, using this both to compare with our model predictions (section \ref{ssec:ResbolLF}) and as an input when forward modelling the SMBH population (sections \ref{ssec:rhoBH} \& \ref{ssec:Ceq}). Similar to \citet{2025LapiGWBarXiv250715436L}, we modify the parameters of the \citet{Shen_2020} LF to match the compilation of estimates of the BL AGN LF displayed in top panel of Figure \ref{fig:K24bolLFfit}. 

\citet{Shen_2020} parameterise their LF as a double power law 
\begin{equation}\label{eq:dblPL}
    \Phi(\Lbol) = \frac{\Phi_{*}}{(\Lbol/L_{*})^{\gamma_{1}} + (\Lbol/L_{*})^{\gamma_{2}}}~,
\end{equation}
where $\Phi_{*}$ is the comoving number density normalisation, $L_{*}$ is the break luminosity, and $\gamma_{1}$ and $\gamma_{2}$ are the faint-end and bright-end slopes, respectively.

In addition to the polished fits at specific redshift intervals, \citet{Shen_2020} provide two global fits to describe the evolution of the quasar bolometric LF. In this work, we adopt their global fit A and modify the parameters controlling the redshift evolution of the normalisation and the bright-end slope to match the high-$z$ observations. The redshift evolution of the normalisation is parameterised as 
\begin{equation}\label{eq:phistar}
    \log_{10}(\Phi_{*}(z)) = d_{0} + d_{1}(1+z)~,
\end{equation}
with best-fit values of $\{d_{0},d_{1}\}=\{-3.5426,-0.3936\}$. The redshift evolution of the bright end slope is instead parameterised as a double power law
\begin{equation}
    \gamma_2 = \frac{2b_0}{[(1+z)/3]^{b_1} + [(1+z)/3]^{b_2}}~,
\end{equation}
with best-fit values of $\{b_{0},b_{1},b_{2}\}=\{2.5375,-1.0425,1.1201\}$.

In order to match the high-$z$ data displayed in the top panel of Figure \ref{fig:K24bolLFfit} at $z=5.5$, we increase the value of $d_{1}$ to $d_{1}=-0.2436$ and decrease $b_{2}$ to $b_{2} = 0.9$, thereby slowing the evolution of the normalisation and slope of the bright end. To ensure a smoothly evolving LF, we linearly interpolate the values of $d_{1}$ and $b_{2}$ as
\begin{equation}\label{eq:d1}
    d_{1}(z) = \left\{
    \begin{array}{ll}
        -0.3936 & z\leq z_{*} \\[0.5ex]
        -0.3936+0.15\frac{z-z_{*}}{5.5-z_{*}} ~~ & z_{*}<z\leq 5.5\\[0.5ex]
        -0.2436 & z>5.5
    \end{array}
    \right.
\end{equation}
\begin{equation}
    b_{2}(z) = \left\{
    \begin{array}{ll}
        1.1201 & z\leq z_{*} \\[0.5ex]
        1.1201-0.2201\frac{z-z_{*}}{5.5-z_{*}} ~~ & z_{*}<z\leq 5.5\\[0.5ex]
        1.1201 & z>5.5
    \end{array}
    \right.
\end{equation}
where $z_{*}\,(=2)$ is the redshift at which we assume the LF returns to the fiducial global fit A from \citet{Shen_2020}.

These new prescriptions lead to a steady increase in normalisation at $z>2$ relative to the fiducial \citet{Shen_2020} fit and provides good agreement with the $z\sim5-6$ UV-based number densities from \citet{2023ApJ...959...39H,maiolino2024jadesdiversepopulationinfant,2025IgnasarXiv250403551J} at the faint end, while the steepening of the bright-end slope provides good agreement with the UV and X-ray-based estimates from \citet{2011GlikmanApJ...728L..26G,2022GrazianQUBRICsApJ...924...62G,2023GrazianRubiconApJ...955...60G,2021WolfA&A...647A...5W,2023BarlowHallMNRAS.519.6055B,2025BarlowHallarXiv250616145B}, as well as ensuring that the modified LF converges to the fiducial fit of \citet{Shen_2020} at $\Lbol\sim10^{50}~{\rm erg\,s^{-1}}$. As shown in the top panel of Figure \ref{fig:K24bolLFfit} our modified version of the \citet{Shen_2020} LF is in very good agreement with the intermediate LF of \citet[][their Option 2]{2024GrazianApJ...974...84G}.

\subsection{Empirical $\Muv-\Lbol$ mapping from the LRD Luminosity Function}\label{ssec:LRDLF}

As an additional test, in this work we also make use of an empirical mapping between $\Muv$ and $\Lbol$ which we will use, alongside the other bolometric corrections to compute the AGN UV LF from our models (section \ref{sssec:MethodAGNUVLF}). We derive this empirical correction from the \citet{Kokorev2024_PhotoConcensusOfLRDs} estimates of the UV and bolometric LFs for the same sample of LRDs via abundance matching, on the assumption that LRDs follow similar scaling relations to those of the global population of BL AGN at $z\sim5-6$. We note that the LRDs in \citet{Kokorev2024_PhotoConcensusOfLRDs} are only reddened AGN candidates as they are photometrically selected, and so contamination by objects that have a v-shaped SED due to bursty star formation cannot be ruled out without spectroscopic follow up. We find that the \citet{Kokorev2024_PhotoConcensusOfLRDs} sample displays a similar distribution in the $\Muv-\Lbol$ plane to spectroscopically confirmed LRDs \citep{2024arXiv240403576K} and low-luminosity BL AGN \citep{maiolino2024jadesdiversepopulationinfant}. Furthermore, the distribution functions measured in \citet{Kokorev2024_PhotoConcensusOfLRDs} from their photometrically selected sample agree well with those of \citet{Greene2024_LRDs} and \citet{Mathee2024_LRDs} which are computed from a spectroscopically selected sample of LRDs. \citet{Kokorev2024_PhotoConcensusOfLRDs} parameterise their UV LF as a Schechter function 
\begin{equation}
    \phi(L) = \frac{\phi_{*}}{L_*} \left(\frac{L}{L_*}\right)^{\alpha} e^{-(L/L_*)^{\beta}}~,
\end{equation}
where $\phi(L)$ is the comoving number density\footnote{Throughout this work we use $\phi(X)$ to denote the distribution function of reference quantity $X$ in linear space and $\Phi(X)=\ln(10)X\phi(X)$ to denote the distribution function of $X$ in logarithmic space.}, $L_*$ is a characteristic luminosity, $\alpha$ is the index of the power law at low $L$, and $\beta$ controls the exponential cut-off at high $L$. They find the UV LF at $z\sim5.5$ to be well fitted by $\phi_* = (8\pm3)\times10^{-6}~{\rm Mpc}^{-3}$, $M_{\rm UV,*} = -20.64\pm0.67$, $\alpha = -1.76\pm0.67$, and $\beta$ is assumed to be unity. On the other hand, as the bolometric LF in is only computed in four bins, one of which is an upper limit, a Schechter function cannot be directly fit to this. Instead, we fit a single Schechter function to the bolometric LF derived in \citet{akins2024LRDs} for their MIRI subsample and offset it by altering the $\phi_*$ (moving vertically) and $L_{*}$ (moving horizontally) parameters to match the \citet{Kokorev2024_PhotoConcensusOfLRDs} data points. 

We fit a Schechter function to the bolometric LF from \citet{akins2024LRDs} following a Markov Chain Monte Carlo (MCMC) approach implemented with \texttt{emcee} \citep[][]{2013emceeForemanMackey}. We fix $\beta = 1$ and find the luminosity function to be well described by a Schechter function where $\log_{10}(\phi_*/{\rm Mpc^{-3}}) = -4.98^{+0.27}_{-0.25}$, $\log_{10}(L_{\rm bol,*}/[{\rm erg.s^{-1}}]) = 46.76^{+0.16}_{-0.21}$, $\alpha = -1.29^{+0.30}_{-0.26}$. We find the \citet{Kokorev2024_PhotoConcensusOfLRDs} data points to be well described by $\log_{10}(\phi_*/{\rm Mpc^{-3}}) = -5.23^{+0.27}_{-0.25}$, $\log_{10}(L_{*}/[{\rm erg\,s^{-1}}]) = 46.21^{+0.16}_{-0.21}$, and $\alpha = -1.29^{+0.30}_{-0.26}$. That's a horizontal and vertical offset of $-0.55~{\rm dex}$ and $-0.25~{\rm dex}$, respectively, from the \citet{akins2024LRDs} LF. As the highest luminosity bin is an upper limit, we bound the uncertainty in the bright-end extrapolation by assuming $\pm0.5$ uncertainty in $\beta$, such that upper uncertainty bound has a sub-exponential fall-off ($\beta = 0.5$) and the lower uncertainty bound has a super-exponential fall-off ($\beta = 1.5$). The resulting fit and its $1\sigma$ uncertainty bounds are displayed in the bottom panel of Figure \ref{fig:K24bolLFfit}.

\section{Method}\label{sec:Method}
The aim of this work is twofold: on one hand, we aim to probe the consistency of the distinct datasets at $z\sim5-6$, and on the other hand, we aim to test the implications of the initial condition on the subsequent evolution of the SMBH demography making use of the continuity equation. In this section we provide a step-by-step description of our methods for both the first (section \ref{ssec:consistencyOfDataSets}) and second (section \ref{ssec:CeqMethod}) points.

Throughout this work we focus on a redshift of $z\sim5.5$ (and then use this as our initial redshift when forward modelling) as it is the centre of the lower redshift bins in \citet{Kokorev2024_PhotoConcensusOfLRDs} and \citet{Greene2024_LRDs}, as well as being comparable to the mean redshift of the BL AGN samples in \citet[][$\langle z\rangle\sim5.18$]{maiolino2024jadesdiversepopulationinfant} and \citet[][$\langle z\rangle\sim5.15$]{2023ApJ...959...39H}. However, we emphasise that the results, analysis, and conclusions presented in the following sections would be unchanged were we to choose a redshift of $5$ (as done by \citealt{inayoshi2024birthrapidlyspinningovermassive}) or $6$ (as done by \citealt{2025OJAp....8E...9J}).

\subsection{The Consistency of Data Sets}\label{ssec:consistencyOfDataSets}

\begin{figure*}
    \centering
    \includegraphics[width=0.9\textwidth]{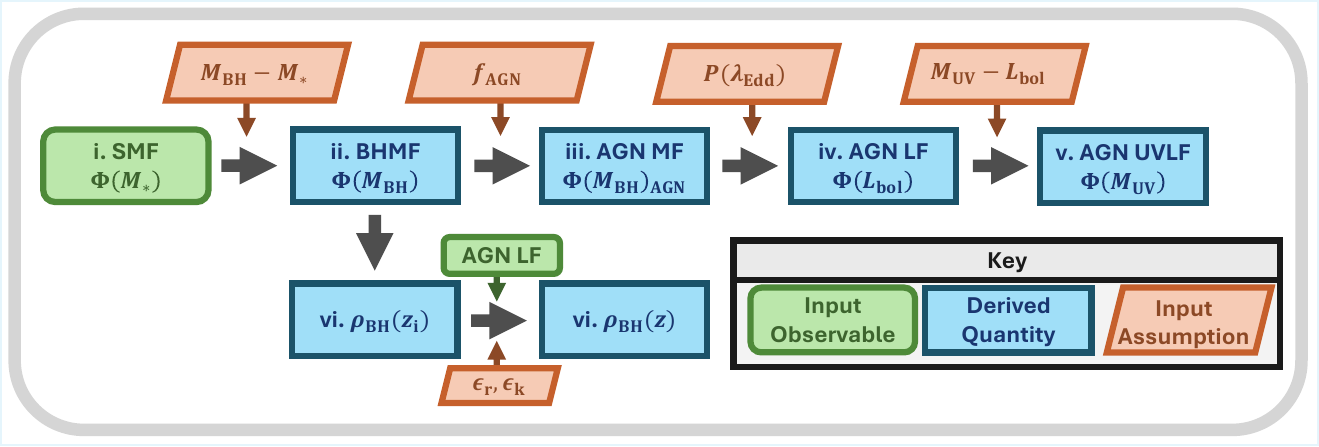}
    \caption{A pictorial representation of our methodology. By moving from left to right, we convert galaxy number densities to AGN demography, which are both independently calibrated quantities. The input ingredients (red boxes) included in each step of the chain are also informed by observations. The match with the observed AGN luminosity function and active SMBH mass function can reveal potential inconsistencies and/or biases in any of the input quantities. The comparison with the SMBH mass densities in the bottom row, can instead constrain the energy efficiencies of SMBHs (Soltan argument, extended to $z\sim10$) and provide an overall and self-consistent view of the accretion histories of SMBHs from very early epochs.}
    \label{fig:MethodCartoon}
\end{figure*}

To examine the consistency of the observed data sets we start from the galaxy stellar mass function (SMF) and, as is commonly done in the local universe, we convert the galaxy statistics to BH statistics via an assumed BH--galaxy scaling relation. From this we can derive quantities that are directly observed (or those easily computed from observations) such as the active BHMF, AGN LF, and SMBH mass density. By comparing the predicted to the observed quantities we can uncover any underlying inconsistencies between the data sets and constrain the conditions necessary to reconcile inputs that are potentially impacted by observational biases (e.g. the $\Mbh-\Mstar$ relation) with the observed AGN demography. This work flow is summarised pictorially in Fig. \ref{fig:MethodCartoon} and each step is described in greater detail below.

\subsubsection*{(i) Galaxy Stellar Mass Function}\label{sssec:MethodSMF}
Under the observationally justified assumption of every massive galaxy hosting a central SMBH, the SMF acts as a natural starting point to test the consistency of current observational data sets. We assume the SMF of \citet{shuntov2024cosmoswebstellarmassassembly} which is computed on the $0.53~{\rm deg^2}$ of the COSMOS field imaged by JWST as part of the COSMOS-Web survey \citep[][]{2023COSMOSWeb}. The SMF is computed in 15 redshift bins in the range $0.2\leq z \leq 12$, mass complete to $\log_{10}(\Mstar/\msun)=7.5-8.8$, and agrees well with the COSMOS-2020 SMF of \citet{Weaver_2023} at $z\leq5.5$. We further assume that the fraction of galaxies that are quenched at this redshift is negligible, as observations suggest that this population should be small and restricted to the highest mass galaxies \citep[][which is consistent with theoretical models \citealt{2015SchayeEAGLEMNRAS.446..521S,2018LagosSHARKMNRAS.481.3573L,2019BehrooziUMMNRAS.488.3143B,2024DeLuciaGAEAQuenchingA&A...687A..68D}]{2013MuzzinApJ...777...18M,Weaver_2023,2024RussellarXiv241211861R,2025MerlinOJAp....8E.170M,2025YangarXiv251012235Y,2025ShuntovarXiv251105259S}.

\subsubsection*{(ii) SMBH Mass Function}\label{sssec:MethodBHMF}
The BHMF is then simply computed from the SMF and an assumed $\Mbh-\Mstar$ relation via the convolution
\begin{equation}\label{eq:BHMFconvolution}
    \BHMF = \int_{M_{\rm *,min}}^{\infty}\!\SMF\, {\rm P}(\Mbh|\Mstar) ~ {\rm d}\log_{10}(\Mstar),
\end{equation}
where $\BHMF$ is the BHMF, $\SMF$ is the SMF, and ${\rm P}(\Mbh|\Mstar)$ is a Gaussian distribution around the mean $\Mbh-\Mstar$ relation. Here, we are implicitly assuming a BH occupation fraction of unity at all masses and redshifts.

\begin{figure}
    \centering
    \includegraphics[width=0.45\textwidth]{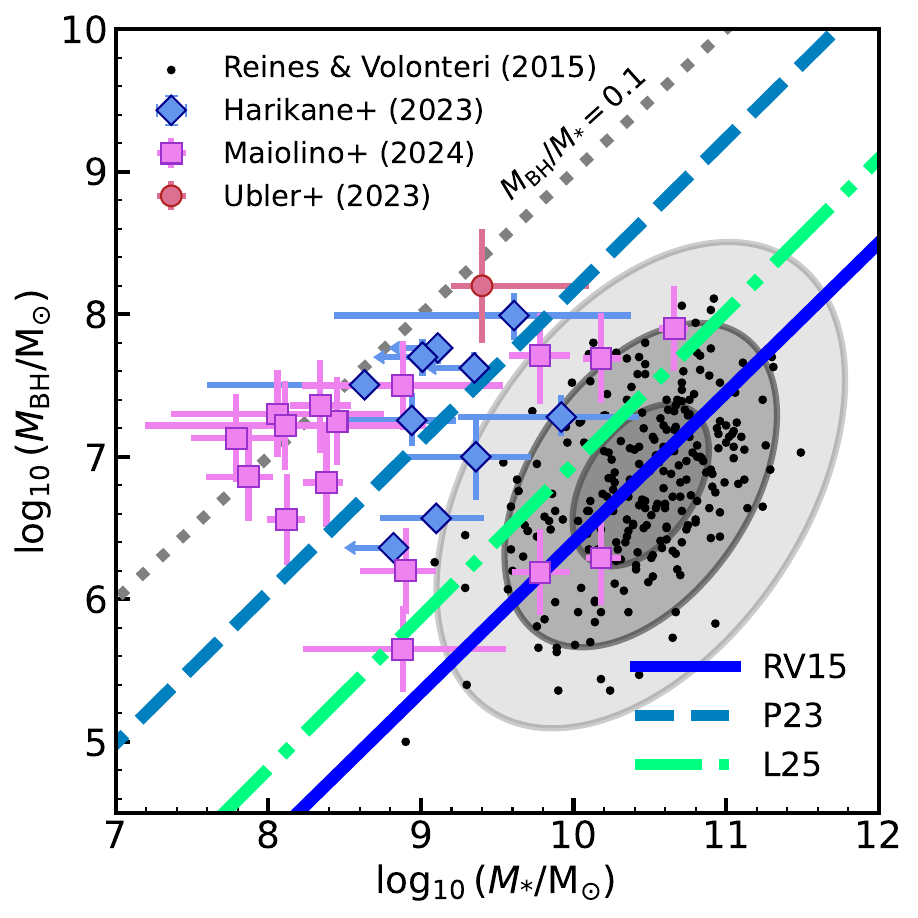}
    \caption{A comparison of the \citet{Reines_2015}, \citet{Pacucci2023_LRDs}, and \citet{2025LiTipOf} $\Mbh-\Mstar$ relations with the local sample of AGN from \citet{Reines_2015} (black points) and the faint AGN samples of \citet{maiolino2024jadesdiversepopulationinfant}, \citet{2023ApJ...959...39H}, and \citet{2023Ubler}. The grey ellipses are the $1\sigma$, $2\sigma$, and $3\sigma$ confidence ellipses for the \citet{Reines_2015} sample.}
    \label{fig:MstMbhSR_LitComp}
\end{figure}

We focus on three different scenarios of how the observed AGN relate to the underlying SMBH population and represent these with different $\Mbh-\Mstar$ relations: the high-$z$ determinations of \citetalias{Pacucci2023_LRDs} and \citetalias{2025LiTipOf}, plus the local AGN determination of \citetalias{Reines_2015} (displayed in Fig. \ref{fig:MstMbhSR_LitComp}). These $\Mbh-\Mstar$ relations bracket the full range of the current data (nearly two orders of magnitude in normalisation).

First, we consider the unrealistic maximal case where there is minimal selection bias in the JWST observations, such that the observed AGN are representative of the underlying population and SMBHs lie systematically above the local $\Mbh-\Mstar$ relation at high redshifts. To represent this scenario we adopt the high-$z$ $\Mbh-\Mstar$ relation from \citetalias{Pacucci2023_LRDs} which is computed from a sample of $21$ BL AGN at $z\sim4-7$ observed by JWST taken from \citet{2023ApJ...959...39H}, \citet{maiolino2024jadesdiversepopulationinfant}, and \citet{2023Ubler}. \citetalias{Pacucci2023_LRDs} determine the high-$z$ $\Mbh-\Mstar$ relation to be $\sim2~{\rm dex}$ above the local relation for AGN \citepalias[][]{Reines_2015} at $\Mbh\sim10^9~\msun$ claiming the increase in normalisation due to selection is $0.2~{\rm dex}$ at most. Interestingly, the noramlisation of the \citetalias{Pacucci2023_LRDs} $\Mbh-\Mstar$ relation is similar to the local relation for dynamically measured SMBHs from \citet{KormendyAndHo2013} as rescaled by \citetalias{Reines_2015}.

Second, we consider the more realistic case where selection effects and measurement uncertainties have biased the observed relation high. To represent this scenario we adopt high-$z$ $\Mbh-\Mstar$ relation from \citetalias{2025LiTipOf} which is computed from a sample of $18$ JWST BL AGN taken from \citet{2023ApJ...959...39H} and \citet{maiolino2024jadesdiversepopulationinfant}. \citetalias{2025LiTipOf} quantify the impact of selection effects and measurement uncertainties by modelling the conditional probability ${\rm P}(\Mbh|\Mstar)$ for a given detection limit. They then infer the intrinsic relation from the observed biased sample. \citetalias{2025LiTipOf} determine the intrinsic high-$z$ $\Mbh-\Mstar$ relation to be $\sim0.5~{\rm dex}$ above the local relation for AGN \citepalias[][]{Reines_2015} at $\Mbh\sim10^9~\msun$ and is similar in normalisation at the high-mass end to the local de-biased relation from \citet{Shankar_2016}.

Finally, given that the BH masses are obtained from locally-calibrated single-epoch virial estimators and there is evidence that at least in some high-$z$ sources these may lead to overestimated BH masses \citep[e.g.][]{2024LupiA&A...689A.128L,2024AbuterNatur.627..281A,2025ElDayemarXiv250913911G,2025ParlantiarXiv251214844P}, we consider the case where the observed relation is biased high due to both selection effects and an overestimation of the BH masses. To represent this case we adopt the local $\Mbh-\Mstar$ relation for AGN from \citetalias{Reines_2015} as it sits $\sim0.5~{\rm dex}$ lower in normalisation than the relation from \citetalias{2025LiTipOf}, which is similar in magnitude to the average overestimation suggested by \citet{Lupi_2024} due to the correlation between the Eddington ratio and the radius of the BLR \citep[][]{2014WangApJ...797...65W,2020MartinezAldamaApJ...903...86M,2024LupiA&A...689A.128L}. It is also approximately inline with the case where the LRDs' BH masses have been significantly and systematically overestimated because the Balmer lines are broadened through scattering \citep[e.g.][]{2025RusakovarXiv250316595R,2025NaiduBHstararXiv250316596N}. \citetalias{Reines_2015} compute the local $\Mbh-\Mstar$ relation from a sample of $244$ local BL AGN selected from SDSS emission-line galaxies.

All of these works obtain BH masses from the broad ${\rm H}\alpha$ emission using the single-epoch virial estimator of \citet{2013ReinesApJ...775..116R} and parameterise the mean $\Mbh-\Mstar$ relation as a linear function in log-space
\begin{equation}\label{eq:MbhMst}
    \log_{10}\left(\frac{\Mbh}{\msun}\right) = \alpha + \beta \log_{10}\left(\frac{\Mstar}{\msun}\right) ,
\end{equation}
with some intrinsic scatter ($\varepsilon$) in the relation. The parameters of these relations are listed in Table \ref{tab:MbhMstParams}. As seen both in the parameter values and Fig. \ref{fig:MstMbhSR_LitComp}, the three determinations display very similar slopes but differ both in normalisation and the magnitude of the intrinsic scatter.

\begin{table}
    \centering
    \begin{tabular}{ccccc}
        \hline
        Work & Redshift & $\alpha$ & $\beta$ & $\varepsilon$ [dex] \\
        \hline\hline
        \citetalias{Pacucci2023_LRDs} & $4<z<7$ & $-2.43\pm0.83$ & $1.06\pm0.09$ & $0.69$\\[0.5ex]
        \citetalias{2025LiTipOf} & $4<z<7$ & ${-3.32}^{+0.56}_{-0.54}$ & $1.02^{+4.56}_{-5.16}$ & $0.97^{+0.52}_{-0.37}$ \\[0.5ex]
        \citetalias{Reines_2015} & $z<0.055$ & $-4.1\pm1.21~^{\rm a}$ & $1.05\pm0.11$ & $0.5$\\[0.5ex]
        \hline
    \end{tabular}
    \caption{The parameters of the three $\Mbh-\Mstar$ relations considered in this work. As described in equation (\ref{eq:MbhMst}), the $\Mbh-\Mstar$ relations are parameterised as a linear function in log-space where $\alpha$ is the intercept, $\beta$ is the slope, and $\varepsilon$ is the intrinsic scatter in the relation.\\
    $^{\rm a}$\citetalias{Reines_2015} normalise their relation to $\Mstar=10^{11}~\msun$ for which $\alpha=7.45\pm0.08$. We have recomputed alpha for consistency with \citetalias{Pacucci2023_LRDs} and \citetalias{2025LiTipOf}, and propagated the uncertainties when doing so.}
    \label{tab:MbhMstParams}
\end{table}

Given that pairwise residual analysis has consistently found the local $\Mbh-\sigma_{\star}$ relation to be more fundamental than the local $\Mbh-\Mstar$ relation, it would be most appropriate to compute the BHMF from the velocity dispersion function (VDF) and $\Mbh-\sigma_{\star}$ relation. However, the VDF is poorly constrained at intermediate to high redshifts, with there being no robust estimates at $z\sim5$. Nevertheless, we offer some insight from a BHMF inferred using the $\Mbh-\sigma_{\star}$ relation via the Faber-Jackson ($\sigma_{\star}-\Mstar$) relation \citep[][]{1976FaberJacksonApJ...204..668F} by deploying the theoretical framework of \citet{Marsden2021_sigma} in Appendix \ref{app:highZVDF}.

\subsubsection*{(iii) Active SMBH Mass Function}\label{sssec:MethodActiveBHMF}
The BHMF obtained from equation (\ref{eq:BHMFconvolution}) can then be converted to an active BHMF ($\Phi(\Mbh)_{\rm AGN}$) via an assumed AGN fraction ($\fAGN$), that we assume to be independent of mass for simplicity. As we will be comparing the derived LFs to those of BL AGN, our AGN fraction is defined as the fraction of BL AGN with respect to the total SMBH population
\begin{equation}
    \fAGN = \frac{\Phi(\Mbh)_{\rm AGN}}{\Phi(\Mbh)} = \frac{N_{\rm BL\,AGN}}{N_{\rm SMBH}} \equiv \frac{N_{\rm BL\,AGN}}{N_{\rm Gal}}~,
\end{equation}
where the the second equivalence follows our assumption that $\fAGN$ is independent of mass and the final equivalence follows our explicit assumption that every galaxy hosts a SMBH. We test values of the AGN fraction in the range $\fAGN \in [0.01,1]$.

\subsubsection*{(iv) AGN Bolometric Luminosity Function}\label{sssec:MethodAGNLF}
The bolometric luminosity function can be computed from the active BHMF via a convolution with an assumed Eddington ratio distribution function (ERDF, ${\rm P}(\lambda_{\rm Edd}|M_{\rm BH},z)$) which is normalised to unity
\begin{equation}\label{eq:bolLF}
    \Phi(L_{\rm bol}) = \int\! {\rm P}\left(\lambda_{\rm Edd}|M_{\rm BH},z\right) \Phi(M_{\rm BH})_{\rm AGN} \, {\rm d}\log_{10}(\lambda_{\rm Edd})~.
\end{equation}

We assume the ERDF to be log-normally distributed as suggested by observations \citep[e.g.][]{2010WillottAJ....140..546W,2013ApJ...764...45K,2022FarinaApJ...941..106F,2022WuMNRAS.517.2659W,2024HeApJ...962..152H}. For the cases where we assume the BH mass estimates to be accurate \citepalias[e.g.][]{Pacucci2023_LRDs,2025LiTipOf}, we assume the ERDF to have a mean of $\langle\fEdd\rangle=0.4$ and a standard deviation $\sigma_{\fEdd} = 0.5~{\rm dex}$ that is independent of mass for simplicity. These values are chosen to approximately mirror our combined sample of high-$z$ quasars and lower-luminosity BL AGN (see Fig. \ref{fig:LRDfEddDist}), which follow a log-normal distribution with a mean of $\langle\fEdd\rangle \sim 0.44$ and a standard deviation of $\sigma_{\fEdd} = 0.5~{\rm dex}$, as well as being similar to the individual populations ($\langle\fEdd\rangle=0.25,~0.57$, $\sigma_{\fEdd}=0.52,~0.40~{\rm dex}$ for the lower-luminosity BL AGN and quasars, respectively). We note that this distribution has not been corrected for the flux limited nature of the sample, so it will be biased toward the brightest sources, and in turn, the resulting luminosity functions will be upper limits. For the case where we assume the BH masses to be systematically overestimated \citepalias[e.g.][]{Reines_2015}, we propagate this overestimation into the ERDF, assuming $\langle\fEdd\rangle=1$ and a standard deviation $\sigma_{\fEdd} = 0.5~{\rm dex}$ (inline with the mean Eddington ratio of the \citealt{2025RusakovarXiv250316595R} sample). 

We can then estimate the necessary AGN fraction as a function of luminosity via direct comparison between our model predictions and the reference LF. In addition, the average AGN fraction ($\langle\fAGN\rangle$) can be obtained from the ratio of the number densities via integration of the LFs 
\begin{equation}\label{eq:avfagn}
    \langle\fAGN\rangle = \frac{N_{\rm BL\,AGN}}{N_{\rm SMBH}} = \frac{\int\! \phi(L)_{\rm Obs.}\,{\rm d}L}{\int\! \phi(L)_{\rm Model}\left.\right|_{\fAGN=1}\,{\rm d}L}~,
\end{equation}
where $\phi(L)_{\rm Model}$ is the model prediction computed via eq. (\ref{eq:bolLF}) with an AGN fraction of unity and $\phi(L)_{\rm Obs.}$ is our reference LF described in section \ref{ssec:refLF}. We adopt integration limits $\{L_{\rm Min},L_{\rm Max}\} = \{10^{43.5},10^{50}\}~{\rm erg\,s^{-1}}$, such that we do not extrapolate the LF in the low-luminosity regime where there are no current observations.

\subsubsection*{(v) AGN UV Luminosity Function}\label{sssec:MethodAGNUVLF}

\begin{figure}
    \centering
    \includegraphics[width=0.45\textwidth]{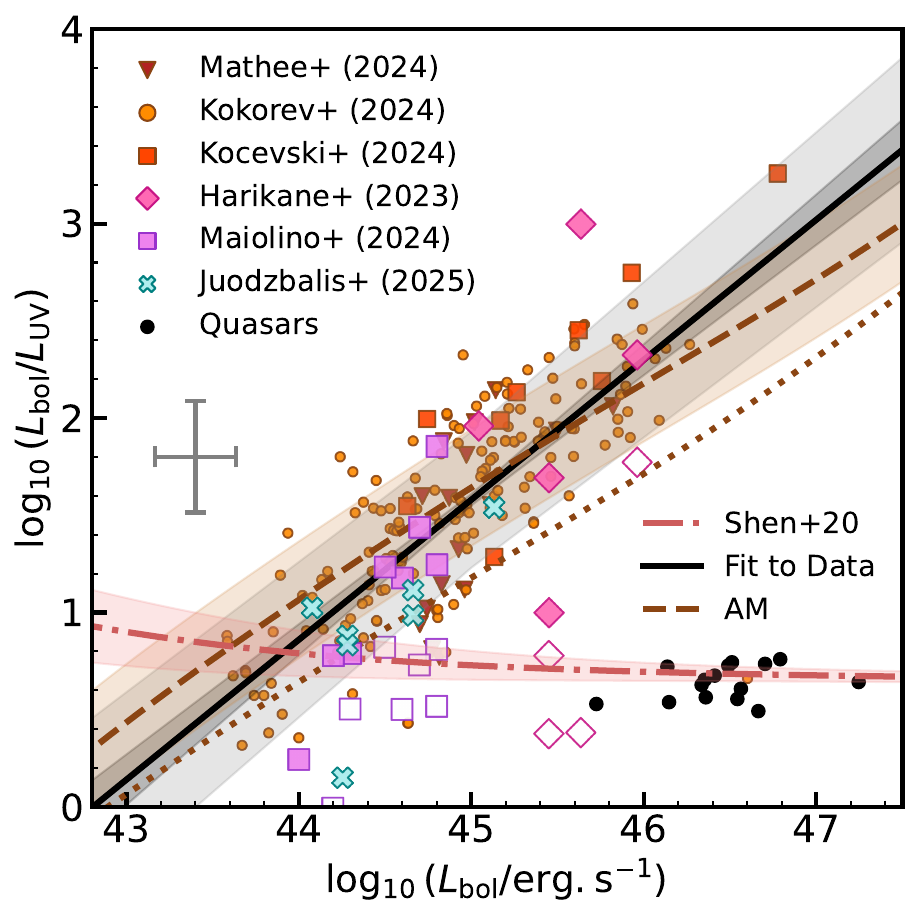}
    \caption{A plot of the $\Lbol-\Lbol/\Luv$ plane comparing the individual BL AGN (points) to the power law fit of this population (black solid line), the relation obtained from abundance matching (AM, brown dashed line), the \citet{Shen_2020} UV-bolometric correction (dark red-dot dashed line). We also include a sample of high-$z$ quasars (black points) from \citet{2019OnoueApJ...880...77O}, \citet{2019MatsuokaApJ...872L...2M}, and \citet{2010WillottAJ....140..546W}. The quasar's $\Lbol$ is computed from the monochromatic luminosity $L_{3000}$ adopting the bolometric correction of \citet{2006RichardsApJS..166..470R}, which matches \citet{2011ShenApJS..194...45S}. The filled BL AGN data points \citep[those of][]{2023ApJ...959...39H,maiolino2024jadesdiversepopulationinfant,2025IgnasarXiv250403551J} use the observed $\Luv$, whereas the open symbols use the intrinsic $\Luv$. The  LRD data points \citep[those of][]{Mathee2024_LRDs,Kokorev2024_PhotoConcensusOfLRDs,2024arXiv240403576K} use the fiducial $\Lbol$. For the clarity of the plot, we have not plotted the uncertainties in the data points, and instead display the median uncertainty as the grey error bar on the centre left. The uncertainties for the individual objects were included when fitting the $\Lbol-\Muv$ relation.}
    \label{fig:kbolFit}
\end{figure}

The UV LF of AGN can then be computed by convolving the bolometric LF given by equation (\ref{eq:bolLF}) with a mapping between bolometric luminosity and UV magnitude. However, at high-z the conversion from bolometric to UV, or vice versa, is non-trivial as all the bolometric corrections are calibrated locally and any possible dust attenuation, host-contamination, or scattered AGN emission also needs to be accounted for, and this is particularly pertinent considering the apparent offset of the lower-luminosity BL AGN and LRDs in the $\Lbol-\Muv$ plane (see Fig. \ref{fig:kbolFit}). We therefore adopt three mappings between $\Lbol$ and $\Muv$: 
\begin{enumerate}
    \item The UV bolometric correction from \citet{Shen_2020}, assuming the bulk of the high-$z$ population are ``typical'' AGN that follow the local relation.
    \item An empirical mapping derived via a linear fit to our sample of BL AGN, assuming that the UV emission is AGN dominated, but there is non-negligible dust attenuation, scattered AGN emission, and/or host contamination. Such that it bypasses some of the uncertainty in the origin of the UV emission.
    \item A statistical mapping derived via abundance matching between the bolometric LF and UV LF of \citet{Kokorev2024_PhotoConcensusOfLRDs}, assuming the LRDs are relatively normal BL AGN and follow the same mapping as the wider population of lower-luminosity BL AGN.
\end{enumerate}
The resulting mappings are displayed in the $\Lbol/L_{\rm UV,obs}-\Lbol$ plane in Figure \ref{fig:kbolFit} and display a significant departure from the local empirical relations of \citet{Shen_2020} and \citet{1994ElvisApJS...95....1E}. We provide full details on the derivation and quantification of these mappings in Appendix \ref{App:MuvLbolMappings}.

\subsubsection*{(vi) SMBH Mass Density}\label{sssec:MethodRhoBH}
The SMBH mass density ($\rho_{\rm BH}$) at $z=5.5$ is computed from the BHMFs that have been obtained from the SMF and $\Mbh-\Mstar$ relations (eq. \ref{eq:BHMFconvolution}) as 
\begin{equation}\label{eq:rhoBH_z5.5}
    \rho_{\rm BH}(z) = \int_{10^{6}~\msun}^{10^{10}~\msun}\! \Mbh\phi(\Mbh,z) \, {\rm d}\Mbh~.
\end{equation}
The subsequent evolution of $\rho_{\rm BH}(z)$ can be simply computed from the AGN LF as 
\begin{equation}\label{eq:rhoBH_z0}
    \rho_{\rm BH}(z) = \rho_{\rm BH}(z_{\rm i}) + \frac{1-\epsilon_{\rm r}-\epsilon_{\rm k}}{\epsilon_{\rm r}\,{\rm c^2}}\int_{z_{\rm i}}^{z}\!\left[\int_{L_{\rm min}}^{L_{\rm max}}\!L\phi(L,z')\,{\rm d}L\right]\frac{{\rm d}t}{{\rm d}z'}\,{\rm d}z'~,
\end{equation}
where $\rho_{\rm BH}(z_{\rm i})$ is the initial value given by equation (\ref{eq:rhoBH_z5.5}), $\epsilon_{\rm r}$ ($=0.10$) is the radiative efficiency, $\epsilon_{\rm k}$ ($=0.0$) is the kinetic efficiency, ${\rm c}$ is the speed of light, and we adopt the limits $L_{\rm min}=10^{42}~{\rm erg\,s^{-1}}$, $L_{\rm max}=10^{50}~{\rm erg\,s^{-1}}$. As in \citet{2025OJAp....8E...9J}, we have neglected any kinetic power, however, adopting $\epsilon_{\rm k}\sim{\rm few}\%$ \citep[as suggested by][]{2008ShankarApJ...676..131S} would only lessen the computed SMBH mass density slightly (by $\sim0.01~{\rm dex}$). For the luminosity function in equation (\ref{eq:rhoBH_z0}) we adopt our modified version of the \citet{Shen_2020} bolometric luminosity function (see section \ref{ssec:refLF}).

The SMBH mass density at $z=5.5$ computed from equation (\ref{eq:rhoBH_z5.5}) can then be compared to that inferred from direct integration of the AGN LFs from \citet{Shen_2020}, our modified version of \citet{Shen_2020}, \citet{Kokorev2024_PhotoConcensusOfLRDs}, \citet{akins2024LRDs}, and \citet{2025BarlowHallarXiv250616145B} via equation (\ref{eq:rhoBH_z0}). For this, we compute $\rho_{\rm BH}$ at $z=5.5$ from the high-$z$ AGN LFs following equation (\ref{eq:rhoBH_z0}), neglecting any possible contribution of BH seeds ($\rho_{\rm BH}(z_{\rm i}) = 0$) and assuming $\epsilon_{\rm r}=0.1$, $\epsilon_{\rm k}=0$. We adopt an upper redshift limit of $z_{\rm i} = 10$ and luminosity limits $[L_{\rm min},L_{\rm max}] = [42,50]~{\rm erg \, s^{-1}}$, extrapolating the \citet{Kokorev2024_PhotoConcensusOfLRDs} and \citet{akins2024LRDs} LFs using the Schechter fits described above. 

Here, we have chosen to extrapolate the LRD LFs \citep[e.g.][]{Kokorev2024_PhotoConcensusOfLRDs,akins2024LRDs} in redshift and luminosity, but we note that there is no significant change ($\sim0.1~{\rm dex}$) were we to not extrapolate in either luminosity or redshift.

\subsection{Forward Modelling the SMBH Demography}\label{ssec:CeqMethod}
To predict the evolution of the BHMF from the initial conditions displayed in Fig. \ref{fig:BHMF} we employ the continuity equation \citep[][]{1971ApJ...170..223C,1992MNRAS.259..725S,2013CQGra..30x4001S}
\begin{equation}\label{eq:CEqOriginal}
    \frac{\partial\Phi(M_{\rm BH})}{\partial t} = -M_{\rm BH} \frac{\partial}{\partial M_{\rm BH}}\left[\frac{\langle\dot{M}_{\rm BH}\rangle\Phi(M_{\rm BH})}{M_{\rm BH}}\right],
\end{equation}
where $\Phi(\Mbh)$ is the BHMF and $\langle\dot{M}_{\rm BH}\rangle$ is the average accretion rate of BHs of a given mass. Parameterising the average specific BH accretion rate (sBHAR) as 
\begin{equation}
    \frac{\langle\dot{M}_{\rm BH}\rangle}{M_{\rm BH}} = \frac{1-\epsilon_{\rm r}-\epsilon_{\rm k}}{\epsilon_{\rm r} {\rm c}^2} l \langle\lambda_{\rm Edd}\rangle U(M_{\rm BH}), 
\end{equation}
where ${\rm c}$ is the speed of light, $\epsilon_{\rm r}~(=0.10)$ is the radiative efficieny, $\epsilon_{\rm k}~(=0.05)$ is the kinetic efficiency, $l = 1.26\times10^{38}~{\rm erg\, s^{-1}\, \msun^{-1}}$, $\langle\fEdd\rangle$ is the mean Eddington ratio, and $U(\Mbh)$ is the duty cycle. The continuity equation can then be expressed as 
\begin{equation}\label{eq:CEq}
    \begin{split}
         \frac{\partial\Phi(M_{\rm BH})}{\partial z} = -\frac{l}{{\rm c}^2 \ln(10)} \frac{{\rm d}t}{{\rm d}z}   \frac{\partial}{\partial\log_{10}(M_{\rm BH})}~~~~~~~~~~~~\\[2ex]
         \times\left[\frac{1-\epsilon_{\rm r}-\epsilon_{\rm k}}{\epsilon_{\rm r}}\langle \lambda_{\rm Edd}\rangle \Phi(M_{\rm BH})_{\rm AGN}\right],
    \end{split}
\end{equation}
where $\Phi(M_{\rm BH})_{\rm AGN} = \Phi(M_{\rm BH})\times U(M_{\rm BH})$ is the active BHMF. 

We follow the procedure outlined in \citet{2017A&A...600A..64T} to solve equation (\ref{eq:CEq}). In their methodology the duty cycle is parameterised as a double powerlaw for which the parameters are determined via an MCMC fit of the bolometric LF computed from equation (\ref{eq:bolLF}) to our reference LF (section \ref{ssec:refLF}), where we adopt the ERDF of \citet{2017A&A...600A..64T} as they demonstrate that it is in good agreement with the measurements of \citet{2013ApJ...764...45K}. 

The ERDF of type-1 AGN is parameterised as a log-normal distribution with a mean given by
\begin{equation*}
    \log_{10}(\lambda_{\rm c}) = {\rm min}[{\rm max}[-1.9+0.45z,\,\log_{10}(0.03)],\,-0.25],
\end{equation*}
which we have limited to $-0.25$ (i.e. the value at $z\sim 3.67$, which is consistent with quasars at $z\sim5-7$) to avoid unrealistic extrapolations at high redshift and the width is given by
\begin{equation*}
    \sigma_{\rm c} = {\rm max}[1.03-0.15z,\,0.6].
\end{equation*}
This is consistent with the ERDF adopted in section \ref{sssec:MethodAGNLF} (which had $\log_{10}(\lambda_{\rm c})\sim-0.4$, $\sigma_{\rm c}=0.5$) to approximate the distribution of quasars and lower-luminosity BL AGN at $z\sim5.5$.

The ERDF of type-2 AGN is parameterised as a powerlaw with exponential cut-off
\begin{equation}
    {\rm P}_2(\lambda_{\rm Edd},z)= \Phi_{\rm c} \,\lambda_{\rm Edd}^{\alpha}\,e^{-\lambda_{\rm Edd}/\lambda_0}
\end{equation}
where
\begin{equation}
    \alpha = \left\{
    \begin{array}{ll}
        -0.6 & z\leq0.6 \\
        -0.6/(0.4+z) & z>0.6
    \end{array}
    \right.~,
\end{equation}
$\lambda_0=1.5$ (or $2.5$ when $\epsilon_{\rm r}\geq0.1$), and $\Phi_c$ is chosen such that the distribution is normalised. The relative contribution of Type-1 and 2 AGN is given by the obscured fraction ($f_{\rm obs}$) from \citet{Ueda2014} which is constant for $z\geq2$, such that our total ERDF is given by
\begin{equation}
    {\rm P}(\lambda_{\rm Edd}|M_{\rm BH},z) = A[(1-f_{\rm obs}){\rm P}_1(\lambda_{\rm Edd},z) \,+\, f_{\rm obs}{\rm P}_2(\lambda_{\rm Edd},z)],
\end{equation}
where $A$ is a constant that normalises the distribution in $\log_{10}(\lambda_{\rm Edd})$ over the range $\log_{10}(\lambda_{\rm Edd})\in[-3,2]$.

The initial conditions are the BHMFs obtained from equation (\ref{eq:BHMFconvolution}). The continuity equation is then run from $z=5.5$ to $z=0$ with time steps of $\Delta z = -0.02$. We will discuss the predicted evolution of the BHMF and compare with the local estimates of the BHMF in section \ref{ssec:Ceq}.

\subsubsection{The Average Duty Cycle}

In addition to the evolution of the BHMF, this methodology also predicts the duty cycle as a function of BH mass and redshift. We compute the mean duty cycle of the SMBH population as 
\begin{equation}\label{eq:meanDC}
    \langle U(z)\rangle = \frac{\int_{10^{6}~\msun}^\infty\!U(M_{\rm BH},z)\Phi(M_{\rm BH},z)\,{\rm d}\log_{10}(M_{\rm BH})}{\int_{10^{6}~\msun}^\infty\!\Phi(M_{\rm BH},z)\,{\rm d}\log_{10}(M_{\rm BH})},
\end{equation}
i.e. the ratio of the number density of AGN to the number density of the total BH population. We can then compare this to the predictions of theoretical models and observational estimates. However, we note that a robust direct comparison with observational estimates is not possible due to differing definitions of the duty cycle. For instance, works such as \citet{2023ScholtzarXiv231118731S}, \citet{maiolino2024jadesdiversepopulationinfant}, and \citet{2025IgnasarXiv250403551J} estimate the duty cycle from the difference in normalisation between the galaxy and AGN UV LFs, whereas \citet{2023AritaApJ...954..210A,2025MNRAS.536.3677A} compute the duty cycle as the ratio of the number density of AGN from the UV LF to the number density of dark matter halos.

\subsubsection{The Evolution of the $\Mbh-\Mstar$ Relation}

While our initial conditions are derived from the SMF, the subsequent evolution of the BHMF is independent of the SMF determinations at lower redshifts. Therefore, we can then employ abundance matching \citep[as in sec. \ref{sssec:MethodAGNUVLF} we follow][their eq. 37]{Aversa_2015} between the observed SMF and our BHMF derived from the continuity equation to predict the evolution of the $\Mbh-\Mstar$ relation for completeness. We assume that the intrinsic scatter is initially that of the relation used to obtain the initial conditions (i.e. $0.5~{\rm dex}$ for \citetalias{Reines_2015}, $0.69~{\rm dex}$ for \citetalias{Pacucci2023_LRDs}, and $0.97~{\rm dex}$ for \citetalias{2025LiTipOf}) and that the magnitude of this scatter decreases with time to that of the local relation \citep[as may be expected due to subsequent mergers][]{2010HirschmannScatterMNRAS.407.1016H}. To this purpose, we linearly interpolate in redshift between the initial value at $z=5.5$ and $0.5~{\rm dex}$ (i.e. the intrinsic scatter of the \citetalias{Reines_2015} relation) at $z=0$. Assuming the scatter remains constant to present day would only reduce the normalisation of the predicted $z=0$ relations from the \citetalias{Pacucci2023_LRDs} and \citetalias{2025LiTipOf} relations.

\section{Results}\label{sec:Results}
In this section we present our results examining the consistency of the current observational data sets (sections \ref{ssec:BHMF} - \ref{ssec:ResUVLF}) and predicting the evolution of the SMBH demography (sections \ref{ssec:rhoBH} \& \ref{ssec:Ceq}). First, we compare the BHMFs inferred from the SMF to estimates of the high-$z$ active BHMF and local BHMF, which will test viable duty cycles for a given choice of underlying $\Mbh-\Mstar$ relation (\ref{ssec:BHMF}). Second, we examine the consistency of the derived AGN bolometric LFs (\ref{ssec:ResbolLF}) and UV LFs (\ref{ssec:ResUVLF}) with current observational determinations. Third, we compare the SMBH mass density to both high and low-$z$ estimates (\ref{ssec:rhoBH}). Finally, we turn our attention to forward modelling the SMBH demography within the continuity equation and predicting the evolution of the $\Mbh-\Mstar$ relation (\ref{ssec:Ceq}).

\subsection{The BHMF at $z\sim5.5$}\label{ssec:BHMF}
\begin{figure}
    \centering
    \includegraphics[width=0.45\textwidth]{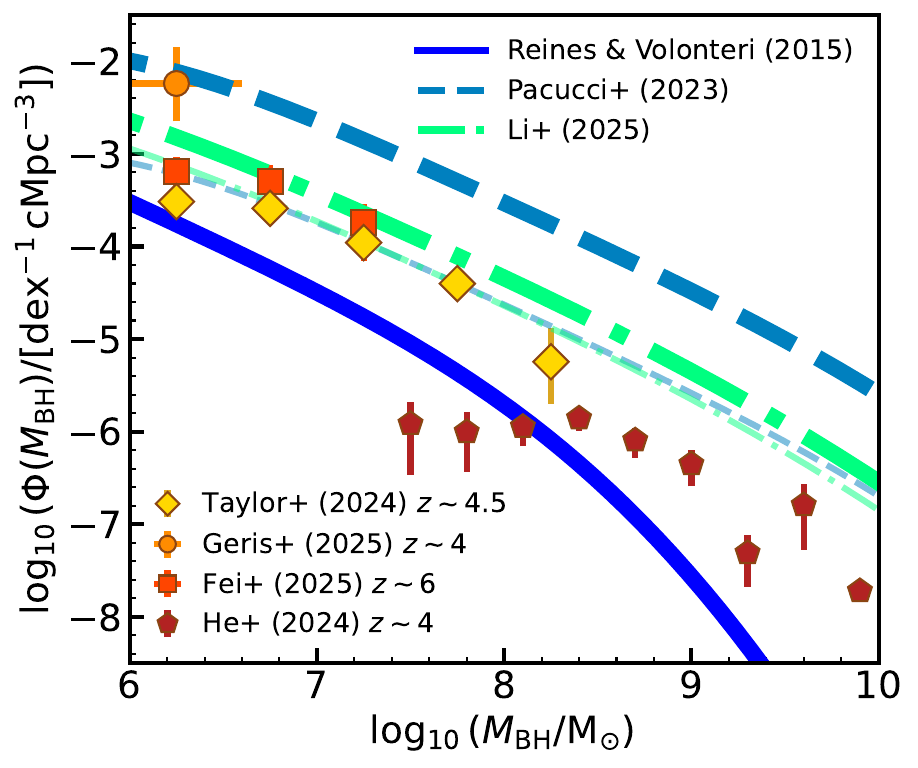}
    \caption{A plot of the BHMF at $z=5.5$ obtained from the SMF assuming the three $\Mbh-\Mstar$ relations: \citetalias{Reines_2015} (solid dark blue line), \citetalias{Pacucci2023_LRDs} (dashed light blue line), \citetalias{2025LiTipOf} (dot-dashed mint green solid line). These total BHMFs are compared to the observed active BHMFs of BL AGN. We have also included the BHMFs scaled by the AGN fraction that best aligns them with the active BHMF of \citet{taylor2024broadlineagn35z6black}, these AGN fractions are $\fAGN\sim0.08,~0.5$ for the \citetalias{Pacucci2023_LRDs}, and \citetalias{2025LiTipOf} relations, respectively, and an AGN fraction in excess of unity is required for \citetalias{Reines_2015}. These scaled BHMFs are denoted by the lower-opacity, thinner lines of the same color and linestyle. The high-$z$ active BHMFs included are those of \citet{taylor2024broadlineagn35z6black}, \citet{2025GerisarXiv250622147G}, \citet{2025FeiarXiv250920452F}, and \citet{2024HeApJ...962..152H}. A comparison between the BHMFs obtained from the SMF assuming the three $\Mbh-\Mstar$ relations and several theoretical models can be found in Fig. \ref{fig:SAMcomp_MFLF}.}
    \label{fig:BHMF}
\end{figure}

The total BHMF is difficult to reliably determine, even in the local universe, as it is prone to observational biases. Yet, clues to the demography of SMBHs at high-$z$ can still be extracted by comparing with the active BHMFs of BL AGN, as well as the local estimates of the total BHMF. In Figure \ref{fig:BHMF} we aim to set constraints on the viable duty cycles by comparing the BHMFs at $z=5.5$ estimated from the SMF and the different $\Mbh-\Mstar$ relations (as described in section \ref{sssec:MethodBHMF} step {\it ii}) with the high-$z$ data \citep{taylor2024broadlineagn35z6black,2024HeApJ...962..152H,2025GerisarXiv250622147G,2025FeiarXiv250920452F}, as labelled.

By comparing the total BHMFs from our models with the observed active BHMFs in Figure \ref{fig:BHMF}, we observe that the BHMFs constructed using the \citetalias{Pacucci2023_LRDs} and \citetalias{2025LiTipOf} relations sit above the high-$z$ estimates of the active BHMFs of BL AGN from \citet{taylor2024broadlineagn35z6black} and \citet{2025FeiarXiv250920452F}. To estimate the AGN fraction ($\fAGN$), we rescale our model BHMFs to match the \citet{taylor2024broadlineagn35z6black} active BHMF of BL AGN, finding $\fAGN\sim0.08$ and $\fAGN\sim0.5$ aligns the models using the \citetalias{Pacucci2023_LRDs} and \citetalias{2025LiTipOf} $\Mbh-\Mstar$ scaling relations, respectively, with the observed active BHMF\footnote{If we compute the BHMF from the SMF at $z\sim4.5$, the average redshift of the AGN sample in \citet{taylor2024broadlineagn35z6black}, the required AGN fractions approximately reduce by half ($\fAGN\sim0.05,\,0.25$ for the \citetalias{Pacucci2023_LRDs}, \citetalias{2025LiTipOf} relations, respectively), however, the \citetalias{Reines_2015} relations still requires $\fAGN\gtrsim1$.}. If we were to ignore the completeness correction in \citet{taylor2024broadlineagn35z6black}, this would lead to a $\sim0.6~{\rm dex}$ decrease in the necessary AGN fraction ($\fAGN\sim0.02,\,0.12$ for \citetalias{Pacucci2023_LRDs}, \citetalias{2025LiTipOf}, respectively).

In contrast, the BHMF constructed using the \citetalias{Reines_2015} $\Mbh-\Mstar$ relation lies below the active BHMF of \citet{taylor2024broadlineagn35z6black}, as well as that of lower-luminosity quasars from \citet{2024HeApJ...962..152H} at $\Mbh\gtrsim10^{8.5}~\msun$, indicating an inconsistency between the data sets under this $\Mbh-\Mstar$ relation. Forcing the \citetalias{Reines_2015}-based BHMF to match the current high-$z$ data with a physical AGN fraction ($\fAGN\leq1$) requires translating the measured active BHMF of \citet{taylor2024broadlineagn35z6black} to the left by $\geq0.8~{\rm dex}$. This corresponds a systematic overestimation of BH masses by single-epoch methods that is similar to that found by GRAVITY+ for a quasar at $z=4$ \citep[][see also the $\sim1~{\rm dex}$ overestimation found in \citealt{2025ParlantiarXiv251214844P}]{2025ElDayemarXiv250913911G}. Even considering the \citet{taylor2024broadlineagn35z6black} active BHMF without the incompleteness correction requires a leftward shift of $\sim0.5~{\rm dex}$ to align observations with the \citetalias{Reines_2015}-based BHMF \citep[consistent with the average overestimation suggested in ][]{2024LupiA&A...689A.128L}.

The results presented in this section set initial constraints on the required duty cycles of active high-$z$ SMBHs. The observed active BHMF can be reproduced from the SMF with either a relation much higher in normalisation paired with a low duty cycle or a relation that is more moderate paired with a higher duty cycle. In particular, the \citetalias{Reines_2015}-based BHMF would fall short in matching the \citet{taylor2024broadlineagn35z6black} data even with no incompleteness corrections in the active BHMF and $\fAGN=1$. We find BH masses at high-$z$ to be systematically overestimated by $\sim0.8~{\rm dex}$ in order to reconcile the two.  In the next section we examine whether the inferred duty cycles of $\fAGN\sim0.08,~0.5,~>1$ for the \citetalias{Pacucci2023_LRDs}, \citetalias{2025LiTipOf}, and \citetalias{Reines_2015} relations, respectively, are consistent with those required to reproduce the observed AGN LFs.

\subsection{The Bolometric Luminosity Function of BL AGN}\label{ssec:ResbolLF}
\begin{figure*}
    \centering
    \includegraphics[width=0.9\textwidth]{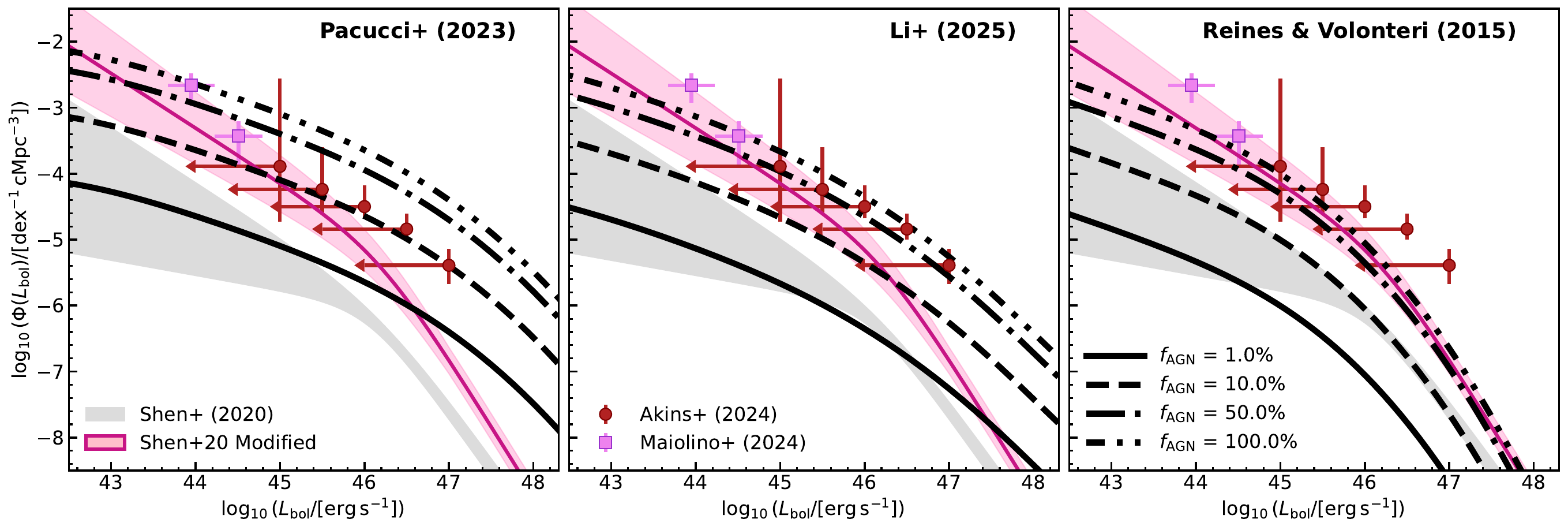}
    \caption{A comparison of the observed AGN luminosity functions with the AGN luminosity functions derived from the $z=5.5$ SMF assuming the $\Mbh-\Mstar$ relation of \citetalias{Pacucci2023_LRDs} (\emph{left}), \citetalias{2025LiTipOf} (\emph{centre}), and \citetalias{Reines_2015} (\emph{right}). The derived luminosity functions are displayed as the black lines, where the differing line styles denote the $\fAGN$ values assumed. The pre-JWST LF of \citet{Shen_2020} is denoted by the grey shaded region, whereas our reference LF is denoted by the pink line and the corresponding shaded region. As points, we also include the UV-based estimate from \citet{maiolino2024jadesdiversepopulationinfant} and the LRD estimate of \citet{akins2024LRDs}. The left-pointing arrows of the \citet{akins2024LRDs} points denote the LF with the $1~{\rm dex}$ correction suggested by \citet{2025GreenearXiv250905434G}.}
    \label{fig:bolLFcomp}
\end{figure*}

Here we present the AGN bolometric LF predicted by our models. As described in section \ref{sssec:MethodAGNLF} step ({\it v}), the bolometric LF is computed by convolving the active BHMFs obtained from the SMF with an assumed ERDF that is informed by our sample of high-$z$ AGN. The resulting bolometric LFs are displayed in Figure \ref{fig:bolLFcomp}, where the left-hand panel displays the LFs derived assuming the \citetalias{Pacucci2023_LRDs} relation, the middle panel assumes the \citetalias{2025LiTipOf} relation, and the right-hand panel assumes the \citetalias{Reines_2015} relation. The black lines denote the derived LFs for different values of the AGN fraction as labelled.

We find average AGN fractions of $\langle\fAGN\rangle\sim0.17,~0.56,~1.00$ for the \citetalias{Pacucci2023_LRDs}, \citetalias{2025LiTipOf}, \citetalias{Reines_2015} $\Mbh-\Mstar$ relations, respectively, reproduce our reference LF. This is within a factor of two of the estimate from the BHMF for the \citetalias{Pacucci2023_LRDs} relation, consistent with the estimate from the BHMF in the cases of the \citetalias{2025LiTipOf} and \citetalias{Reines_2015} relations. We note that an AGN fraction of unity, as is required when adopting the \citetalias{Reines_2015} relation, implies that every galaxy hosts a BL AGN. However, BL AGN are a subpopulation of the total AGN population where the BLR is observed, with NL AGN being more numerous than BL AGN \citep[e.g.][]{2023ScholtzarXiv231118731S}. This suggests that the \citetalias{Reines_2015} relation is inconsistent with the total AGN LF at this epoch, based on the current observational estimates. If the LFs are revised lower in the future, then this may change, and the \citetalias{Reines_2015} relation may become feasible once more.

At the faint end, we observe that only the \citetalias{Pacucci2023_LRDs} $\Mbh-\Mstar$ relation can reproduce the lowest luminosity bin of the \citet{maiolino2024jadesdiversepopulationinfant} LF and even then, still requires a maximal AGN fraction of $\fAGN\sim1$. The fact that $\fAGN\sim1$ is necessary to reproduce the low luminosity bin of \citet{maiolino2024jadesdiversepopulationinfant}, even under this extreme $\Mbh-\Mstar$ relation, may be an indication that the number density of low-luminosity AGN observed by JWST is inflated, possibly due to significant host contamination in some sources. 

At the bright end, we observe that both the \citetalias{Pacucci2023_LRDs} and \citetalias{2025LiTipOf} $\Mbh-\Mstar$ relations predict a low AGN fractions of $\fAGN\lesssim0.01$, qualitatively in agreement with the estimates from the clustering of UV-bright quasars \citep[e.g.][]{2023AritaApJ...954..210A,2024EilersApJ...974..275E,2024PizzatiMNRAS.534.3155P,2025SchindlerarXiv251008455S}. Contrastingly, the \citetalias{Reines_2015} $\Mbh-\Mstar$ relation requires a much higher AGN fraction, suggesting a possible tension. We will explore this in more detail in section \ref{ssec:DiscussionClustering}.

Briefly turning our attention to the LRDs. In the most extreme case, where all the LRDs in the \citet{akins2024LRDs} sample are AGN and that their bolometric luminosities are accurate and AGN dominated, we find the LRDs can be reconciled with the \citetalias{Pacucci2023_LRDs} and \citetalias{2025LiTipOf} $\Mbh-\Mstar$ relations, but not the \citetalias{Reines_2015} relation. However, applying a correction of $\sim1~{\rm dex}$ to the bolometric luminosities \citep[as suggested by][]{2025GreenearXiv250905434G,2025UmedaarXiv251204208U} would reconcile the \citet{akins2024LRDs} sample with the \citetalias{Reines_2015} relation. \\

The results presented in this section indicate that, in line with what was inferred from the BHMFs, in order to reconcile the JWST measured SMF with the bolometric LF of the high-$z$ AGN population whilst maintaining duty cycles of $\fAGN\lesssim0.5$, an $\Mbh-\Mstar$ relation that is higher in normalisation than that of local AGN (e.g. \citetalias{Reines_2015}) is favoured. While a large systematic overestimation of the BH masses could counterbalance the unphysical duty cycles required by the active BHMF to reconcile the \citetalias{Reines_2015} $\Mbh-\Mstar$ relation with the JWST active BHMF and LF of BL AGN, based on the current estimates of the number density of BL AGN, the \citetalias{Reines_2015} $\Mbh-\Mstar$ does not provide sufficient leeway in the BHMF or LF to accomodate the NL AGN population.

\subsection{The UV Luminosity Function of AGN}\label{ssec:ResUVLF}
As the wavelength range covered by NIRcam allows for the selection of AGN in the rest-frame UV at $z\gtrsim3.15$, the UV LF becomes another independent route, characterised by different systematics, to probe for the demography of AGN at high redshifts. We thus extend our method to predict the UV LF of BL AGN from our bolometric AGN LFs presented in Figure \ref{fig:bolLFcomp}. 

Prior to investigating how well the active BHMFs derived from the SMF can reproduce the BL AGN UV LF, we show in Appendix C a key consistency check. We convert our reference SMF, which we use as input to generate BHMFs, to a star formation rate (SFR) function via convolution with the star-forming main sequence of \citet{2024ClarkeMSApJ...977..133C}. The resulting SFR function is consistent with the UV LFs of \citet{2022HarikaneUVLF} and \citet{2022FinkelsteinUVLF}, with only a slight overprediction at bright luminosities which is expected due to dust extinction. This test ensures that the SMF is representative of the full population of UV galaxies at $z\sim5$, consistent with the vast dominance of star-forming galaxies at this redshift \citep[e.g.][]{2025MerlinOJAp....8E.170M}, and thus it can be safely used as a starting point to infer the UV LF of AGN. 

\begin{figure*}
    \centering
    \includegraphics[width=0.9\textwidth]{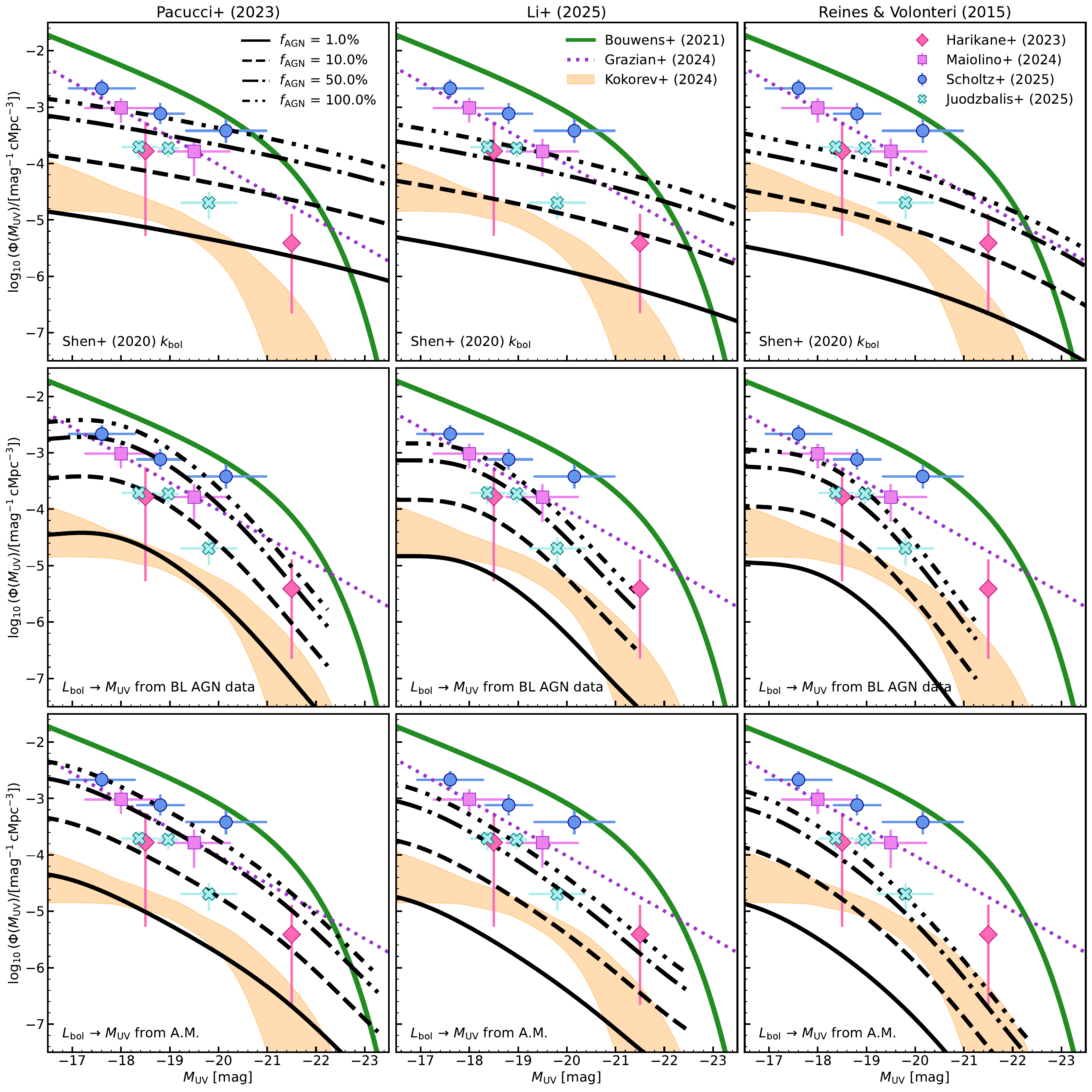}
    \caption{A comparison of the $z=5.5$ UV luminosity function obtained from the SMF to that of the LRDs and the galaxy UV LF. As in Fig. \ref{fig:bolLFcomp}, each of the columns uses a different $\Mbh-\Mstar$ relation when computing the UV LF, these are: \citetalias{Pacucci2023_LRDs} (\emph{left}), \citetalias{2025LiTipOf} (\emph{middle}), and \citet{Reines_2015} (\emph{right}). In each a different mapping from $\Lbol$ to $\Muv$ is used, these are: the UV-bolometric correction from \citet[][\emph{top}]{Shen_2020}, the power law fitted to the data in the $\Lbol-\Muv$ plane (\emph{middle}), and the $\Lbol-\Muv$ relation from abundance matching (\emph{bottom}). The UV LF obtained from the SMF assuming different values of $\fAGN$ are displayed as the black lines, where the line styles denote the assumed AGN fraction ($\fAGN$). The galaxy UV LF from \citet{2021BouwensUVLF} is denoted by the solid greenline. The UV LFs of BL AGN \citep[][]{2023ApJ...959...39H,maiolino2024jadesdiversepopulationinfant,2025IgnasarXiv250403551J} and NL AGN \citep[][]{2023ScholtzarXiv231118731S} are displayed as points. Whereas, the orange shaded region denotes the LRD UV LF of \citet{Kokorev2024_PhotoConcensusOfLRDs} and the purple dotted line denotes the intermediate UV LF of \citet[][their option 2]{2024GrazianApJ...974...84G}.}
    \label{fig:UVLFcomp}
\end{figure*}

As set forth in section \ref{sssec:MethodAGNUVLF} step ({\it vi}), we test three mappings between $\Lbol$ and $\Muv$ due to the uncertainties surrounding this mapping at high $z$, and the resulting UV LFs are displayed in Figure \ref{fig:UVLFcomp}. In the top row, we test the UV bolometric correction from \citet{Shen_2020} under the assumption that the majority of the high-$z$ population are ``typical'' AGN. In the middle row, we test an empirical mapping derived from our sample of high-$z$ BL AGN, under the assumption that the UV emission is AGN dominated, but there is non-negligible dust attenuation, scattered AGN emission, and/or host contamination. In the bottom row, we use abundance matching to derive a statistical mapping between the bolometric LF and UV LF of \citet{Kokorev2024_PhotoConcensusOfLRDs} under the assumption that LRDs and the wider lower-luminosity BL AGN population follow the same mapping. We display the UV LFs derived using the \citetalias{Pacucci2023_LRDs}, \citetalias{2025LiTipOf}, and \citetalias{Reines_2015} $\Mbh-\Mstar$ relations in the left-hand, middle, right-hand columns, respectively.

When adopting the \citet{Shen_2020} bolometric correction, we find that the that AGN fractions of $\fAGN\sim0.5,~1,~\gtrsim1$ are necessary to reproduce the determination from \citet{maiolino2024jadesdiversepopulationinfant} for the \citetalias{Pacucci2023_LRDs}, \citetalias{2025LiTipOf}, \citetalias{Reines_2015} relations, respectively. The empirical mapping boosts the UV LF at the faint end and suppresses it at the bright end, requiring lower AGN fractions to reproduce observations ($\fAGN\sim0.2,~0.8,~1$ for the \citetalias{Pacucci2023_LRDs,2025LiTipOf,Reines_2015} relations, respectively). Whereas, the mapping from abundance matching requires approximately the same AGN fraction to match the \citet{maiolino2024jadesdiversepopulationinfant} UV LF as the \citet{Shen_2020} bolometric correction. 

The \citet{Shen_2020} bolometric correction produces a much flatter UV LF for a mass-independent AGN fraction than the empirical and abundance-matched mappings, which are much steeper at the bright end, better matching the shape of the UV LFs of lower-luminosity BL AGN. From this we would infer that when adopting the \citet{Shen_2020} bolometric correction, the AGN fraction would depend strongly on mass / luminosity. Whereas, the empirical and abundance-matched mappings would suggest the AGN fraction is weakly dependent on mass / luminosity. 

As the high-$z$ quasar population follows the \citet{Shen_2020} bolometric correction (see Fig. \ref{fig:kbolFit}), we can estimate the average duty cycle of the total UV-luminous BL AGN population by comparing the model predictions to the UV LF of \citet{2024GrazianApJ...974...84G}. We compute the average duty cycle from the ratio of the number densities of the \citet{2024GrazianApJ...974...84G} UV LF to the model prediction (as in eq. \ref{eq:avfagn}) with integration limits \{$M_{\rm Min}$, $M_{\rm Max}$\} = \{$-30$,$-18$\} ${\rm mag}$ and find average AGN fractions of $\fAGN\sim0.3,~0.8,~1$ for the \citetalias{Pacucci2023_LRDs}, \citetalias{2025LiTipOf}, \citetalias{Reines_2015} relations, respectively. 

Contrastingly, while all three $\Mbh-\Mstar$ relations can broadly reproduce the \citet{maiolino2024jadesdiversepopulationinfant} UV LF when adopting the empirical mapping, none can reproduce the quasar population (even when using the fit that excludes the LRD data), indicating that this mapping can only be applicable for a subsample of BL AGN. To reproduce the total BL AGN UV LF, different mappings are likely needed for quasars and lower-luminosity AGN. However, we do not pursue this, as it would require knowledge of the relative abundance of these populations, as well as an understanding of where / why the transition happens between these populations. Furthermore, it is unclear whether the dust extinction inferred from SED fitting (which broadly aligns these lower-luminosity BL AGN with the \citealt{Shen_2020} bolometric correction) is truly tracing dust, as the trend of dust extinction with luminosity conflicts with that observed locally. 

Finally, when using the mapping from abundance matching, we are assuming that the LRDs' UV emission is AGN dominated. If instead this is host dominated \citep[as suggested by the BH* / BH envelope interpretation][]{2025NaiduBHstararXiv250316596N,2025GreenearXiv250905434G,2025DeGraaffarXiv251121820D,2025UmedaarXiv251204208U}, the LRD UV LF should lie much lower that the determination from \citet{Kokorev2024_PhotoConcensusOfLRDs}. This would remove the tension with the pre-JWST determinations of the faint end of the UV LF, but also results in the model predictions from the abundance matched mapping having little physical meaning.\\

In summary, under all three mappings, the AGN fraction necessary to reproduce the \citet{2022HarikaneUVLF} and \citet{2025IgnasarXiv250403551J} UV LFs is approximately consistent with the estimate from the BHMF, whereas a larger AGN fraction is necessary to reproduce the \citet{maiolino2024jadesdiversepopulationinfant} UV LF. If the current observational determinations are accurate, they would favour a high-normalisation $\Mbh-\Mstar$ relation, with only the \citetalias{Pacucci2023_LRDs} $\Mbh-\Mstar$ relation being able to reproduce the \citet{maiolino2024jadesdiversepopulationinfant} and \citet{2023ScholtzarXiv231118731S} UV LFs under all three mappings.

\subsection{The SMBH Mass Density}\label{ssec:rhoBH}
Under the assumption that SMBHs primarily grow via radiatively efficient accretion, the So\l{}tan Argument \citep[][]{Soltan1982_QuasarMasses} allows us to predict the evolution of the SMBH mass density ($\rho_{\rm BH}$) from the luminosity density ($\rho_{L}$) integrated across cosmic time. Recently, \citet{2025OJAp....8E...9J} suggested that the UV-luminous quasars made up $\leq10\%$ of the total BH mass density at $z\sim6$ derived from the SMF assuming the $\Mbh-\Mstar$ relation of \citetalias{2025LiTipOf}. This allows for the heavily obscured growth mode seen in the LRDs to be dominant at this epoch. However, this conclusion rests on a number of assumptions, starting from the choice of scaling relations and reference AGN LF. As we highlighted in the previous section, different data sets must be closely interconnected in specific ways to ensure consistency, for example high-normalisation scaling relations and low duty cycles, or vice versa. In this section we revisit the overall cumulative accretion from integrated AGN LFs and SMBH mass densities at $z\sim5.5$ within the framework of our comprehensive approach (section \ref{sssec:rhoBH_hiz}), and then use our reference models at $z\sim5.5$ as initial conditions to predict the implied SMBH mass densities down to $z\sim0$ (section \ref{sssec:rhoBH_loz}).

\begin{table}
    \centering
    \begin{tabular}{ccc}
        \hline
        Input & Redshift & $\log_{10}(\rho_{\rm BH}/[\msun\,{\rm cMpc^3}])$ \\
        \hline\hline
        Luminosity Function & & \\
        \hline
        \citet{Shen_2020} & 5.5 & $2.8_{-0.3}^{+0.3}$\\[0.5ex]
        Modified Shen & 5.5 & $4.4_{-0.3}^{+0.3}$\\[0.5ex]
        \citet{2025BarlowHallarXiv250616145B} & 5.5 & 4.68\\[0.5ex]
        \citet{akins2024LRDs} & 5.5 & $4.8_{-0.11}^{+0.30}$ \\[0.5ex]
        \citet{Kokorev2024_PhotoConcensusOfLRDs} & 5.5 & $4.4_{-0.5}^{+0.4}$ \\[0.5ex]
        \hline
        SMBH Mass Function & & \\
        \hline
        \citetalias{Reines_2015} & 5.5 & 2.84 \\
        \citetalias{Pacucci2023_LRDs} & 5.5 & 5.04 \\
        \citetalias{2025LiTipOf} & 5.5 & 4.21 \\
        \hline
        BHMF + LF & & \\
        \hline
        \citetalias{Reines_2015} & 0 & $5.73_{-0.12}^{+0.15}$ \\[0.5ex]
        \citetalias{Pacucci2023_LRDs} & 0 & $5.81_{-0.10}^{+0.12}$ \\[0.5ex]
        \citetalias{2025LiTipOf} & 0 & $5.74_{-0.12}^{+0.14}$ \\[0.5ex]
        \hline
    \end{tabular}
    \caption{The estimates of the SMBH mass density at $z=5.5$ obtained from the AGN LFs (\emph{top}) and the BHMFs presented in section \ref{ssec:BHMF} (\emph{middle}), as well as the estimates at $z=0$ using the BHMF as initial conditions and the AGN LF to give the subsequent evolution (\emph{bottom}). Applying the \citet{2025GreenearXiv250905434G} correction to the \citet{Kokorev2024_PhotoConcensusOfLRDs} and \citet{akins2024LRDs} LFs would correspondingly reduce the SMBH mass density at $z=5.5$ by an order of magnitude.\\
    \emph{Note}: In eq. (\ref{eq:rhoBH_z5.5}), we have chosen an upper integration mass limit of $10^{10}~\msun$. However, if we impose the lower value of $10^9~\msun$ used in \citet{2025OJAp....8E...9J}, we find that the \citetalias{Reines_2015} value is insensitive to the change, whereas the values obtained from the \citetalias{Pacucci2023_LRDs} and \citetalias{2025LiTipOf} decrease by $\sim0.1~{\rm dex}$.}
    \label{tab:rho_BH}
\end{table}

\subsubsection{The SMBH Mass Density at $z\geq5.5$}\label{sssec:rhoBH_hiz}

\begin{figure*}
    \centering
    \includegraphics[width=0.9\textwidth]{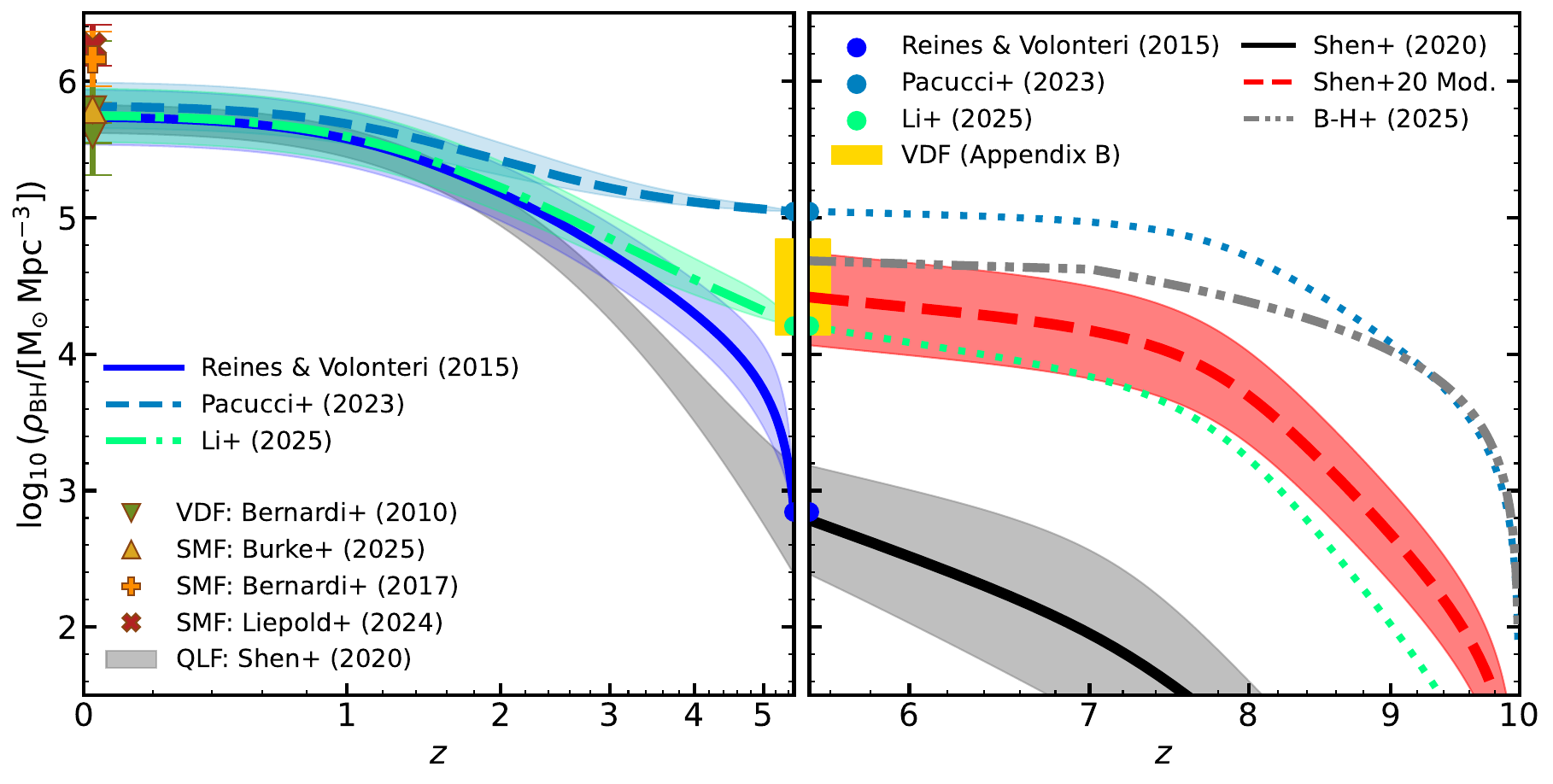}
    \caption{\emph{Left:} The SMBH mass density from $z=5.5$ to $z=0$ computed from our reference luminosity function and with the initial conditions obtained using the $\Mbh-\Mstar$ relation of \citetalias{Reines_2015} (solid dark blue line), \citetalias{Pacucci2023_LRDs} (dashed light blue line), and \citetalias{2025LiTipOf} (dot-dashed mint green line). These curves are compared to the prediction from the \citet{Shen_2020} quasar luminosity function (grey shaded region). The local estimates displayed are the estimate derived from the VDF \citet{2010MNRAS.404.2087B} assuming the $\Mbh-\sigma_{\star}$ relation of \citet{2019deNicola} (lower green upturned-triangular marker) and \citet{2021ApJ...921...36B} (upper green upturned-triangular marker), the estimate of \citet{2025Burke} (tan triangular marker), the estimate of \citet{2024ApJ...971L..29L} (red x-shaped marker), and the estimate from the \citet{2017MNRAS.467.2217B} SMF and the \citet{KormendyAndHo2013} $\Mbh-\Mstar$ relation (orange plus-shaped marker). If we adopt only the \citet{Shen_2020} LF the \citetalias{Pacucci2023_LRDs} and \citetalias{2025LiTipOf} curves are impacted negligibly, where as, because the \citetalias{Reines_2015}-based $\rho_{\rm BH}(z=5.5)$ is consistent with the prediction of \citet{Shen_2020}, the \citetalias{Reines_2015} curve would follow the same evolution (grey shaded region). \emph{Right:} The SMBH mass density from $z=10$ to $z=5.5$ computed using the \citet{Shen_2020} LF (black line), our reference LF (red dashed line), and the \citet{2025BarlowHallarXiv250616145B} X-ray LF (grey double-dot-dashed line). These are compared to the estimate from the VDF (yellow shaded region) and the mass densities obtained from the SMF with the three $\Mbh-\Mstar$ relations (circular markers). In addition we show the show the mass density obtained by tuning the high-$z$ evolution of the LF to match the estimates from the SMF at $z=5.5$ (dotted lines, see text for details). In Fig. \ref{fig:SAMcomp_rhoBH}, we compare our predicted evolution of the SMBH mass density to the predictions of several theoretical models.}
    \label{fig:rhoBH_comp_toZ0}
\end{figure*}

As described in section \ref{sssec:MethodRhoBH} step ({\it vi}), we compute the SMBH mass density at $z=5.5$ from the BHMFs presented in the previous section via equation (\ref{eq:rhoBH_z5.5}). We then compare these value to SMBH mass density at $z=5.5$ predicted by the So\l{}tan argument, computed via equation (\ref{eq:rhoBH_z0}) for several LFs. The resulting values are listed in Table \ref{tab:rho_BH} and their evolution displayed in the right-hand panel of Figure \ref{fig:rhoBH_comp_toZ0}.

From the values listed in Table \ref{tab:rho_BH} (as well as the right-hand panel of \ref{fig:rhoBH_comp_toZ0}) we see that the \citet{Shen_2020} quasar LF can fully account for the BH mass density inferred from the \citetalias{Reines_2015}-based BHMF, whereas the BH mass density obtained from our reference LF is in good agreement with the BH mass density derived from the \citetalias{2025LiTipOf}-based BHMF, and the BH mass density implied by the \citetalias{Pacucci2023_LRDs} is even higher than this, aligning with the $1\sigma$ upper bound of the mass density derived from the \citet{akins2024LRDs} LF. However, applying the correction of $1~{\rm dex}$ to the bolometric luminosities suggested by \citet{2025GreenearXiv250905434G}, would reduce the mass densities obtained from the LRD LFs by an order of magnitude, consistent with the LRDs being a small subpopulation of BL AGN even when accounting for selection effects in the $\Mbh-\Mstar$ relation. From this direct comparison it emerges that the fraction of implied obscured sources, i.e. of sources not recorded in the quasar LF, will depend on the class of model considered, being equal to the obscured fractions implied by the \citet{Shen_2020} quasar LF in the \citetalias{Reines_2015} model, and significantly larger when considering the other two models, in line with \citet{2025OJAp....8E...9J}. 

Interestingly, we find that there is good consistency at $z=5.5$ between the \citetalias{2025LiTipOf} estimate, the estimate from the VDF (see Appendix \ref{app:highZVDF}), and the mass density computed from direct integration of our reference LF. While there are uncertainties originating from the assumptions necessary to compute the SMBH mass density from the VDF and LF at this epoch, the consistency between multiple observables suggests a preference for the \citetalias{2025LiTipOf} relation over the \citetalias{Pacucci2023_LRDs} relation, which would lead to an inconsistency between the SMBH mass density predicted by the $\Mbh-\Mstar$ relation and the $\Mbh-\sigma_{\star}$ relation or integrated AGN emissivity. On the other hand, an overestimation of the BH masses at high-$z$ would also lead to a lower-normalisation $\Mbh-\sigma_{\star}$ relation, and an approximate consistency between the \citetalias{Reines_2015} relation, the VDF, and the \citet{Shen_2020} LF.

The $z=5.5$ \citetalias{Pacucci2023_LRDs}-based mass density is similar to the prediction of the CAT super-Eddington model \citep[$\log(\rho_{\rm BH})=5.3$;][]{2024TrincaLRDs} and the {\sc A-SLOTH} heavy-seeding model from \citet[][$\log(\rho_{\rm BH})\sim5.3$]{2025JeonApJ...988..110J}, whereas the \citetalias{2025LiTipOf}-based mass density (as well as the prediction of the reference LF) is consistent with the CAT Eddington-limited model \citep[$\log(\rho_{\rm BH})=3.8$ for $\Mbh\geq10^5~\msun$;][]{2022MNRAS.511..616T}, the {\sc Shark} v2.0 model \citep[$\log(\rho_{\rm BH})=3.9$][]{2018LagosSHARKMNRAS.481.3573L,2024LagosSharkV2MNRAS.531.3551L}, and the {\sc A-SLOTH} light-seeding model from \citet[][$\log(\rho_{\rm BH})\sim4.5$]{2025JeonApJ...988..110J} and the CAT Eddington-limited model is more consistent with the \citetalias{Reines_2015}-based estimate when making a more stringent mass cut ($\log(\rho_{\rm BH})=3.2$ for $\Mbh\geq10^6~\msun$). A comparison plot of the predicted SMBH mass density with the predictions of several theoretical models can be found in Fig. \ref{fig:SAMcomp_rhoBH}.

\begin{figure}
    \centering
    \includegraphics[width=0.45\textwidth]{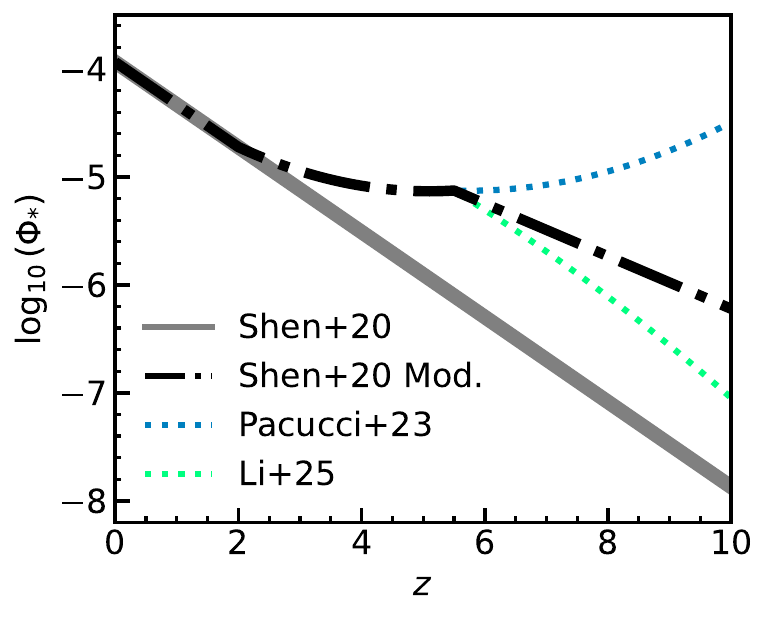}
    \caption{The evolution of the normalisation ($\Phi_{*}$) at the break luminosity ($L_{*}$) for the fiducial \citet{Shen_2020} global fit A (grey solid line), our reference LF (black dot-dashed line), and the evolution necessary to match the estimates from the SMF with an initial redshift of $z_{\rm i}=10$ (dotted lines).}
    \label{fig:NormEvolution}
\end{figure}

In the right-hand panel of Figure \ref{fig:rhoBH_comp_toZ0} we show the integrated SMBH mass density obtained from our reference AGN LF (red dashed line), which is comparable (within a factor of $2$) to the mass density obtained from the integration of the \citet{2025BarlowHallarXiv250616145B} X-ray LF (grey double-dot-dashed line), which has been corrected to bolometric following the \citet{Shen_2020} correction. For completeness, we also include the mass density from the fiducial \citet{Shen_2020} global fit A (solid black line). The dotted blue and mint green lines show the SMBH mass density when we tune the evolution of the LF to match the \citetalias{Pacucci2023_LRDs} and \citetalias{2025LiTipOf}-based estimates at $z=5.5$, respectively. For this we modify the high-$z$ evolution of the LF normalisation by setting $d_1(z=10)=-0.086,~-0.319$ for the \citetalias{Pacucci2023_LRDs} and \citetalias{2025LiTipOf} relations, respectively, and linearly interpolate these to $d_1(z=5.5)=-0.2436$. As clearly displayed in Figure \ref{fig:NormEvolution}, while the \citetalias{2025LiTipOf}-based mass density at $z=5.5$ can be reproduced with a high-$z$ evolution midway between that of our reference LF and the fiducial \citet{Shen_2020} fit, a much shallower evolution is necessary to reproduce the \citetalias{Pacucci2023_LRDs}-based mass density (even if we were to start the integration at $z_{\rm i}=20$).

\subsubsection{The SMBH Mass Density at $z<5.5$}\label{sssec:rhoBH_loz}

The left-hand panel of Figure \ref{fig:rhoBH_comp_toZ0} displays the evolution of the BH mass density from $z=5.5$ to present day. We observe that, despite the initial conditions spanning over two orders of magnitude, the predicted BH mass density at $z=0$ are within $0.1~{\rm dex}$ of each other, as well as being within $0.1~{\rm dex}$ of the local estimate of \citet{2025Burke} and $0.2~{\rm dex}$ of the value obtained from the \citet{2010MNRAS.404.2087B} VDF with either the \citet{2019deNicola} or the \citet{2021ApJ...921...36B} $\Mbh-\sigma_{\star}$ relation \citep[assuming the aperture correction of ][]{2021deGraaffApJ...913..103D}. While the \citetalias{Reines_2015} and \citetalias{2025LiTipOf} curves suggest that only a small fraction of the present day BH mass density had already accumulated by $z=5.5$ (specifically $\sim0.1\%$ for \citetalias{Reines_2015} and $\sim3\%$ for \citetalias{2025LiTipOf}), the \citetalias{Pacucci2023_LRDs} curve implies that $\sim18\%$ of the present day BH mass density was already in place at this epoch, leading to the relatively small amount of evolution displayed in Figure \ref{fig:rhoBH_comp_toZ0}. 

All our accretion models, characterised by the same radiative efficiency ($\epsilon_{\rm r}=0.1$), are roughly consistent with the $z=0$ SMBH mass density derived from the VDF and $\Mbh-\sigma_{\star}$ relation (downward green triangles in Figure \ref{fig:rhoBH_comp_toZ0}) but fall short in reproducing the estimate from the SMF and $\Mbh-\Mstar$ relation (the ``x'' and ``+''-shaped markers in Fig. \ref{fig:rhoBH_comp_toZ0}). This discrepancy, was already noted in previous works \citep[e.g.][]{Tundo_2007} and, in particular, \citet{Shankar2020NatAs...4..282S} and \citet{2024ApJ...976....6Z} have highlighted the inconsistency between accretion models and the BHMF obtained from the $\Mbh-\Mstar$ relation which would require extremely low values of the radiative efficiency, possibly an indication of a bias more pronounced in the $\Mbh-\Mstar$ relation \citep[e.g.][]{Shankar_2016,2025ShankarMNRAS.tmp..713S}. \citet{2025LubertoarXiv250814164L} recently investigated the radiative efficiency necessary to reconcile a high-$z$ estimate of $\rho_{\rm BH}$ based on the \citetalias{Pacucci2023_LRDs} relation with the local estimate, finding a need for an even higher radiative efficiency of $\epsilon_{\rm r}\sim0.2$, further corroborating the results presented here and by \citet{Shankar2020NatAs...4..282S} and \citet{2024ApJ...976....6Z}.

We note that pairwise residual analysis consistently indicates that stellar velocity dispersion is the galactic variable most strongly correlated with SMBH mass, thus, supporting the VDF-based BHMF. In this logic, the initial condition should also be based on the $\Mbh-\sigma_{\star}$ relation, which as discussed earlier on, appears to be consistent between high-$z$ and local AGN. As a reliable estimate of the $z\sim5$ VDF is not available, in Appendix \ref{app:highZVDF} we deploy the framework of \citet{Marsden2021_sigma} and find the predicted VDF-based $\rho_{\rm BH}$ at $z=5.5$ to lie below the estimate based on the \citetalias{Pacucci2023_LRDs} relation, more aligned with the estimate based on the \citetalias{2025LiTipOf} relation, which would create more room for a steadier growth of SMBHs down to $z=0$.

\subsection{Forward Modelling the SMBH Demography}\label{ssec:Ceq}

\begin{figure*}
    \centering
    \includegraphics[width=0.9\textwidth]{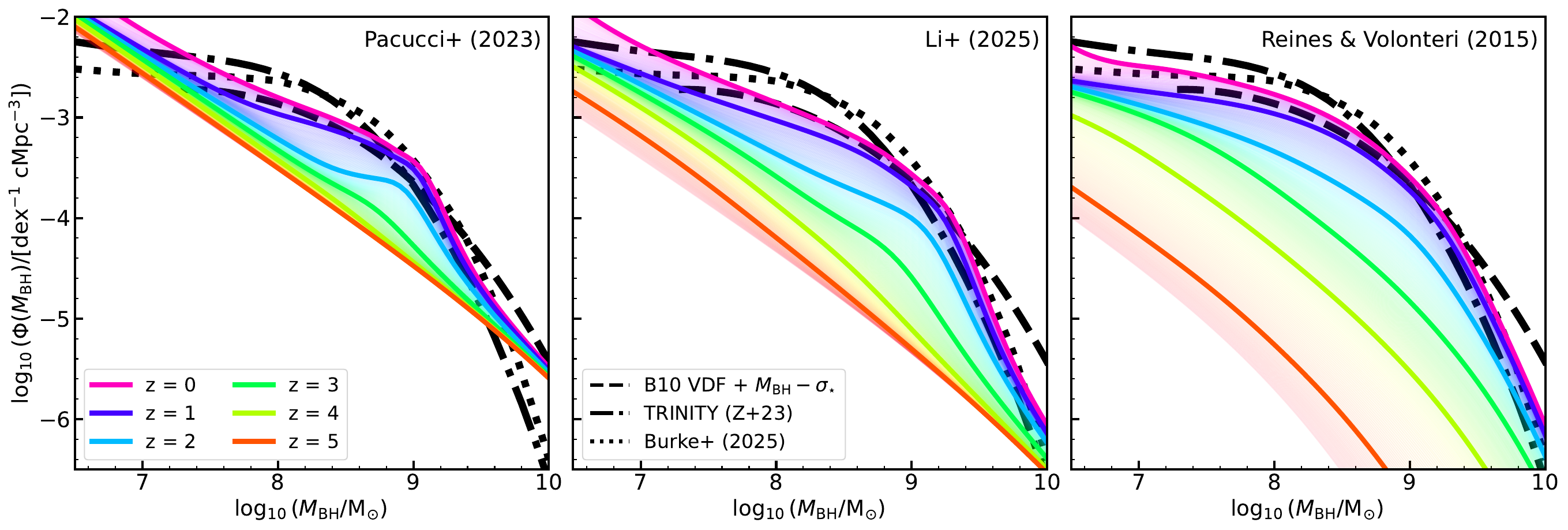}
    \caption{comparison of the evolution of the BHMF predicted by the continuity equation for the \citetalias{Pacucci2023_LRDs} (left-hand panel), \citetalias{2025LiTipOf} (middle panel), and \citetalias{Reines_2015} (right-hand panel) models. The colour gradient denotes the continuous redshift evolution, with the solid coloured lines denoting the BHMF at integer steps in redshift. The black dashed, dot-dashed, and dotted lines denote the local estimates from the \citet{2010MNRAS.404.2087B} VDF using the $\Mbh-\sigma_{\star}$ relation of \citet{2019deNicola} and the aperture correction of \citet{2021deGraaffApJ...913..103D}, from the TRINITY model \citet{2023MNRAS.518.2123Z}, and from the BH occupation fraction \citet{2025Burke}.}
    \label{fig:BHMF Ceq}
\end{figure*}

In this section we present the evolution of the BHMF predicted by the continuity equation technique. As described in detail in section \ref{ssec:CeqMethod}, we employ the continuity equation following the methodology of \citet{2017A&A...600A..64T}, taking the BHMFs inferred from the SMF using the three $\Mbh-\Mstar$ relations (Fig. \ref{fig:BHMF}) as initial conditions, and adopting the ERDF of \citet{2017A&A...600A..64T} but adjusting its parameters at high redshift to approximately reproduce the ERDF of quasars at $z\sim5-7$. The resulting BHMFs are displayed in Fig. \ref{fig:BHMF Ceq} where the colour gradient denotes the continuous redshift evolution and the solid coloured lines display the BHMF at integer steps in redshift. The aim of this section is not to provide a comprehensive and final model for the evolution of the BHMF with time, but rather to merely check the consistency with the local estimates in Figure \ref{fig:BHMF Ceq}. There are, of course, uncertainties that arise due to some of the input assumptions in our modelling (e.g. the ERDF, obscured fraction, AGN LF, neglecting mergers, etc.), however, our conclusions are minimally impacted by these.

The left-hand panel of Figure \ref{fig:BHMF Ceq} displays the resulting evolution of the BHMF from the \citetalias{Pacucci2023_LRDs} initial conditions. We observe that the majority of the evolution occurs areound the knee of the mass function ($\Mbh\sim10^{8-9.5}~\msun$), with minimal evolution outside of this domain. The resulting local BHMF is broadly consistent with the VDF-based estimate, but with a significant departure and steepening below $\Mbh\sim10^{7.5}~\msun$. This discrepancy could of course mirror an incompleteness in the local VDF and/or too high an initial condition. Including BH mergers may redistribute some of this mass but would not improve this mismatch significantly, as also recently shown by \citet{2025LapiGWBarXiv250715436L} who deployed cosmologically motivated BH mergers in the continuity equation using the Smoluchowski formalism \citep[][]{1916SmoluchowskiEqZPhy...17..557S}. The inclusion of mergers may also lead to the high-mass end of the BHMF exceeding the VDF-based estimate.

The middle panel of Figure \ref{fig:BHMF Ceq} displays the results of the accreted mass functions using the \citetalias{2025LiTipOf} BHMF as initial conditions. Similarly to the \citetalias{Pacucci2023_LRDs} BHMF, the majority of the evolution from the \citetalias{2025LiTipOf} initial conditions occurs around the knee. However, as the initial conditions aren't as steep, there is more evolution at all masses than in the \citetalias{Pacucci2023_LRDs} case. The resulting local BHMF is consistent with the \citet{2025Burke} SMF-based estimate at high masses, the VDF-based estimate at the knee, but still displays a steepening below $\Mbh\sim10^{7.5}~\msun$ (albeit, slightly less significant than in the \citetalias{Pacucci2023_LRDs} case). Including BH mergers may redistribute some mass, particularly impacting the high-mass end and possibly bringing the BHMF in line with the VDF-based estimate.

The right-hand panel of Figure \ref{fig:BHMF Ceq} displays the resulting evolution of the BHMF from the \citetalias{Reines_2015} initial conditions. We observe that there is significant redshift evolution of the BHMF across the mass range, with the $z=0$ prediction mostly aligning with the SMF and VDF-based BHMFs at the high-mass end and knee, respectively, whereas it tends toward the TRINITY \citep{2023MNRAS.518.2123Z} estimate in the low masses limit. By $z=3$ the BHMF is $\sim10\%$ of its value at $z=0$, but still lies just below the $z=5.5$ BHMF obtained from the SMF and \citetalias{Pacucci2023_LRDs} relation. The level of evolution implied by this model is consistent with that typically predicted by other accretion models \citep[e.g.][]{2008MerloniMNRAS.388.1011M,Shankar2009,2010CaoApJ...725..388C,Aversa_2015,2017A&A...600A..64T,2022SiciliaApJ...934...66S} as well as cosmological hydrodynamic simulations \citep[e.g.][]{Habouzit2021_SMBHsInSims}.

We note that a BH occupation fraction below unity at moderate to low stellar masses \citep[such as those proposed  by][]{2015MillerApJ...799...98M,2025ZouApJ...992..176Z} would induce a more marked flattening of the local BHMF at $\Mbh\lesssim10^{6-7}~\msun$. In turn, this flattening would exacerbate the tension with the steep low-mass end of the accreted BHMF. While mergers may redistribute some of this mass, the \citetalias{Pacucci2023_LRDs}-based BHMF would be in tension with this already at $z=5.5$, requiring extremely high merger rates that are not supported by current models \citep[e.g.][]{2025LapiGWBarXiv250715436L}. Exploring this in detail is beyond the scope of this work, however we plan to address this in future work.

\subsubsection{The Average Duty Cycle}

\begin{figure}
    \centering
    \includegraphics[width=0.45\textwidth]{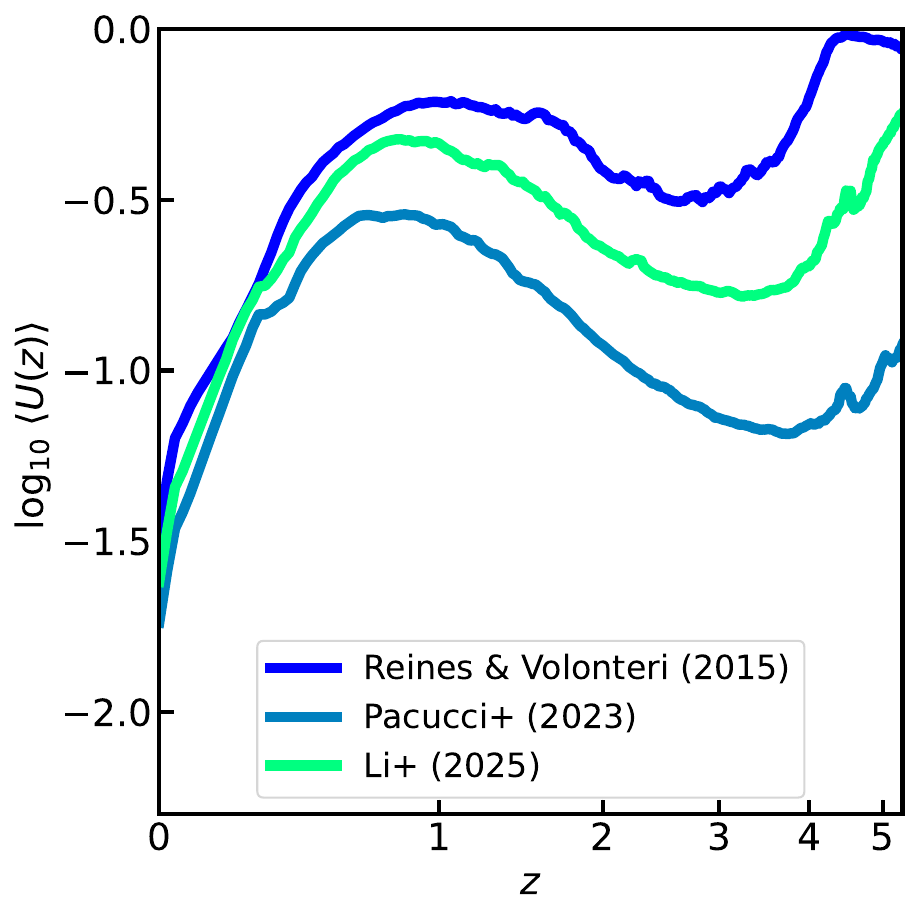}
    \caption{A comparison of the average duty cycle (computed via equation \ref{eq:meanDC}) for the \citetalias{Reines_2015} (upper dark blue line), \citetalias{Pacucci2023_LRDs} (lower light blue), and \citetalias{2025LiTipOf} (middle green line) models.}
    \label{fig:DC}
\end{figure}

In addition to the evolution of the BHMF, the continuity equation approach also predicts the duty cycle as a function of BH mass and redshift including all active SMBH shining above the minimum luminosity recorded in the input AGN LF. The redshift evolution of the mean duty cycle of the SMBH population computed from equation (\ref{eq:meanDC}) for the three sets of initial conditions, $\langle U(z)\rangle$, is displayed in Figure \ref{fig:DC}. The average duty cycle is initially in good agreement with the values obtained from the BHMF in section \ref{ssec:BHMF} ($\fAGN\sim0.08,\,0.5,\,\gtrsim1$ from the BHMF for the \citetalias{Pacucci2023_LRDs}, \citetalias{2025LiTipOf}, and \citetalias{Reines_2015} relations, respectively). At later times the duty cycle declines from $z=5.5$ to $z\sim3$ for all three $\Mbh-\Mstar$ relations. This is the result of the weak evolution of the LF normalisation over this period (see Fig. \ref{fig:NormEvolution}). The duty cycle then increases in all three models, peaking at $z\sim1$, before declining to present day and converging to similar values at $z\lesssim0.5$.

At $z\sim4$, the \citetalias{Pacucci2023_LRDs} model predicts an average duty cycle of $\sim5\%$, broadly consistent with the estimate from the predictions of the super-Eddington CAT model \citep[$\sim1-4\%$][]{2024TrincaLRDs}, but lower than is typically predicted by cosmological hydrodynamic simulations \citep[see Fig. 8 of][]{Habouzit2021AGNInSims}. Whereas, the \citetalias{2025LiTipOf} model predicts an average duty cycle of $\sim20\%$, consistent with the predictions of some accretion models and the Eagle simulation \citep[e.g.][]{2017A&A...600A..64T,Habouzit2021AGNInSims}, and the \citetalias{Reines_2015} model predicts an average duty cycle of $\sim40\%$ \citep[similar to TNG100 and below the prediction of Horizon AGN][]{Habouzit2021AGNInSims}. Even at $z\sim1$ there is a factor of $\sim2$ difference between the average duty cycle of the \citetalias{Pacucci2023_LRDs} and \citetalias{2025LiTipOf} models which could potentially be investigated through clustering analysis by future deep, wide-field surveys.

\subsubsection{The Evolution of the $\Mbh-\Mstar$ Relation}

\begin{figure*}
    \centering
    \includegraphics[width=0.9\textwidth]{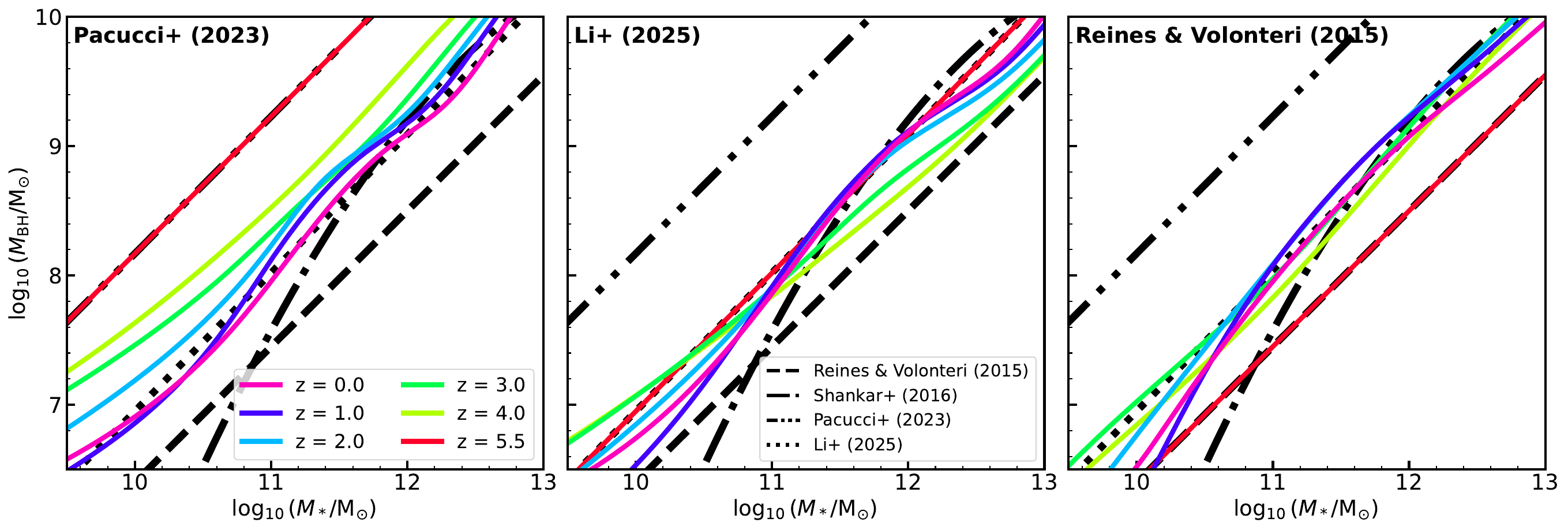}
    \caption{A comparison of the evolution of the $\Mbh-\Mstar$ relation obtained from abundance matching between the observed SMF and the BHMF derived from the continuity equation. The \emph{left} panel displays the results of the \citetalias{Pacucci2023_LRDs} model, the \emph{middle} panel displays the results of the \citetalias{2025LiTipOf} model, and the \emph{right} panel displays the results of the \citetalias{Reines_2015} model. The solid coloured lines denote the derived relation at integer steps in redshift and the black dashed, dot-dashed, double-dot-dashed, and dotted lined denote the local relations of \citetalias{Reines_2015}, \citet{Shankar_2016}, \citetalias{Pacucci2023_LRDs}, and \citetalias{2025LiTipOf}, respectively.}
    \label{fig:MstarMbhEvo}
\end{figure*}

For completeness, we also compute the predicted $\Mbh-\Mstar$ relation at each redshift via abundance matching between the observed SMF and our BHMF derived from the continuity equation. The resulting evolution of the $\Mbh-\Mstar$ relation for the three cases is displayed in Figure \ref{fig:MstarMbhEvo}.

As expected from the agreement displayed in Figure \ref{fig:BHMF Ceq}, we find the local $\Mbh-\Mstar$ relations obtained from all three sets of initial conditions to be in good agreement with one another, as well as with the observed local relations. Specifically, the relations agree approximately in both slope and normalisation with the \citetalias{Reines_2015} local relation at $\Mstar\lesssim10^{11}~\msun$ and with the local determination of \citet{Shankar_2016} above this. Of course, the implementation of the continuity equation only includes the growth of BHs via accretion, whereas growth via SMBH mergers is expected to steepen the relation at the high-mass end, but not significantly impact the overall evolution in normalisation. The most striking difference among the three cases is the level of evolution. We observe that for the \citetalias{Reines_2015} and the \citetalias{2025LiTipOf} models, the predicted $\Mbh-\Mstar$ relations display little evolution, being within the parameter space of local determinations from $z=5.5$ to present. However, the \citetalias{Pacucci2023_LRDs} case displays significant and rapid evolution at early times, due to the BHMF evolving very little while the SMF evolves more significantly, implying a low $\langle{\rm sBHAR}/{\rm sSFR}\rangle$ ratio which dictates the motion in the $\Mbh-\Mstar$ plane. In addition, the evolution characterising the \citetalias{Pacucci2023_LRDs} model, which predicts a BH mass density comparable to the local estimate already by $z\sim2.5$, is approximately consistent with the observations of \citet{2025IgnasarXiv250403551J} who find their lowest redshift bin ($2<z<3.5$) to have converged to the local relation.  However, this evolution may also be somewhat driven by the differing level of observational biases at different redshifts.

\begin{figure}
    \centering
    \includegraphics[width=0.45\textwidth]{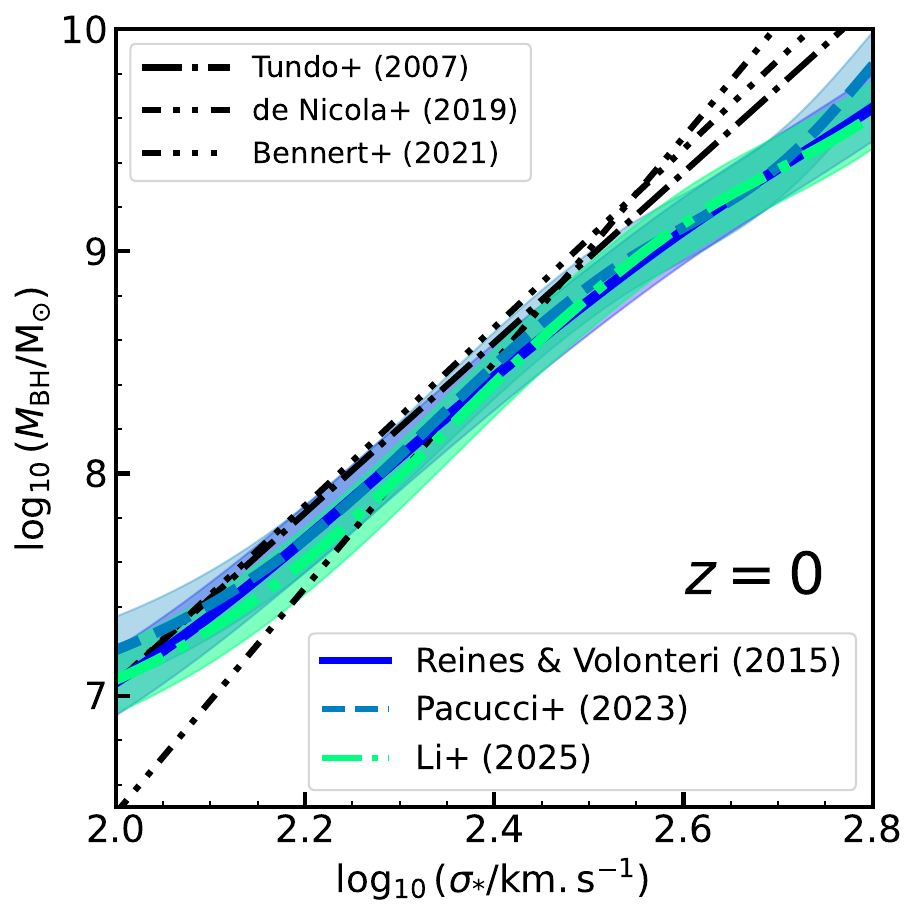}
    \caption{A comparison of the $\Mbh-\sigma_{\star}$ relation obtained by abundance matching between the local BHMF derived using the continuity equation and the VDF of \citet{2010MNRAS.404.2087B} to the literature relations of \citet{Tundo_2007}, \citet{2021ApJ...921...36B}, and \citet{2019deNicola}. The \citet{2021ApJ...921...36B} and \citet{2019deNicola} relations have been aperture corrected to $R_{\rm ap} = R_{\rm e}/8$ in order to match the \citet{2010MNRAS.404.2087B} VDF and the \citet{Tundo_2007} relation. We have assumed the intrinsic scatter of \citet{Tundo_2007} as it uses stellar velocity dispersions computed in the same aperture as the \citet{2010MNRAS.404.2087B} VDF ($R_{\rm ap} = R_{\rm e}/8$). The relation using the BHMF from the \citetalias{Reines_2015} model is denoted by the solid dark blue line, the relation using the BHMF of the \citetalias{Pacucci2023_LRDs} model is denoted by the light blue dashed line, and the relation using the BHMF of the \citetalias{2025LiTipOf} model. The $1\sigma$ uncertainty region is denoted by the corresponding shaded region of the same color. We observe that the $\Mbh-\sigma_{\star}$ relation of all three models are in good agreement with one another and with the relation of \citet{Tundo_2007}.}
    \label{fig:MbhSig}
\end{figure}

We carry out a similar exercise for the $\Mbh-\sigma_{\star}$ relation at $z=0$ via abundance matching between the SDSS VDF of \citet{2010MNRAS.404.2087B} and the BHMF from the continuity equation, and adopt the intrinsic scatter found by \citet{Tundo_2007} of $0.15~{\rm dex}$. The resulting $\Mbh-\sigma_{\star}$ relation is displayed in Figure \ref{fig:MbhSig}. We find the local relation in all three cases to be consistent with the \citet{Tundo_2007} relation and the \citet{2021ApJ...921...36B} relation, with only a slight deviation at the highest masses, which would may reduce through the inclusion of mergers. In Appendix \ref{app:highZVDF} we predict the evolution of the $\Mbh-\sigma_{\star}$ relation by adopting the VDF predicted by the theoretical framework of \citet{Marsden2021_sigma}. We find that, depending on the exact inputs, both the \citetalias{Pacucci2023_LRDs} and \citetalias{2025LiTipOf} BHMFs predict a $\Mbh-\sigma_{\star}$ that is almost constant in redshift and consistent with the local relation. \\

To summarise, by forward modelling the BHMF using the continuity equation, we have demonstrated that all three high-$z$ BHMF constructed using the $\Mbh-\Mstar$ relations can be reconciled with the local estimates of the BHMF, although with an upturn at low masses for the \citetalias{Pacucci2023_LRDs} and \citetalias{2025LiTipOf} models. In addition, the \citetalias{Pacucci2023_LRDs} model shows very little redshift evolution in the BHMF, but a strong evolution in the $\Mbh-\Mstar$ relation \citep[similar to the zevo seed case in][]{2024ApJ...976....6Z}. On the other hand, the \citetalias{Reines_2015} and \citetalias{2025LiTipOf} models predict more evolution in the BHMF and a $\Mbh-\Mstar$ with minimal evolution.

\section{Discussion}\label{sec:Discussion}

In this Section we first briefly describe AGN clustering as an additional powerful probe to discern among equally successful models. We briefly discuss how the LRDs fit into this work and then move on to discuss the possible observational systematics affecting our data-driven methodology, and conclude with a comparison to cosmological galaxy evolution models and a discussion of the wider implications.

\subsection{AGN Clustering as an Independent Constraint}\label{ssec:DiscussionClustering}

In this work, we have presented an investigation of the consistency of different data sets relating to the high-$z$ SMBH population. Specifically, we have built from the JWST $z=5.5$ SMF a corresponding BHMF via three $\Mbh-\Mstar$ relations that bracket the (large) systematics present. We found that to simultaneously reproduce the active BHMF and bolometric/UV AGN LFs, the duty cycles or fractions of active SMBHs should be $\sim~10\%$, $50\%$, and $\gtrsim100\%$, respectively, for the three reference $\Mbh-\Mstar$ scaling relations of \citetalias{Pacucci2023_LRDs}, \citetalias{2025LiTipOf}, and \citetalias{Reines_2015}.

Such a degeneracy could be in principle broken via independent AGN clustering measurements that set constraints on the host dark matter halo masses and thus on the implied duty cycles in terms of fraction of active haloes. As the quasar population appear to reliably follow the local bolometric correction of \citet{Shen_2020}, we can compare the predictions from the three $\Mbh-\Mstar$ relations with the observationally determined duty cycle of UV-luminous quasars from clustering analysis. \citet{2024EilersApJ...974..275E} estimate the number density of UV-luminous quasars ($\Muv\lesssim-26.5~{\rm mag}$) using the UV LF of \citet{2023SchindlerApJ...943...67S} and from the ratio to number density of host halos with $\Mhalo\geq10^{12.43}$, infer the duty cycle to be $\fduty=0.004_{-0.003}^{+0.012}$.

To make sure this is a fair comparison, we run a Monte Carlo model in which we sample the host halos with $\Mhalo\geq10^{12.43}$ from the halo mass function, assign stellar masses from the $\Mstar-\Mhalo$ relation (computed via abundance matching), and BH masses via the three $\Mbh-\Mstar$ relations. To each of these BHs we assign a bolometric luminosity via sampling our ERDF and assuming $\fAGN=1$, then convert this to a UV magnitude using the \citet{Shen_2020} bolometric correction. We then compute the UV LF and compute the duty cycle as the ratio of the \citet{2023SchindlerApJ...943...67S} UV LF to our model UV LFs in the range $\Muv\leq-26.5~{\rm mag}$. We find duty cycles of $\fduty=0.004^{+0.008}_{-0.003},~0.023^{+0.041}_{-0.015},~0.26^{+0.24}_{-0.14}$ for the \citetalias{Pacucci2023_LRDs}, \citetalias{2025LiTipOf}, \citetalias{Reines_2015} relations, respectively, where the uncertainties originate from adopting the $1\sigma$-upper and lower bounds on the minimum halo mass. These results indicate that a higher normalisation $\Mbh-\Mstar$ relation is favoured, hinting at a preference for the \citetalias{Pacucci2023_LRDs} relation at least in the quasar population. However, these findings are sensitive to the high-mass extrapolation and the intrinsic scatter of the relation, with more detailed future studies needed to robustly draw conclusions.

\subsection{The Little Red Dots in the Context of This Work}\label{ssec:DiscussionLRDs}

Throughout this work we have assumed that the LRDs are a small subpopulation of the wider BL AGN population, such that they contribute minorly (or even negligibly) to the BL AGN demographics and therefore, do not need to be carefully treated via separate prescriptions within our framework. Here, we briefly justify this assumption.

We estimate the relative abundance of LRDs from the ratio of the number density of LRDs to the number density of BL AGN. Specifically, we estimate the number density of LRDs by integrating our Schechter fit to the \citet{Kokorev2024_PhotoConcensusOfLRDs} LF, which we correct following \citet[][denoted by the left hand arrows in Fig. \ref{fig:K24bolLFfit}]{2025GreenearXiv250905434G} in the range $\Lbol\geq10^{42.5}~{\rm erg\,s^{-1}}$ and obtain the number density of BL AGN from our reference LF in the same range. From this we estimate that the LRDs only compose $2.1\%$ of the BL AGN population.

Similarly, we can gauge the LRDs contribution to the SMBH population by comparing the \citet{Kokorev2024_PhotoConcensusOfLRDs} LF which has been corrected following \citet{2025GreenearXiv250905434G} to our model predictions with $\fAGN=1$ via eq. (\ref{eq:avfagn}) for $\Lbol\geq10^{42.5}~{\rm erg\,s^{-1}}$. From this we estimate that LRDs compose $\sim1,~3,~4.5\%$ of the SMBH population for the \citetalias{Pacucci2023_LRDs}, \citetalias{2025LiTipOf}, and \citetalias{Reines_2015} relations, respectively. This suggests that the LRDs compose only a small fraction of the SMBH population ($\lesssim5\%$) independently of how the observed AGN relate to the underlying SMBH population.

\subsection{Systematics}\label{ssec:DiscussionSystematics}

The main systematics that may impact our findings are the reliability of the SMBH mass, luminosity, and number density estimates, as well as the assumptions within our methodology. We tackle each of these in turn and discuss to what extent they could impact our findings.

\subsubsection{SMBH Mass and Luminosity Estimates}

The SMBH mass measurements of the BL AGN have been obtained from single-epoch virial estimators using the broad Balmer lines. These estimators rely on the observed widths of the broad Balmer lines tracing the motion of the BLR clouds, as well as locally calibrated empirical relations. Recent direct measurements of the BLR dynamics from GRAVITY+ have suggested that single epoch methods may overestimate the SMBH masses in at least some high-$z$ systems \citep[][see also \citealt{2025ParlantiarXiv251214844P}]{2024AbuterNatur.627..281A,2025ElDayemarXiv250913911G}. Furthermore, not accounting for secondary correlations within these estimators -- particularly when working with small, flux-limited samples -- can potentially also bias your BH masses high at a population level \citep[][]{2024LupiA&A...689A.128L}. In addition, at least in the LRD subpopulation, there is growing evidence for the central SMBH being enshrouded in dense neutral gas \citep[e.g.][]{2025InayoshiApJ...980L..27I,2025NaiduBHstararXiv250316596N,2025deGraaffarXiv250316600D,2025DEugenioarXiv250311752D,2025DEugenioarXiv250614870D}, which would be responsible for the v-shaped SED, the SMBH mass being significantly overestimated by single-epoch virial estimators, and a large deviation from the locally calibrated bolometric corrections. 

In our framework, the \citetalias{Reines_2015} $\Mbh-\Mstar$ relation, representative of local AGN, has been assumed as the possible underlying scaling relation when both selection effects and SMBH mass overestimates are accounted for. However, this relation still requires very large, and sometimes unphysical, duty cycles of $\fAGN\gtrsim1$ to reproduce the large number densities of AGN at $z\sim5-6$, suggesting that either these estimates are also overestimated, as discussed below, and/or that the UV--bolometric corrections require further tuning.

\subsubsection{Number Density Estimates}
In section \ref{sec:Results} we tested our predicted AGN demography against several observational estimates of the number density of high-$z$ AGN which may, in turn, suffer from systematics. The \citet{taylor2024broadlineagn35z6black} active BHMF may be biased high as 21 of their sample of 62 spectroscopically confirmed BL AGN are LRDs. If the BH*/BH envelope interpretation is found to be widely applicable to the LRDs, then the BH masses of these objects would likely be overestimated by $\gtrsim2~{\rm dex}$, significantly impacting the resulting number density estimate. Furthermore, when computing the active BHMF, \citet{taylor2024broadlineagn35z6black} apply an incompleteness correction obtained through a Monte Carlo approach, which is significant at low masses ($\sim1~{\rm dex}$ at $\Mbh=10^{6.5}~\msun$). While removing the incompleteness correction alone does not reconcile the active BHMF with the total BHMF predicted by the \citetalias{Reines_2015} $\Mbh-\Mstar$ relation, if the LRD masses also get revised lower, then this may well reconcile the two. However, there would still be a discrepancy at the high-mass end between the BHMF inferred using the \citetalias{Reines_2015} relation and the active BHMF of \citet{2024HeApJ...962..152H}.

The UV LF of BL AGN may also be biased high due to host-galaxy contamination, which appears to be present in several sources of the \citet{maiolino2024jadesdiversepopulationinfant} and \citet{2025IgnasarXiv250403551J} samples (see Fig. \ref{fig:kbolFit}). An overestimation of these UV LFs would also impact our reference LF, which is tuned only to UV-based estimates at the faint end. However, we have neglected correcting these UV LF for dust extinction when converting them to bolometric, which would somewhat counterbalance such an effect. By using, among others, an empirical $\Muv-\Lbol$ mapping, we overcame some of these uncertainties, though some uncertainty in the UV LF of BL AGN remains (exemplified by the tension between these UV LFs and the recent upper limit of \citealt{2025LinarXiv250407196L}).

Finally, due to the small field of view of JWST, the nominal surveyed area is still quite small. So, a simple overestimation of the number density due to cosmic variance or uncertainty in the volume definition remains a possibility.

\begin{figure*}
    \centering
    \includegraphics[width=0.7\textwidth]{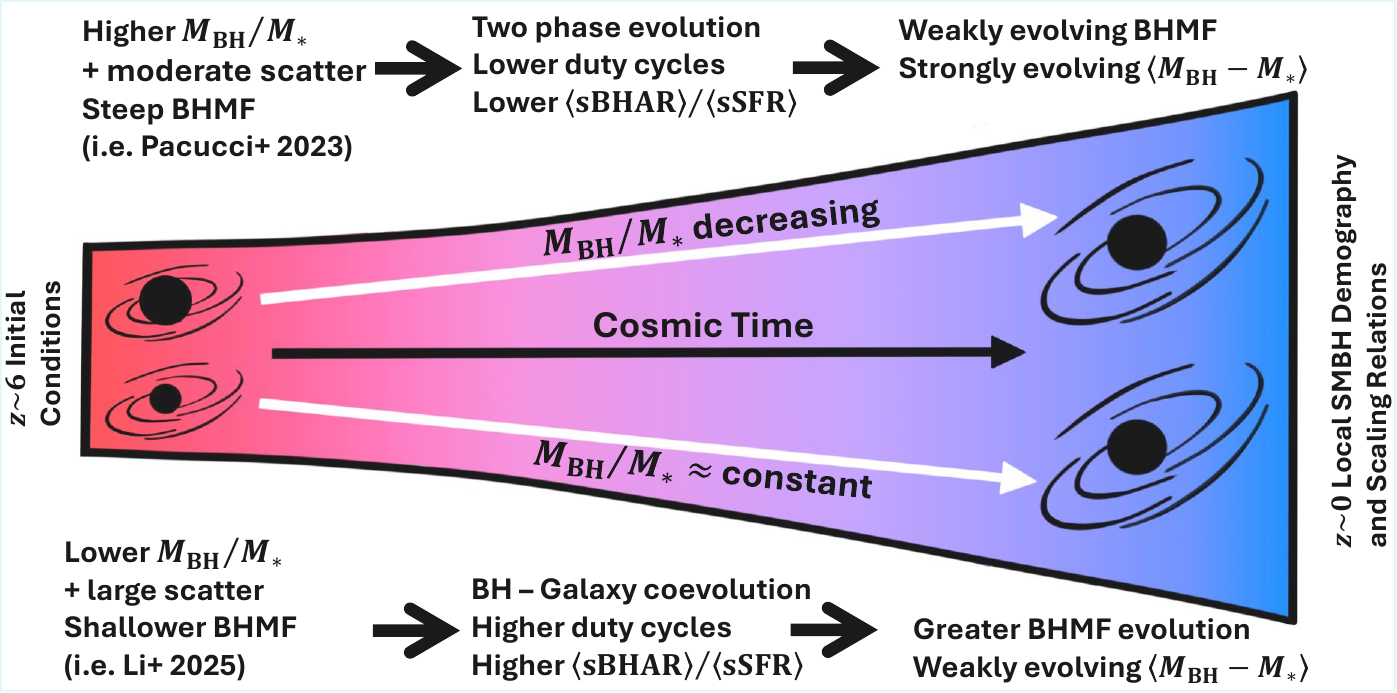}
    \caption{A diagram summarising the two broad classes of models that we find are consistent with the current observational data sets.}
    \label{fig:ViableModelsCartoon}
\end{figure*}

\subsection{Comparison to Theoretical Models}\label{ssec:DiscussionTheoreticalModels}

Theoretical modelling is key to conclusively decipher where these faint BL AGN sit in relation to the total underlying SMBH population, the implications for SMBH seeding, and the implications for the origin, evolution, and driving forces behind the SMBH--galaxy connection. Tackling this problem with cosmological hydrodynamic simulations, semi-analytic models (SAMs), and semi-empirical models (SEMs) in tandem offers the most promising approach, as they are complementary techniques, allowing different aspects of the problem to be probed. 

There are ongoing efforts to understand the JWST observations within theoretical frameworks. For instance, \citet{2024TrincaLRDs} recently showed that, in the framework of the Cosmic Archaeology Tool \citep[CAT;][]{2022MNRAS.511..616T,Trinca2023LRDs}, both the over-massive nature of the low-luminosity BL AGN and the \citet{akins2024LRDs} LF could be reproduced via episodic periods of super-Eddington accretion triggered by galaxy mergers, whereas their Eddington-limited model is consistent with both the $\Mbh-\Mstar$ relation of local AGN \citepalias[e.g.][]{Reines_2015} and the quasar luminosity function at $z\sim6$. However, due to the limited redshift range of the model, it is unclear what signatures this evolutionary pathway may leave on the local SMBH demography. \citet{2025PorrasValverderaXiv250411566P} presented a complementary study comparing SAMs that at $z\sim0$ are consistent with the SMF, local estimates BHMF, and the $\Mbh-\Mstar$ relation, examining how they differ at high redshift, if they can reproduce the distribution functions of LRDs and high-$z$ AGN, and what mechanisms are responsible. They found that, on the whole, the SAMs' total BHMF were consistent with the high-$z$ active BHMFs of \citet{taylor2024broadlineagn35z6black}, but were unable to reproduce the bolometric LF of \citet{akins2024LRDs}, with the exception of {\sc DarkSage} \citep[][]{2016StevensDarkSage1}. In addition, the majority of the SAMs' predicted LFs fall below, but close to our reference LF.

At $z\sim5.5$, we find our reference LF to be in good agreement with the prediction from the CAT super-Eddington model at the bright end ($\Lbol\gtrsim10^{46.5}~{\rm erg\,s^{-1}}$)
, but sits mid-way between the CAT super-Eddington and Eddington-limited models at $\Lbol\lesssim10^{44}~{\rm erg\,s^{-1}}$. At $z\sim5.5$, the BHMFs of the CAT super-Eddington model and the {{\sc A-SLOTH}} heavy-seeding model \citep{2025JeonApJ...988..110J} are in better agreement with that inferred from the SMF using the \citetalias{Pacucci2023_LRDs} $\Mbh-\Mstar$ relation. Whereas, the CAT Eddington-limited model and the light-seeding model from \citet{2025JeonApJ...988..110J} are in better agreement with the estimates using the \citetalias{Reines_2015} and \citetalias{2025LiTipOf} relations, respectively. At $z\sim10$ we find that in the range $\Mbh\geq10^5~\msun$, the SMBH mass density of the CAT super-Eddington model and the {\sc A-SLOTH} heavy-seeding model are already consistent with the VDF-based estimate at $z=5.5$ and exceeds the estimate from the \citetalias{2025LiTipOf} $\Mbh-\Mstar$ relation. Restricting this to the same mass range as eq. (\ref{eq:rhoBH_z5.5}) ($\Mbh\geq10^6~\msun$), we find the $z=9.5$ prediction of the heavy-seeding model to be in good agreement with the prediction from our reference LF (see Fig. \ref{fig:SAMcomp_rhoBH}). While in the range $\Mbh\geq10^5~\msun$, the SMBH mass density of the CAT Eddington-limited model displays little evolution from $z\sim10-5$ being broadly consistent with the VDF-based estimate and the value inferred using the \citetalias{2025LiTipOf} $\Mbh-\Mstar$ relation at $z=5.5$ across this redshift range, restricting this to $\Mbh\geq10^6~\msun$ reduces the mass density such that it evolves similarly to the prediction from the \citet{Shen_2020} LF. Notably, the GAlaxy Evolution and Assembly semi-analytic model \citep[GAEA][]{2014DeLuciaGAEAMNRAS.445..970D,2016HirschmannGAEAMNRAS.461.1760H} was recently demonstrated to be in agreement with the \citetalias{2025LiTipOf} relation at $z\sim4-7$ for a number of seeding prescriptions \citep[][]{2025Vieri}, suggesting that the seeding mechanism alone may be insufficient to explain the AGN observed by JWST. Their predicted $\Mbh-\Mstar$ relation also shows weak evolution across the total redshift range ($z\sim0-10$) consistent with our model that adopts the initial conditions from the \citetalias{2025LiTipOf} relation.

\subsection{Wider Implications}\label{ssec:DiscussionWiderImplications}

Our results suggest that the currently available observational estimates of large number densities of the high-$z$ AGN observed by JWST, coupled to their large masses and mildly sub-Eddington accretion rates, would favour a high-normalisation in the $\Mbh-\Mstar$ relation, lying above the one measured for local AGN. 

On one hand, a much higher-normalisation relation \citepalias[e.g.][]{Pacucci2023_LRDs} is required to reproduce the UV-based estimates of the AGN space densities from \citet{maiolino2024jadesdiversepopulationinfant} and \citet{2023ScholtzarXiv231118731S}, at face value indicating that the observed lower-luminosity AGN are representative of the underlying SMBH population. These findings would in turn imply that most of the SMBHs would preceed the growth of their stellar hosts, which would then catch up at lower redshifts to settle onto the local relations, in broad agreement with the popular two-stage evolutionary scenario envisaging a fast assembly of the central regions of the galaxies followed by a more prolonged stellar growth at later epochs \citep[e.g.][]{2009CookTwoPhase1MNRAS.397..534C,2010CookTwoPhase2MNRAS.402..941C,2010OserApJ...725.2312O,2018LapiApJ...857...22L,2024MoTwoPhase1MNRAS.532.3808M,2024ChenTwoPhase2MNRAS.532.4340C,Silk2024_LRDs,2024LaiMNRAS.531.2245L,2025ChenTwoPhase4arXiv250903283C}, also supported by the recent observations of \citet{2025BillandarXiv250704011B}.

On the other hand, accounting for selection effects and measurement uncertainties, would imply that the mean relation should lie significantly below the determination of \citetalias{Pacucci2023_LRDs}, as argued by \citetalias{2025LiTipOf} \citep[as well as][]{2025SilvermanarXiv250723066S,2025RenarXiv250902027R}. Indeed, we find that the $\sigma_{\star}$-based BHMF is in good agreement with the one extracted from the \citetalias{2025LiTipOf} $\Mbh-\Mstar$ relation, suggesting the latter to be a more robust tracer of SMBH mass.

In addition, assuming that also SMBH mass estimates are overestimated, would imply an intrinsic $\Mbh-\Mstar$ relation comparable to the determination of \citetalias{Reines_2015} for local AGN. Besides the fact that this relation requires sometimes unphysical ($>1$) duty cycles to reproduce the AGN LFs, the $z=5.5$ SMBH mass density computed from the time-integration of our reference AGN LF is more than an order of magnitude larger than what is obtained from the \citetalias{Reines_2015} BHMF, as shown in the right panel of Figure \ref{fig:rhoBH_comp_toZ0}. Even on the assumption of maximally spinning SMBHs ($\epsilon_{\rm r}=0.42$) a large discrepancy ($\sim0.8~{\rm dex}$) between the prediction from our reference LF and the \citetalias{Reines_2015}-based BHMF remains, implying that the AGN LF at $z\gtrsim5$ is necessarily overestimated to reconcile the two.


All in all, we find that at face value, assuming the observed AGN are representative of the underlying population, the high-$z$ AGN population are as massive as observations suggest and can be as numerous as indicated by the current measurements of the AGN luminosity function. If selection effects are accounted for, such that the observed AGN are the most massive objects in the tail of the distribution, then the high-$z$ AGN population can be as massive as observations suggest but not as numerous as indicated by the \citet{maiolino2024jadesdiversepopulationinfant} and \citet{2023ScholtzarXiv231118731S} UV LFs. If there is also a systematic overestimation of SMBH masses, then the high-$z$ AGN population are neither as massive or as numerous as current observations suggest, as it would be extremely challenging to account for a large number density of AGN starting from comparatively low average $\Mbh-\Mstar$ relations. Given the above considerations, we can broadly define two classes of viable models (summarised in Fig. \ref{fig:ViableModelsCartoon}):
\begin{itemize}
    \item The first class follows a strong two-phase growth scenario, represented in this work by the \citetalias{Pacucci2023_LRDs} model, as well as in other works by, for example, the {\sc CAT} super-Eddington model and {\sc DarkSage}. These models are characterised by steep initial conditions, such as a high-normalisation $\Mbh-\Mstar$ relation at high-$z$ coupled with relatively lower duty cycles, leading to a weakly evolving BHMF as lower redshifts, a low $\langle{\rm sBHAR}\rangle/\langle{\rm sSFR}\rangle$ and a strongly evolving $\Mbh-\Mstar$ relation.
    \item The second class of viable models follows the scenario where BHs roughly coevolve with their host galaxies, represented in this work by the \citetalias{2025LiTipOf} model, as well as in other works by, for example, {\sc GAEA}. These models are characterised by shallower initial conditions, such as a modest offset from the $\Mbh-\Mstar$ relation of local AGN coupled with a larger intrinsic scatter and high duty cycles, leading to a BHMF that evolves more steadily and a $\Mbh-\Mstar$ relation that displays little evolution.
\end{itemize}

There are a number of potential observations that could help shed light on the demography of SMBHs at high-$z$. In particular, the local BHMF predicted via the continuity equation (Fig. \ref{fig:BHMF Ceq}) displays an upturn at low masses in the \citetalias{Pacucci2023_LRDs} and \citetalias{2025LiTipOf} models, an artifact of the initial conditions that persists to $z=0$. Therefore, pinning down the low-mass end of the local BHMF will not only give constraints to SMBH seeding models, but help in constraining the demography of SMBHs at high-$z$. Furthermore, AGN clustering measurements from large and deep surveys such as Euclid and LSST will help to set independent and precious constraints on the viable duty cycles of AGN at moderate to high redshifts, which may help discriminate between these two classes of models (e.g. Fig. \ref{fig:DC}). Finally, high resolution spectroscopy of lower redshift LRD candidates will help in constraining the prevalence of rest-frame absorbers among the LRD population \citep[][]{2025DEugenioarXiv250311752D}, offering improved constraints on their masses and accretion rates, as well as gaining insights into whether these absorbers are in stationary or oscillatory states \citep[e.g.][]{2025DEugenioarXiv250614870D}.

\section{Conclusions}\label{sec:Conclusions}

The demography of $z\sim5$ AGN as revealed by JWST is allegedly showing a large population of accreting SMBHs with number densities and masses orders of magnitude above what was measured by pre-JWST observations. In this work, we accurately analyse the consistency of such diverse and perplexing data sets by starting from the galaxy stellar mass function which we first convert into a SMBH mass function via a chosen $\Mbh-\Mstar$ relation, and then into an AGN luminosity function via an assumed AGN fraction and an Eddington ratio distribution consistent with the observed high-$z$ BL AGN. By comparing with current estimates of the SMBH mass function and BL AGN luminosity functions in the bolometric and UV planes, we have set well-grounded conditions on the viability of the input high-$z$ SMBH scaling relations and their role as initial conditions for SMBH growth models at $z\lesssim5$. Our results can be summarised as follows:

\begin{itemize}
    \item We find that to reconcile the JWST galaxy stellar mass function with the high-$z$ active SMBH mass function, an $\Mbh-\Mstar$ relation that is higher in normalisation than that of local AGN is favoured (Figure \ref{fig:BHMF}). More specifically, the stellar mass function can be reconciled with the high-$z$ active SMBH mass function of braod-line AGN via either an $\Mbh-\Mstar$ relation that is much higher in normalisation \citepalias[e.g.][]{Pacucci2023_LRDs} paired with a low duty cycle ($\fAGN\sim0.08$), or a more moderate relation \citepalias[e.g.][]{2025LiTipOf} paired with a higher duty cycle ($\fAGN\sim0.5$). Assuming an $\Mbh-\Mstar$ relation consistent with local AGN (e.g., \citetalias{Reines_2015}) would require $\fAGN>1$ to reconcile with current estimates of the active SMBH mass function (Figure \ref{fig:BHMF}).
    \item From the SMBH mass function we derive the bolometric AGN luminosity function adopting the observed Eddington-ratio distribution at $z\sim5$ and assuming different AGN fractions. We find that current number densities of broad-line AGN can be broadly matched with $\fAGN\sim0.1,~0.5$ for the \citetalias{Pacucci2023_LRDs}, \citetalias{2025LiTipOf} $\Mbh-\Mstar$ relations, respectively, and $\fAGN\sim1$ assuming local $\Mbh-\Mstar$ relation for AGN (RV15), in line with what estimated from the SMBH mass functions (Figure \ref{fig:bolLFcomp}).
    \item Similar results are retrieved when examining the UV luminosity functions derived with an empirical $\Lbol-\Muv$ mappings (Figure \ref{fig:UVLFcomp}). In all cases, the faint end of the AGN luminosity functions are extremely challenging to match even with maximal values of the duty cycles, pointing to a possible overestimation of these data (Figures \ref{fig:bolLFcomp} \& \ref{fig:UVLFcomp}).
    \item We carry out an extensive So\l{}tan-type argument integrating in time from $z\sim10$ to $z\sim5.5$ a new estimate of the bolometric AGN luminosity function that takes into account all the latest current measurements. With standard values of the radiative efficiency ($\epsilon_{\rm r}\sim10\%$), we find a SMBH mass density in line with the SMBH mass density implied by the $\Mbh-\Mstar$ relation of moderate normalisation by \citetalias{2025LiTipOf} and the one derived from the $\Mbh-\sigma_{\star}$ and the high-$z$ velocity dispersion function (extracted from the galaxy stellar mass function and the Faber-Jackson relation; Fig. \ref{ssec:rhoBH}). Such a convergence of results further supports the view that $\Mbh-\Mstar$ relations with extreme normalisations (e.g., \citetalias{Pacucci2023_LRDs}) are biased high.
    \item Via a continuity equation approach we predict the evolution of the SMBH mass function from $z=5.5$ down to $z=0$ using in input our updated bolometric AGN luminosity function and observed Eddington-ratio distributions, and starting from the BHMFs implied by the three $\Mbh-\Mstar$ relations. SMBH mass functions consistent with very high normalisation $\Mbh-\Mstar$ relations (e.g., \citetalias{Pacucci2023_LRDs}) imply weak evolution at later times, and very steep low-mass BHMFs, in tension with local estimates. Vice versa, more moderate or low normalisation $\Mbh-\Mstar$ relations (e.g. \citetalias{2025LiTipOf}, \citetalias{Reines_2015}) generate accretion histories and local BHMFs more aligned with previous estimates (Figure \ref{fig:BHMF Ceq}).
    \item Starting from a high-normalisation in the $\Mbh-\Mstar$ relation (e.g., \citetalias{Pacucci2023_LRDs}) leads to a strong evolution of the scaling relation, gradually settling on the local relation from above, while it remains nearly constant when starting from more moderate initial conditions (e.g., the \citetalias{2025LiTipOf} and \citetalias{Reines_2015} cases; Figure \ref{fig:MstarMbhEvo}).
\end{itemize}

The data-driven investigation put forward in this work offers valuable insights on both 1) the consistency of current high-$z$ data and 2) the co-evolution of SMBHs and their host galaxies (Figure \ref{fig:ViableModelsCartoon}). At face value, the high number densities of AGN measured at $z\sim5.5$ necessarily require underlying $\Mbh-\Mstar$ relations that are higher than those measured for local AGN to avoid unphysical duty cycles ($\fAGN>1$), favouring a two-phase evolution where SMBHs grow faster than their hosts at early epochs (Figures \ref{fig:MstarMbhEvo} and \ref{fig:ViableModelsCartoon}). However, with an $\Mbh-\Mstar$ relation only modestly above that of local AGN (e.g. \citetalias{2025LiTipOf}), we can still achieve full self-consistency with all current data, including the $\Mbh-\sigma_{\star}$ relation, moderate duty cycles ($\sim50\%$), and a more steady evolution in the $\Mbh-\Mstar$ evolution with improved match with the local SMBH mass function.

\section*{Acknowledgements}
The authors thank the referee, Knud Jahnke, for their insightful comments which have helped strengthen the manuscript. We thank the whole CAT team for providing the results from the CAT models and their useful discussions. We also thank Hannah {\"U}bler for their useful comments. DR acknowledges support from the University of Southampton via the Mayflower studentship. VC thanks the BlackHoleWeather project and PI Prof. Gaspari for salary support. AT acknowledge financial support by the PRIN MUR “2022935STW" (CUP C53D23000950006) funded by European Union-Next Generation EU, and by the Bando Ricerca Fondamentale INAF 2023, Mini-grant ``Cosmic Archaeology with the first black hole seeds'' (RSN1 1.05.23.04.01). HF acknowledges support at Fudan University from the Shanghai Super Post-doctoral Excellence Program grant No. 2024008.

\section*{Data Availability}
The data used within this paper is publicly available in the referenced papers. The code underlying this paper will be shared upon reasonable request to the corresponding author.


\bibliographystyle{mnras}
\bibliography{references.bib}



\appendix
\section{Faint BL AGN Sample}\label{app:LRDsample}
In this work we make use of a sample of faint BL AGN observed by JWST at $4.5\leq z\leq6.5$ to derive a mapping between $\Lbol$ and $\Muv$ and another sample to gain a sense of the distribution of Eddington ratios of these objects. Here we briefly summarise the selection criteria and key quantities of the parent samples and refer the interested reader to the corresponding source paper for the full methodology used to derive the quantities used.

\begin{figure}
    \centering
    \includegraphics[width=0.45\textwidth]{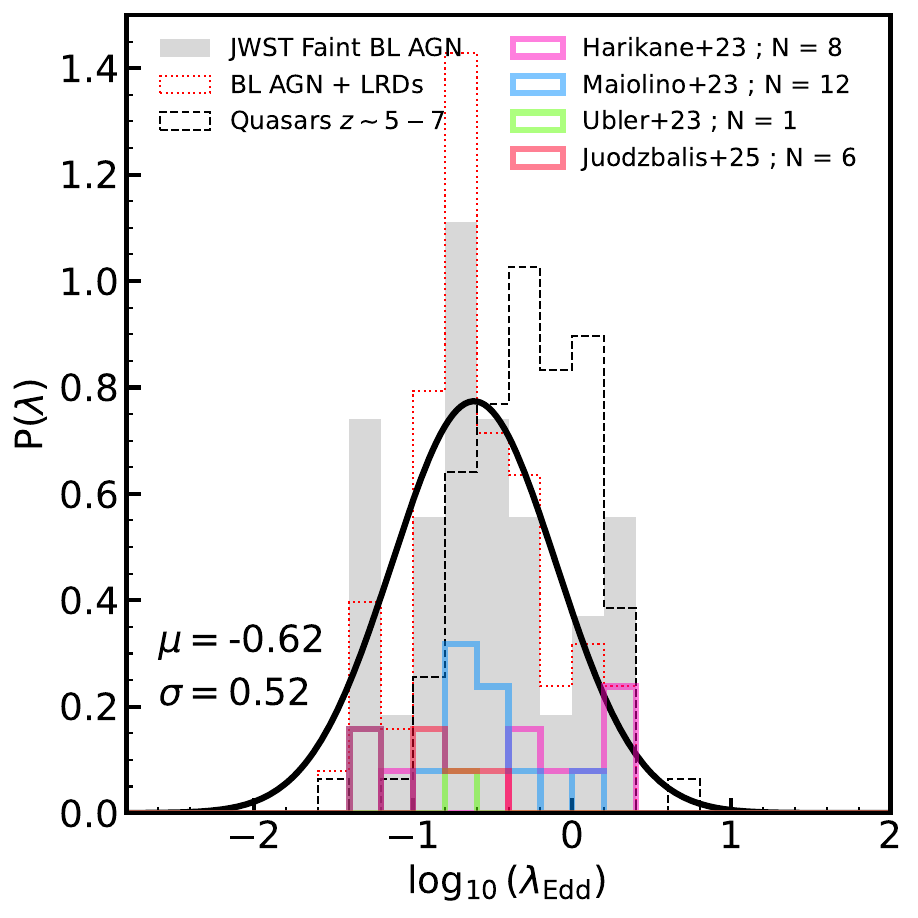}
    \caption{The distribution of Eddington ratios for our sample of faint BL AGN observed by JWST (grey histogram), as well as the distributions of the sample broken down by source paper (coloured histograms), compared to a sample of quasars at $z\sim5-7$ (black dashed histogram) obtained from \citet{2010WillottAJ....140..546W,2011TrakhtenbrotApJ...730....7T,2019MatsuokaApJ...872L...2M,2019OnoueApJ...880...77O,2019ShenApJ...873...35S}, as well as a BL AGN + LRD sample \citep[with the LRDs from][red dotted histogram]{Greene2024_LRDs,2024arXiv240403576K,Mathee2024_LRDs}. The best fit Gaussian to our BL AGN sample is denoted by the solid black line which has a mean of $\langle\log_{10}(\fEdd)\rangle = -0.62$ and a dispersion of $\sigma_{\fEdd} = 0.52$. Whereas, if we fit the combined BL AGN + quasar sample we find the best fit values $\langle\log_{10}(\fEdd)\rangle = -0.35$, $\sigma_{\fEdd} = 0.50$ \citep[broadly consistent with the mean $\fEdd$ values at $z\sim4-5$ found by][]{2024HeApJ...962..152H,2024LaiMNRAS.531.2245L}. The ERDF assumed in section \ref{ssec:ResbolLF} is chosen to agree with the BL AGN + quasar sample.  We note that several papers derive both the black hole mass and bolometric luminosity from the broad component of the \halpha{} line, which can lead to spurious correlations. However, the two properties have differing dependencies, with the BH mass primarily dependent on the width and not the luminosity.}
    \label{fig:LRDfEddDist}
\end{figure}

The AGN samples used in this work are composed of objects from:
\begin{itemize}
    \item \citet{2023ApJ...959...39H}: They select objects observed in the GLASS \citep[][]{2022GLASSApJ...935..110T} and CEERS \citep[][]{2025CEERSApJ...983L...4F} with NIRSpec from the sample of \citet{2023NakajimaApJS..269...33N} with broad H$\alpha$ and/or H$\beta$ and narrow forbidden [OIII] and [NII] emission lines. BH masses and bolometric luminosities are estimated from the H$\alpha$ emission using the estimator of \citet{2005GreeneApJ...630..122G} with the bolometric correction of \citet{2006RichardsApJS..166..470R}, and the extinction values are taken from \citet{2023NakajimaApJS..269...33N}. From these we select the objects in the range $4.5\leq z\leq6.5$.
    \item \citet{2024arXiv240403576K}: They photometrically select objects from NIRCam imaging from CEERS \citep[][]{2025CEERSApJ...983L...4F}, NGDEEP \citep[][]{2024NGDEEPApJ...965L...6B}, JADES \citep[][]{2023EisensteinarXiv230602465E}, and UNCOVER \citep[][]{2024BezansonApJ...974...92B} with high signal-to-noise ratios in F444W (SNR>12), reddened rest-frame optical slopes ($\beta_{\rm opt}>0$), blue rest-frame UV slopes ($-2.8<\beta_{\rm UV}<-0.37$), and that are compact $r_{\rm h}<1.5r_{\rm h,\, stars}$. Two further cuts are imposed to remove sources whose optical continuum slope is boosted by strong line emission ($\beta_{\rm F277W-F356W}>-1$ and $\beta_{\rm F277W-F410M}>-1$). They also spectroscopically confirm 15 LRDs using NIRSpec data and estimate the BH masses from the H$\alpha$ or H$\beta$ emission for all but one source which uses the Pa$\delta$ line and the extinction is obtained from SED fitting. We select the spectroscopically confirmed LRDs in the range $4.5\leq z\leq6.5$.
    \item \citet{Kokorev2024_PhotoConcensusOfLRDs}: They photometrically select objects from NIRCam imaging of CEERS \citep[][]{2025CEERSApJ...983L...4F}, PRIMER \citep[][]{2021PRIMERjwst.prop.1837D}, FRESCO \citep[][]{2023FRESCOMNRAS.525.2864O}, JADES \citep[][]{2023EisensteinarXiv230602465E}, and JEMS \citep[][]{2023JEMSSci...380..416W} using the selection criteria of one of two color cuts to ensure reddened optical slopes and blue UV slopes, a compactness criterion of $f_{\rm F444w}(0.4'')/f_{\rm F444w}(0.2'')<1.7$, and a final cut to remove brown dwarfs (${\rm F115W-F200W}>-0.5$). The bolometric luminosity is estimated from the continuum of the SED fit and the extinction is computed from the SED fit. We select objects in the range $4.5\leq z\leq6.5$.
    \item \citet{maiolino2024jadesdiversepopulationinfant}: They identify objects from the JADES survey \citep[][]{2023EisensteinarXiv230602465E} in the GOODS North and GOODS South fields using the target selection of \citet{2023EisensteinarXiv230602465E} and \citet{2023BunkerA&A...677A..88B}. They identify BL AGN via the detection of a broad component of either ${\rm H}\alpha$ or ${\rm H}\beta$ (at $z\geq7$) and without a broad component in the forbidden transitions, particularly [OIII]5007\AA. The BH masses are estimated from the H$\alpha$ emission using \citet{2013ReinesApJ...775..116R}, the bolometric luminosity is computed from the broad H$\alpha$ using the estimator of \citet{2012SternMNRAS.426.2703S}, and the extinction is obtained from SED fitting. From these we select the objects in the range $4.5\leq z\leq6.5$.
    \item \citet{Mathee2024_LRDs} : They select objects from EIGER \citep[][]{2023EIGERApJ...950...66K} and FRESCO \citep[][]{2023FRESCOMNRAS.525.2864O} with high signal-to-noise H$\alpha$ emission (${\rm SNR}_{{\rm H}\alpha,{\rm broad}}>5$), $L_{{\rm H}\alpha,{\rm broad}}>2\times10^{42}~{\rm erg/s}$, and $v_{{\rm FWHM},{\rm H}\alpha,{\rm broad}}>1000~{\rm km/s}$. The BH mass is estimated from the broad H$\alpha$ emission using \citet{2013ReinesApJ...775..116R} and the bolometric luminosity is computed from the broad H$\alpha$ emission using \citet{2005GreeneApJ...630..122G} with the bolometric correction of \citet{2006RichardsApJS..166..470R}. We select objects in the range $4.5\leq z\leq6.5$.
    \item \citet{2023Ubler} : They obtain NIRSpec data for GA-NIFS and estimate the BH mass from the broad H$\alpha$ emission using \citet{2013ReinesApJ...775..116R}, the bolometric luminosity is computed using three estimators: H$\beta$ and [OIII] emission using \citet{2009NetzerMNRAS.399.1907N}, the H$\beta$ emission using \citet{2020DallaBontaApJ...903..112D}, and the H$\alpha$ emission using \citet{2012SternMNRAS.426.2703S}.
    \item \citet{2025IgnasarXiv250403551J}:  They identify objects from the JADES survey \citep[][]{2023EisensteinarXiv230602465E} in the GOODS North and GOODS South fields following the approach of \citet{maiolino2024jadesdiversepopulationinfant}. They identify BL AGN via the detection of a broad component of either ${\rm H}\alpha$ and without a broad component in the forbidden transitions, particularly [OIII]5007\AA. The BH masses are estimated from the H$\alpha$ emission using \citet{Reines_2015} and the bolometric luminosity is computed from the broad H$\alpha$ using the estimator of \citet{2012SternMNRAS.426.2703S}. From these we select the objects in the range $4.5\leq z\leq6.5$.
\end{itemize}

The full catalogue of individual AGN can be found at \url{https://github.com/DanR001/JWSTfaintAGN}. To derive an empirical mapping between $\Lbol$ and $\Muv$ we use all the AGN in our redshift range ($4.5\leq z \leq6.5$) where these values have been computed in the parent papers and to get a sense of the ERDF we use all BL AGN (excluding little red dots) in our redshift range where either the Eddington ratio has been computed in the parent paper or the bolometric luminosity and BH mass are provided.

\section{Insights from the high-$z$ VDF}\label{app:highZVDF}

\begin{figure*}
    \centering
    \includegraphics[width=0.9\textwidth]{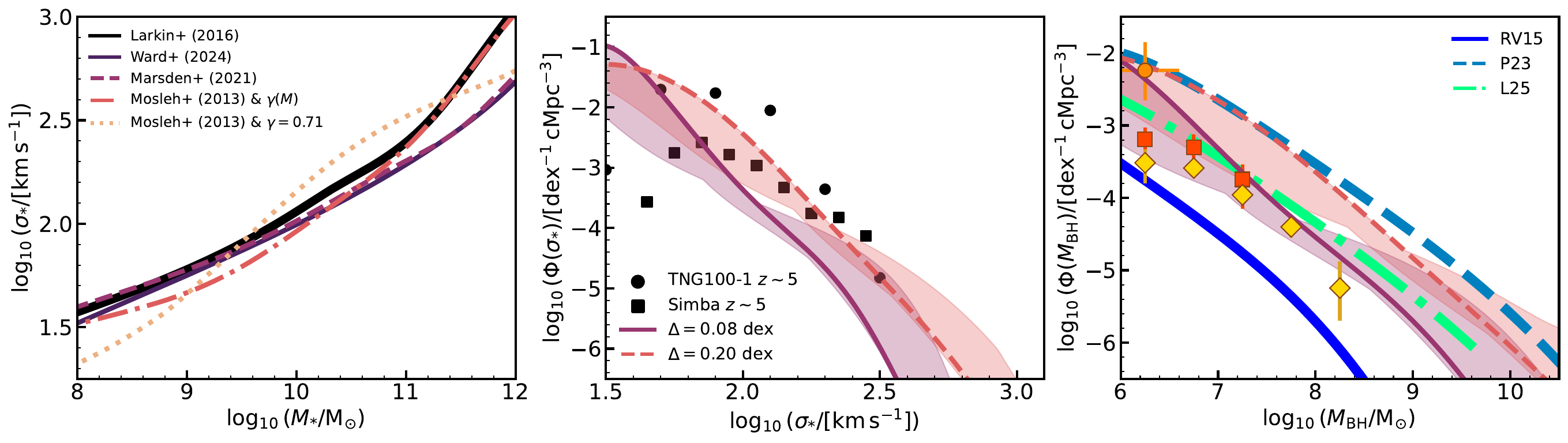}
    \caption{The high-$z$ predictions from the framework of \citet{Marsden2021_sigma}. \emph{Left:} a comparison of the predicted $\sigma_{\star}-\Mstar$ relation at $z=5.5$ with different input size-mass relations, as well as the prediction from \citet{2016LarkinMNRAS.462.1864L}. \emph{Middle:} a comparison of the VDF at $z=5.5$ with two assumed values of the intrinsic scatter ($\Delta$) and the predictions from the TNG100-1 and Simba M50 N512 simulations. The shaded regions denote the range of possible VDFs depending on the $\sigma_{\star}-\Mstar$ relation adopted from those in the left-hand panel. The lines show the prediction from the fiducial model of \citet{Marsden2021_sigma}. \emph{Right:} a comparison of the predicted BHMF at $z=5.5$ from the VDF to those predicted from the SMF and the high-$z$ active BHMFs (as in Fig. \ref{fig:BHMF}). As in the middle panel the lines show the prediction of the \citet{Marsden2021_sigma} framework when adopting two different intrinsic scatters and the shaded regions show the uncertainty associated to the range of $\sigma_{\star}-\Mstar$ relations.}
    \label{fig:SigMstar_VDF_BHMF}
\end{figure*}

\begin{figure*}
    \centering
    \includegraphics[width=0.9\textwidth]{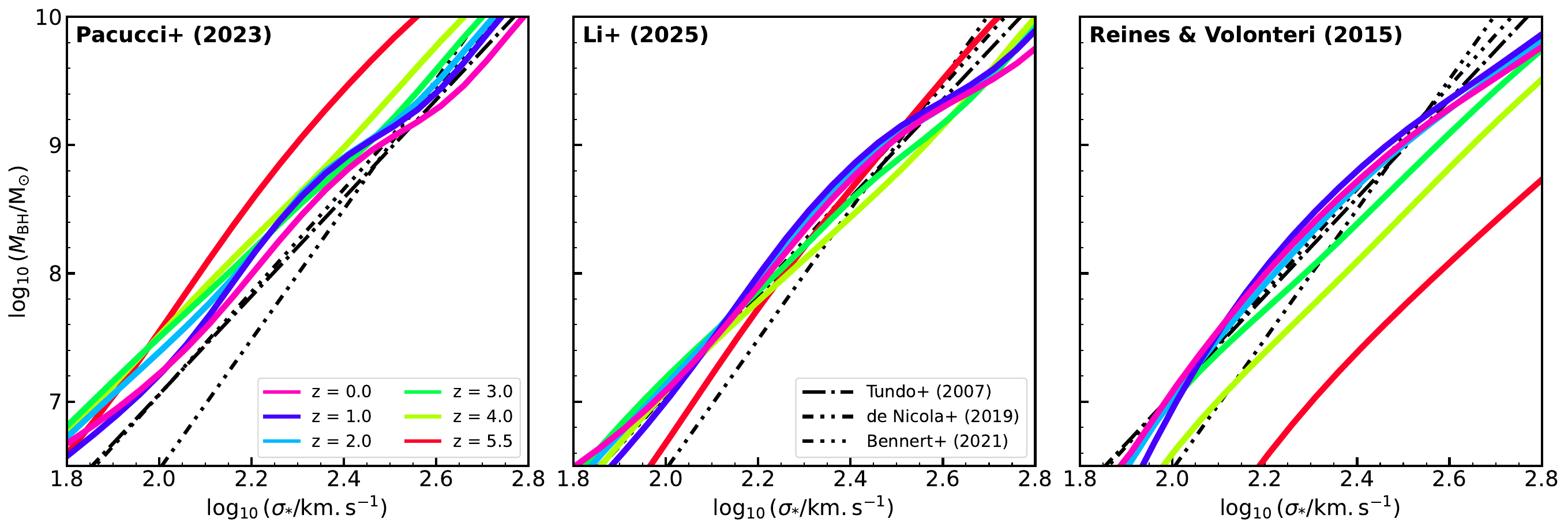}
    \caption{The predicted evolution of the $\Mbh-\sigma_{\star}$ relation predicted via abundance matching between the VDF predicted by \citet{Marsden2021_sigma} and the BHMF predicted via the continuity equation. The left, middle, and right panels display the results from the \citetalias{Pacucci2023_LRDs}, \citetalias{2025LiTipOf}, and \citetalias{Reines_2015} initial conditions, respectively. These are compared to the local relations of \citet{Tundo_2007}, \citet{2019deNicola}, and \citet{2021ApJ...921...36B}, where the latter two have been aperture corrected to $R_{\rm ap}/R_{\rm e}=1/8$ following \citet{2021deGraaffApJ...913..103D}.}
    \label{fig:MbhSigEv}
\end{figure*}

The VDF is poorly constrained at intermediate to high redshifts and so, while the $\Mbh-\sigma_{\star}$ relation is consistently shown to be more fundamental than the $\Mbh-\Mstar$ relation through pairwise residuals analysis, a direct insight into the high-$z$ demography of SMBHs from the VDF is currently not possible. Nevertheless, as has been done in previous works \citep[e.g.][]{2011BezansonApJ...737L..31B,2016LarkinMNRAS.462.1864L,2018RicarteNatarajanMNRAS.474.1995R}, the redshift evolution of the VDF can be estimated from the SMF and a redshift dependent $\sigma_{\star}-\Mstar$ relation. From this we can infer the corresponding BHMF using the local $\Mbh-\sigma_{\star}$ relation, which the observed AGN with extreme $\Mbh/\Mstar$ ratios appear to be consistent with, as well as estimating the SMBH mass density. 

To predict the evolution of the $\sigma_{\star}-\Mstar$ relation, we deploy the theoretical frameworks of \citet{2016LarkinMNRAS.462.1864L} and \citet{Marsden2021_sigma}\footnote{\url{https://github.com/ChrisMarsden833/VelocityDispersion}}, both of which compute the velocity dispersion via Jeans modelling, but use different input data and assumptions when doing so. When adopting the same size-mass relation in input we find good agreement between these two models across redshift ($z\sim0-6$), as well as being consistent with the evolution predicted by \citet{2020CannarozzoMNRAS.498.1101C}. Due to this agreement we focus solely on the \citet{Marsden2021_sigma} model from now on.

To understand the sensitivity of our predicted $\sigma_{\star}-\Mstar$ relation to the input size-mass relation we test a range of size-mass relations and redshift evolutions. As standard we adopt the fit to MaNGA galaxies included in \citet{Marsden2021_sigma}, along with a redshift evolution of the form $R_{\rm e}(\Mstar,z)=R_{\rm e}(\Mstar,0)\times(1+z)^{-\gamma}$ and the $\gamma(\Mstar)$ of \citet{Marsden2021_sigma} which \citep[as in ][]{2018RicarteNatarajanMNRAS.474.1995R} is calibrated to match \citet{2013HuertasCompanyMNRAS.428.1715H}. We compare this to adopting the JWST size-mass relation of \citet{2024WardApJ...962..176W}, as well as the relation for red galaxies from \citet{2013MoslehApJ...777..117M} with both the mass dependent redshift evolution of \citet{Marsden2021_sigma} and the mass independent evolution of \citet{2024OrmerodMNRAS.527.6110O}. The resulting relations are compared to the prediction of \citet{2016LarkinMNRAS.462.1864L} in the left-hand panel of Figure \ref{fig:SigMstar_VDF_BHMF}. There is good agreement between all the relations, particularly between the standard prediction of \citet{Marsden2021_sigma} and when using the JWST size-mass relation. The only exception is the case where we only evolved the size-mass relation in normalisation, which has a differing shape. Due to this agreement we will adopt the fit to MaNGA data from \citet{Marsden2021_sigma} when moving forward, and use the other relations to bound the uncertainty.

Now in possession of redshift dependent $\sigma_{\star}-\Mstar$ relation, we can infer the VDF at high $z$ from the \citet{shuntov2024cosmoswebstellarmassassembly} SMF via the convolution 
\begin{equation}
    \Phi(\sigma_{\star}) = \int\!\Phi(\Mstar){\rm P}(\sigma_{\star}|\Mstar)\,{\rm d}\log_{10}(\Mstar)~,
\end{equation}
where ${\rm P}(\sigma_{\star}|\Mstar)$ is assumed to be a Gaussian distribution and we adopt an intrinsic scatter of $\Delta=0.08~{\rm dex}$ in the $\sigma_{\star}-\Mstar$ relation \citep[as in ][]{2020CannarozzoMNRAS.498.1101C}. However, we test the impact of adopting a scatter of $0.2~{\rm dex}$ at high redshift. The resulting VDFs are displayed in the middle panel of Figure \ref{fig:SigMstar_VDF_BHMF}. We find that adopting a redshift independent scatter of $0.08~{\rm dex}$ leads to a VDF slightly below that predicted by TNG100-1\footnote{\url{https://www.tng-project.org/}} \citep[][]{2019NelsonTNGComAC...6....2N} and Simba\footnote{\url{https://simba.roe.ac.uk/}} \citep{2019DaveSimbaMNRAS.486.2827D}, but adopting a larger intrinsic scatter brings them in line with one another.

Similarly, we can then convert this high-$z$ VDF to a BHMF via the local $\Mbh-\sigma_{\star}$ relation. We adopt the relation of \citet{2021ApJ...921...36B} as it is computed in the same aperture as our VDF ($R_{\rm ap}/R_{\rm e}=1$) and aligns with the high-$z$ BL AGN of \citet{maiolino2024jadesdiversepopulationinfant}. The predicted BHMFs from our VDFs are displayed in the right-hand panel of Figure \ref{fig:SigMstar_VDF_BHMF}. We find that when adopting an intrinsic scatter of $\Delta = 0.08$ in the $\sigma_{\star}-\Mstar$ relation we obtain a BHMF that is consistent with that inferred from the SMF using the \citetalias{2025LiTipOf} relation, whereas when adopting the larger scatter of $\Delta = 0.20$ we obtain a BHMF that is broadly consistent with that inferred from the SMF using the \citetalias{Pacucci2023_LRDs} relation. Interestingly, both are consistent with the recent estimate from \citet{2025GerisarXiv250622147G} at the low-mass end.

In addition, from these BHMFs we can estimate the SMBH mass density at $z=5.5$ via eq. (\ref{eq:rhoBH_z5.5}) and obtain the values $\log_{10}(\rho_{\rm BH}/[{\rm M}_{\odot}\,{\rm cMpc}^{-3}]) = 4.14^{+0.35}_{-0.46}$ which is consistent with the estimate from the SMF using the \citetalias{2025LiTipOf} relation. If we adopt the larger scatter of $\Delta = 0.20$, we obtain a value of $\log_{10}(\rho_{\rm BH}/[{\rm M}_{\odot}\,{\rm cMpc}^{-3}]) = 4.84^{+0.08}_{-0.42}$ which lies in between the estimates from SMF using the \citetalias{Pacucci2023_LRDs} and \citetalias{2025LiTipOf} relations.

Finally, using our predicted VDF we can compute the evolution $\Mbh-\sigma_{\star}$ relation from $z=5.5$ to $z=0$ by abundance matching with the BHMF predicted via the continuity equation. To do this we correct our predicted VDF from $R_{\rm ap}/R_{\rm e}=1$ to $R_{\rm ap}/R_{\rm e}=1/8$ using the correction of \citet{2021deGraaffApJ...913..103D} and we adopt an intrinsic scatter of $0.15~{\rm dex}$ as in the \citet{Tundo_2007} relation. The resulting $\Mbh-\sigma_{\star}$ relations are displayed in Figure \ref{fig:MbhSigEv}. We find that in this framework the BHMF from \citetalias{Reines_2015} initial conditions suggests a significant evolution in the $\Mbh-\sigma_{\star}$ relation at $z\sim2-5.5$, whereas the predictions using the from the BHMFs from \citetalias{Pacucci2023_LRDs} or \citetalias{2025LiTipOf} initial conditions display little redshift evolution. Here, we have adopted the \citet{Marsden2021_sigma} $\sigma_{\star}-\Mstar$ relation with an intrinsic scatter of $0.08~{\rm dex}$ that is independent of redshift and in this case the \citetalias{2025LiTipOf} model shows the least evolution, being consistent with the local $\Mbh-\sigma_{\star}$ relations across the entire redshift range. However, adopting an intrinsic scatter of $0.20~{\rm dex}$ in the $\sigma_{\star}-\Mstar$ relation at high redshift is sufficient to make the \citetalias{Pacucci2023_LRDs} model most consistent with the local relations. Therefore, we conclude that both the \citetalias{Pacucci2023_LRDs} or the \citetalias{2025LiTipOf} relations can be consistent with the $\Mbh-\sigma_{\star}$ relation across cosmic time, depending on the exact redshift evolution of the scaling relations and their intrinsic scatter.

\section{$\Muv-\Lbol$ Mappings}\label{App:MuvLbolMappings}

\subsection{Mapping from Shen et al. (2020)}\label{App:MuvLbolMappings-Shen}

The UV bolometric correction of \citet{Shen_2020} provides a mapping between $\Lbol$ and the intrinsic UV luminosity, that is consistent with the historical correction of \citet{1994ElvisApJS...95....1E}. Under the assumption that the bulk of the high-$z$ AGN population are ``typical'' AGN such that they follow the local relation, we use this mapping to convert our derived bolometric LFs to UV LFs assuming there to be minimal dust attenuation. This relation is compared to a sample of lower-luminosity BL AGN and LRDs in Figure \ref{fig:kbolFit}. From this it is clear that the majority of these lower luminosity high-$z$ AGN differ significantly from the local bolometric correction from \citet{Shen_2020}. Whereas, quasars at the same redshift (black points in Fig \ref{fig:kbolFit}) appear to follow the local bolometric correction.

On one hand, if one interprets this deviation as the result of dust attenuation, it implies that dust extinction increases with luminosity, contrary to the low-redshift trend \citep[e.g.][]{2005SimpsonMNRAS.360..565S,2014MerloniMNRAS.437.3550M,Ueda2014}. Given such high values of $\Av$ implied by Figure \ref{fig:kbolFit} and inferred from SED fitting (the sample in Fig. \ref{fig:kbolFit} has a mean extinction $\langle\Av\rangle\sim1.5~{\rm mag}$ inferred from SED fitting), one would expect the resulting UV LF to be heavily suppressed, in contradiction to the observed UV LF of faint BL AGN. \citet{2025ApJ...978...92L}, \citet{Greene2024_LRDs}, and \citet{Kokorev2024_PhotoConcensusOfLRDs} suggested instead that the observed rest-frame UV can be interpreted as either scattered light from an AGN or contamination from the host-galaxy. On the other hand, at least in the LRD subpopulation of BL AGN, there's growing evidence that these objects are not dusty but instead, their SEDs are intrinsically different \citep[e.g.][]{2025NaiduBHstararXiv250316596N,2025SacchiarXiv250509669S,2025GreenearXiv250905434G}.

\subsection{Empirical Mapping from the AGN sample.}\label{App:MuvLbolMappings-Emp}

We derive an empirical mapping directly between $\Lbol$ and $\Muv$ from the sample of BL AGN displayed in Figure \ref{fig:kbolFit} which naturally encapsulates contributions from the dust correction, the bolometric correction, scattered AGN emission, contamination from the host galaxy, and/or an intrinsic difference in the SED. We find the $\Muv-\Lbol$ relation to be well described by
\begin{equation}\label{eq:Lbol_Muv_Fit}
    \Muv = (-0.81\pm0.09)\log_{10}\left(\frac{\Lbol}{10^{45}~{\rm erg\,s^{-1}}}\right) - (18.63\pm0.06),
\end{equation}
with an intrinsic scatter of $\sim0.79~{\rm mag}$. The resulting mapping is displayed in Figure \ref{fig:kbolFit} as the solid black line. This relation falls below the bolometric correction of \citet{Shen_2020} at low luminosities which may potentially be interpreted as a regime with significant host contamination.

If we correct the bolometric luminosities of the LRDs \citep[those from][]{Mathee2024_LRDs,Kokorev2024_PhotoConcensusOfLRDs,2024arXiv240403576K} by $1~{\rm dex}$ \citep[as suggested by ][]{2025GreenearXiv250905434G}, then the majority of the LRDs lie below the bolometric correction of \citet{Shen_2020}, and many have $\log_{10}(\Lbol/\Luv)<0$. This can be interpreted as the LRDs UV emission being host dominated, consistent with the BH-star scenario \citep[][]{2025NaiduBHstararXiv250316596N}.

Due to the uncertainty in the LRDs, we also fit the relation to only the BL AGN data \citep[those from ][]{2023ApJ...959...39H,maiolino2024jadesdiversepopulationinfant,2025IgnasarXiv250403551J}, excluding the LRDs. In this case we find the $\Muv-\Lbol$ relation to be well described by
\begin{equation}
    \Muv = (-0.21\pm0.35)\log_{10}\left(\frac{\Lbol}{10^{45}~{\rm erg\,s^{-1}}}\right) - (18.95\pm0.30),
\end{equation}
with an intrinsic scatter of $\sim0.97~{\rm mag}$. This is a shallower relation than obtained when including the LRDs and there is larger uncertainty in this fit compared to that including the LRDs due to the small number of data points and the limited luminosity range covered by the sample.

\subsection{Statistical Mapping from Abundance Matching.}\label{App:MuvLbolMappings-AM}
In the last approach, we derive an empirical monotonic relation between $\Lbol$ and $\Muv$ via abundance matching between the measured bolometric and UV AGN LFs of \citet{Kokorev2024_PhotoConcensusOfLRDs}. To this purpose, we deploy the procedure outlined in \citet[][their eq. 37]{Aversa_2015}
\begin{equation}\label{eq:AMofLFs}
    \begin{split}
        \int_{\Muv}^{+\infty} \! \phi(\Muv',z)~{\rm d}\Muv' ~~=~~ \frac{1}{2}\int_{-\infty}^{+\infty} \! \Phi(\Lbol',z) ~~~~~~~~~~~~~~~~~~~~~\\[2ex]
        \times~{\rm erfc}\left\{\frac{\log_{10}(\Lbol(\Muv)/\Lbol')}{\sqrt{2}\tilde{\sigma}_{\Muv}}\right\} ~ {\rm d}\log_{10}(\Lbol'),
    \end{split}
\end{equation}
where $\tilde{\sigma}_{\Muv}=\sigma_{\Muv}/({\rm d}\Muv/{\rm d}\Lbol)$ and $\sigma_{\Muv} = 0.79~{\rm mag}$ is the assumed intrinsic scatter in the $\Muv-\Lbol$ relation, matching that of our sample. Here, we have chosen to use the LFs of \citet{Kokorev2024_PhotoConcensusOfLRDs} in input, as using our reference bolometric LF and the UV LF of \citet{2024GrazianApJ...974...84G} would approximately reproduce the \citet{Shen_2020} bolometric correction, only differing at low luminosities due to the differing faint end slopes. This would offer no new information as we tune the normalisation of our reference LF to the UV estimates under the assumption of the \citet{Shen_2020} bolometric correction.

In using the \citet{Kokorev2024_PhotoConcensusOfLRDs} UV LF in equation (\ref{eq:AMofLFs}), we are implicitly assuming that the UV emission is AGN dominated and so, the resulting UV LFs are upper limits. If the LRD UV emission is in fact host dominated \citep[as suggested by][]{2025NaiduBHstararXiv250316596N,2025GreenearXiv250905434G}, then this mapping no longer describes the wavelength dependence of the AGN emission but instead, describes a correlation between $\dot{M}_{\rm BH}$ and SFR.

The mapping resulting from abundance matching is displayed in Figure \ref{fig:kbolFit} as the dashed brown line and associated brown shaded $1\sigma$ region. The resulting empirical relation from abundance matching, derived from a statistical approach at population level, is reassuringly close to the relations derived from sources on a single basis. The remaining mild discrepancies, visible at the bright and faint ends, could be simply ascribed to the loose constraints at the extremes of the LFs.

We also test the case where the bolometric luminosities of the LRDs have been overestimated by an order of magnitude, as suggested by \citet{2025GreenearXiv250905434G}. This is displayed in Figure \ref{fig:kbolFit} as the dotted brown line and is similar to the linear fit to the sample at low luminosities and $\sim0.5~{\rm dex}$ below it at the bright end, primarily resulting in a moderate boost to the bright end of the model UV LF. As the UV LFs resulting from this mapping are bounded by those from the \citet{Shen_2020} bolometric correction and the linear fit to the data, we do not pursue this mapping further here because we find that it does not impact our findings.

\section{Consistency of SMF and Galaxy UV LF}\label{app:SMFtoUVLF}

\begin{figure}
    \centering
    \includegraphics[width=0.45\textwidth]{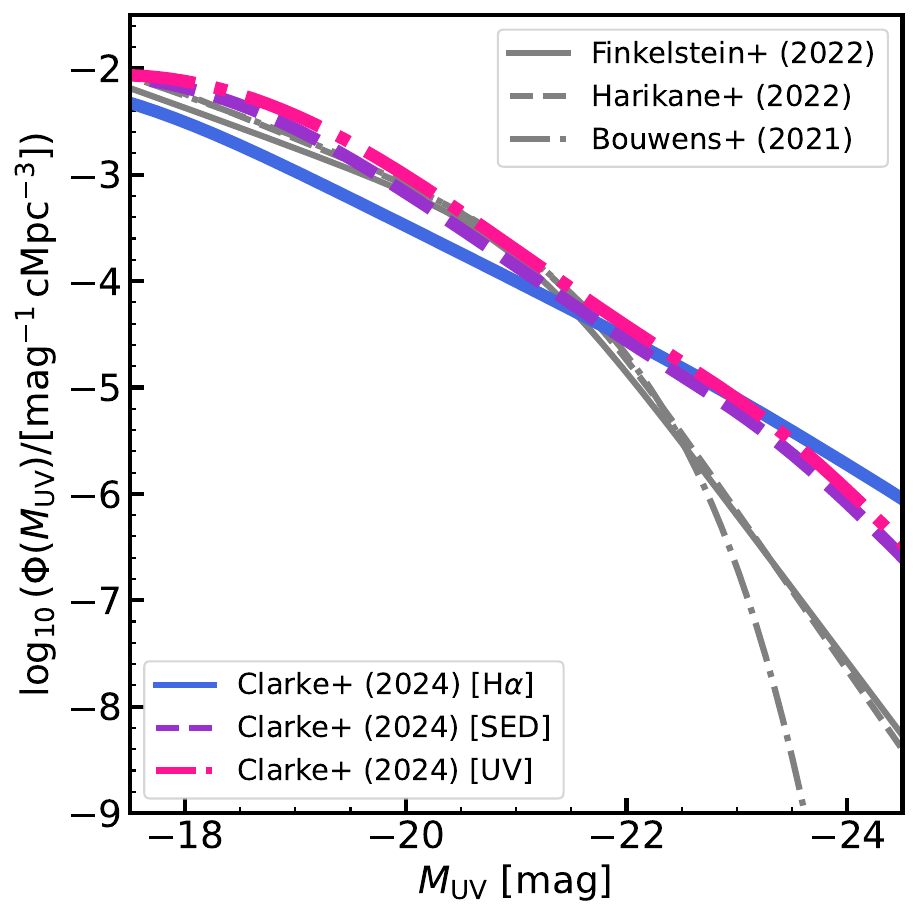}
    \caption{A comparison of the galaxy UV LFs of \citet{2022FinkelsteinUVLF} (thin solid dark red line), \citet{2022HarikaneUVLF} (thin dashed orange line), and \citet{2021BouwensUVLF} (thin dot-dashed yellow / light orange line) with the UV LFs derived from the SMF using the ${\rm H}\alpha$ (thick solid blue line), SED (thick dashed purple line), and UV (thick pink dot-dashed line) determinations of the star-forming main sequence. We observe that, for such a simplistic construction of the UV LF, they agree well with the literature galaxy UV LFs of \citet{2022FinkelsteinUVLF} and \citet{2022HarikaneUVLF} across the $\Muv$ range, with the exception of the UV LF using ${\rm H}\alpha$ main sequence which over-shoots the literature UV LFs at the bright end.}
    \label{fig:SMFtoUVLF}
\end{figure}

Prior to investigating AGN, we need to perform a key consistency check and ascertain whether the SMF is fully consistent with the UV LF of galaxies at high $z$. To this end, we convert our reference SMF \citep[i.e.][]{shuntov2024cosmoswebstellarmassassembly} to a SFR function, assuming a negligible quenched fraction at this redshift \citep[as suggested by observations][]{2013MuzzinApJ...777...18M,Weaver_2023,2024RussellarXiv241211861R}, and then convert this to a UV LF. A number of works have found that, at a given mass and redshift, the distribution of SFRs for star-forming galaxies tend to exhibit a double Gaussian shape \citep[e.g.][]{2012BetherminApJ...757L..23B,2012SargentApJ...747L..31S,2015IlbertA&A...579A...2I,2018BisigelloA&A...609A..82B}, which mirrors the bimodality of the main-sequence and starburst classes of galaxies. We follow the approach of \citet{2012SargentApJ...747L..31S}, parameterising the SFR probability distribution as a double Gaussian to describe the star-forming main sequence and starburst regimes

\begin{equation}\label{eq:Adpdsfr}
    \begin{array}{ll}
         {\rm P}({\rm SFR}|z,\Mstar) = &  \frac{A_{\rm MS}}{\sqrt{2\pi}\sigma_{\rm MS}}e^{-\log_{10}({\rm SFR}/\langle{\rm SFR}\rangle_{\rm MS})^2/2\sigma_{\rm MS}^2}\\[2ex]
         & + \frac{A_{\rm SB}}{\sqrt{2\pi}\sigma_{\rm SB}}e^{-\log_{10}({\rm SFR}/\langle{\rm SFR}\rangle_{\rm SB})^2/2\sigma_{\rm SB}^2}
    \end{array}\, ,
\end{equation}
where we use the parameters from a double Gaussian fit to the ${\rm P}({\rm sSFR})$ distribution of \citet{2017CaputiApJ...849...45C} at $z\sim4-5$: $A_{\rm MS}=0.6$ and $A_{\rm SB}=1-A_{\rm MS}=0.4$ are the fraction of main sequence galaxies and starburst galaxies, respectively, $\langle{\rm SFR}\rangle_{\rm MS}$ is the mean SFR of main sequence galaxies at a given stellar mass, $\langle{\rm SFR}\rangle_{\rm SB} = \langle{\rm SFR}\rangle_{\rm MS} + 1.1~{\rm dex}$ is the mean SFR of starburst galaxies at a given stellar mass, $\sigma_{\rm MS}=0.36~{\rm dex}$ and $\sigma_{\rm SB}=0.24~{\rm dex}$ are the dispersion of the individual main sequence and starburst distributions, respectively. (We note a very similar resulting UV LF is obtained if one includes the mass dependence of the contributions of the main sequence and starburst regimes, as well as their intrinsic scatter, by parameterising the distribution of \citealt{2025RinaldiApJ...981..161R} which is computed in four mass bins.)

The SFR function is then computed via the convolution of the SMF with the SFR probability distribution
\begin{equation}
    \Phi({\rm SFR}) = \int\!\Phi(\Mstar){\rm P}({\rm SFR}|\Mstar)\,{\rm d}\log_{10}(\Mstar)\,,
\end{equation}
where ${\rm P}({\rm SFR}|\Mstar)$ is given by equation (\ref{eq:Adpdsfr}). Here, we are neglecting to consider the fraction quenched of quenched galaxies, however, this is expected to be very low at this epoch \citep[][]{Weaver_2023}. The SFR function is then converted to an intrinsic UV LF following \citet{1998KennicuttARA&A..36..189K} and \citet{OkeAndGunn1983}. While we have assumed the SFR-to-UV relation of \citet{1998KennicuttARA&A..36..189K}, we find a modest variation of $\sim\pm0.3~{\rm dex}$ if we assume the conversion of \citet{2008CorteseMNRAS.390.1282C} or of \citet{2011MurphyApJ...737...67M} and \citet{2011HaoApJ...741..124H} as done in \citet{2024ClarkeMSApJ...977..133C}.

The resulting UV LFs assuming the SMF of \citet{shuntov2024cosmoswebstellarmassassembly} and the three star-forming main sequences derived using JWST observations from the JADES and CEERS surveys \citep[][]{2024ClarkeMSApJ...977..133C} are displayed in Figure \ref{fig:SMFtoUVLF}. The three determinations of the star-forming main sequence presented in \citet{2024ClarkeMSApJ...977..133C} use different tracers of star formation: the ${\rm H}\alpha$ luminosity, the UV luminosity, and the results from SED fitting. We observe that the all three UV LFs are consistent with literature galaxy UV LFs at the faint end through to the knee, with variance within the uncertainties of the SFR-to-UV conversion and UV LF. Of course, our derived UV LFs are the distribution of {\it intrinsic} UV luminosity and not {\it observed} UV luminosity. Therefore, the dust attenuation would reduce them in UV LF in the bright end, as massive galaxies will typically display the largest $\Av$, and the required mean $\Auv$ values of $\gtrsim0.7$ for $\Muv\lesssim-23~{\rm mag}$ is consistent with those of massive high-z galaxies \citep[e.g.][]{2025FisherMNRAS.539..109F}. If one were to use the parameters for $z\sim2$ from \citet{2012SargentApJ...747L..31S}, the resulting UV LF is consistent with the galaxy UV LFs of \citet{2022FinkelsteinUVLF} and \citet{2022HarikaneUVLF} across the luminosity range, including the bright end, however, this would only allow for minimal dust attenuation in massive galaxies. Therefore, we conclude that the SMF of \citet{shuntov2024cosmoswebstellarmassassembly} is not inconsistent with the literature galaxy UV LFs, representing the full census of UV luminous galaxies at this redshift, and can therefore be used as an anchor point to investigate the UV LF of AGN.

\section{Supplementary Comparison Plots With Theoretical Models}\label{App:SAMcomp}

\begin{figure*}
    \centering
    \includegraphics[width=0.9\textwidth]{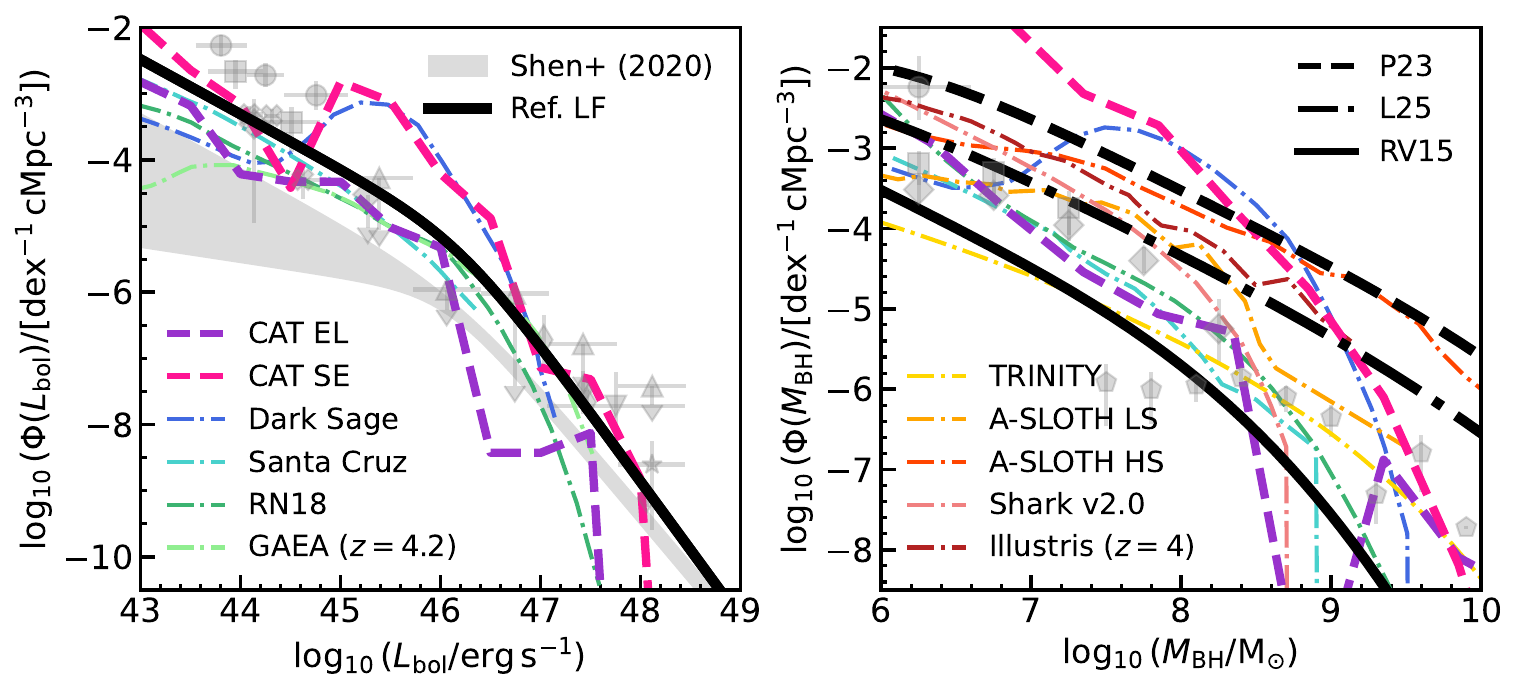}
    \caption{{\emph Left:} A comparison of our reference LF (black solid line), the \citet{Shen_2020} LF (grey shaded region), and the observational data from Fig. \ref{fig:K24bolLFfit} with the CAT Eddington-limited model (dashed purple line), CAT super-Eddington model (dashed pink line) and several other theoretical models at $z\sim5$ (dot-dashed lines). From this comparison, we see that our adopted reference luminosity function is broadly consistent with the predictions of semi analytic models. {\emph Right:} A comparison of the BHMFs inferred from the \citetalias{Pacucci2023_LRDs}, \citetalias{2025LiTipOf}, and \citetalias{Reines_2015} $\Mbh-\Mstar$ relations (black lines) and the observational data shown in Fig. \ref{fig:BHMF} with the CAT Eddington-limited and super-Eddington models, as well as several other theoretical models (dot-dashed lines). From this comparison, we observe that the predicted BHMFs from theoretical models span the region between the \citetalias{Reines_2015} and \citetalias{Pacucci2023_LRDs}-based BHMFs. The other theoretical models included are {\sc DarkSage} \citep[][]{2016StevensDarkSage1}, the Santa Cruz SAM \citep[][]{1999SomervilleSantaCruzSAMMNRAS.310.1087S,2022GabrielpillaiSantaCruzSAMMNRAS.517.6091G}, the SAM of \citet[][RN18]{2018RicarteNatarajanMNRAS.474.1995R}, {\sc GAEA} \citep[][]{2020FontanotMNRAS.496.3943F}, {\sc TRINITY} \citep[][]{2023MNRAS.518.2123Z}, the {\sc A-SLOTH} light and heavy-seeding models \citep[][]{2025JeonApJ...988..110J}, {\sc Shark} v2.0 \citep[][]{2018LagosSHARKMNRAS.481.3573L,2024LagosSharkV2MNRAS.531.3551L}, and the Illustris simulation \citep[][]{2015NelsonIllustrisA&C....13...12N,2015SijackiIllustrisMNRAS.452..575S}.}
    \label{fig:SAMcomp_MFLF}
\end{figure*}

\begin{figure*}
    \centering
    \includegraphics[width=0.9\textwidth]{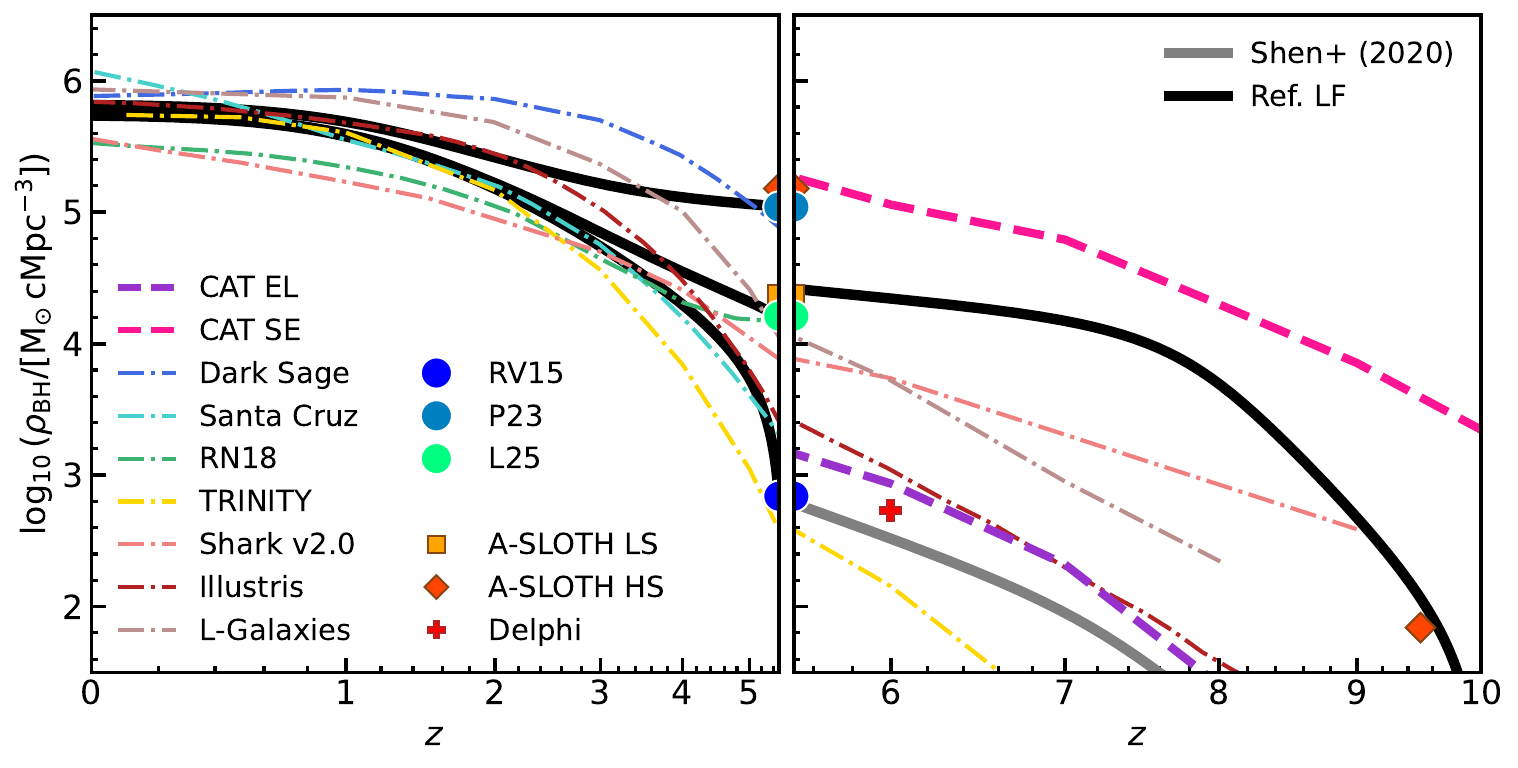}
    \caption{The evolution of the SMBH mass density as in Fig. \ref{fig:rhoBH_comp_toZ0}, with the $z=5.5$ values inferred using the three $\Mbh-\Mstar$ relations denoted by the circular markers and the prediction from the {\sc A-SLOTH} light and heavy-seeding models displayed by the square and diamond markers, respectively. In the left-hand panel, the black solid lines display the evolution of the SMBH mass density from the initial conditions inferred from the SMF via the three $\Mbh-\Mstar$ relations the evolution obtained using the reference luminosity function, whereas the coloured, dot-dashed lines show the predictions of several theoretical models (as in Fig. \ref{fig:SAMcomp_MFLF}), some of which were taken from the compilation in \citet[][the left-hand panel of their Figure 6]{2025PorrasValverderaXiv250411566P}. In the right-hand panel, the SMBH mass density obtained from our reference LF and the \citet{Shen_2020} LF are shown as the solid black and grey lines, respectively. These are compared to the CAT Eddington-limited and super-Eddington models \citet{2022MNRAS.511..616T,2024TrincaLRDs}, the {\sc A-SLOTH} models, {\sc Delphi} \citep[][]{2014DayalDelphiMNRAS.445.2545D}, {\sc Shark} v2.0, {\sc TRINITY}, {\sc L-Galaxies} \citep[][]{2025BonoliLGalarXiv250912325B}, and the Illustris simulation.}
    \label{fig:SAMcomp_rhoBH}
\end{figure*}


\bsp	
\label{lastpage}
\end{document}